\documentclass[longauth]{aa}

\usepackage{graphicx}

\usepackage{natbib}
\usepackage{scalerel}

\usepackage[table]{xcolor}
\usepackage{color}

\usepackage{txfonts}
\usepackage[pdfencoding=auto,psdextra]{hyperref}
\hypersetup{
    colorlinks=true,
    linkcolor=blue,
    filecolor=magenta,      
    urlcolor=blue,
    citecolor=blue
}
\makeatletter
\renewcommand*\aa@pageof{, page \thepage{} of \pageref*{LastPage}}
\makeatother

\usepackage[utf8]{inputenc}

\usepackage[switch, modulo]{lineno}
\usepackage{threeparttable} 
\linenumbers

\renewcommand{\linenumbers}[0]{}

\usepackage{euclid}
\usepackage{orcidlink}
\let\orcid\orcidlink

\begin{document}
%
%
\title{Euclid Data Release 1 (DR1)}
\subtitle{The MERge processing}


\author{Euclid Collaboration: M.~K\"ummel\orcid{0000-0003-2791-2117}\thanks{\email{mkuemmel@usm.lmu.de}}\inst{\ref{aff1}}
\and E.~Romelli\orcid{0000-0003-3069-9222}\inst{\ref{aff2}}
\and J.~Gracia-Carpio\orcid{0000-0003-4689-3134}\inst{\ref{aff3}}
\and D.~Hern\'andez~Lang\orcid{0000-0003-0194-6423}\inst{\ref{aff1}}
\and E.~Merlin\orcid{0000-0001-6870-8900}\inst{\ref{aff4}}
\and F.~Caro\orcid{0009-0003-1053-0507}\inst{\ref{aff5}}
\and E.~Soubrie\orcid{0000-0001-9295-1863}\inst{\ref{aff6}}
\and H.~Dole\orcid{0000-0002-9767-3839}\inst{\ref{aff6}}
\and S.~Galeotta\orcid{0000-0002-3748-5115}\inst{\ref{aff2}}
\and Y.~Fang\orcid{0000-0002-0334-6950}\inst{\ref{aff1}}
\and T.~Gasparetto\orcid{0000-0002-7913-4866}\inst{\ref{aff5}}
\and H.~Kong\orcid{0000-0001-8731-1212}\inst{\ref{aff7}}
\and T.~Dumerchat\orcid{0009-0008-0196-2953}\inst{\ref{aff7}}
\and C.~Tortora\orcid{0000-0001-7958-6531}\inst{\ref{aff8}}
\and L.~K.~Hunt\orcid{0000-0001-9162-2371}\inst{\ref{aff9}}
\and L.~Quilley\orcid{0009-0008-8375-8605}\inst{\ref{aff10}}
\and M.~Castellano\orcid{0000-0001-9875-8263}\inst{\ref{aff5}}
\and M.~D.~Lepinzan\orcid{0000-0003-1287-9801}\inst{\ref{aff2}}
\and P.~Dimauro\orcid{0000-0001-7399-2854}\inst{\ref{aff11},\ref{aff5}}
\and R.~Cabanac\orcid{0000-0001-6679-2600}\inst{\ref{aff12}}
\and L.~Maurin\orcid{0000-0002-8406-0857}\inst{\ref{aff6}}
\and T.~Vassallo\orcid{0000-0001-6512-6358}\inst{\ref{aff2},\ref{aff13}}
\and I.~A.~Zinchenko\orcid{0000-0002-2944-2449}\inst{\ref{aff14}}
\and N.~Aghanim\orcid{0000-0002-6688-8992}\inst{\ref{aff6}}
\and B.~Altieri\orcid{0000-0003-3936-0284}\inst{\ref{aff15}}
\and S.~Andreon\orcid{0000-0002-2041-8784}\inst{\ref{aff16}}
\and N.~Auricchio\orcid{0000-0003-4444-8651}\inst{\ref{aff17}}
\and H.~Aussel\orcid{0000-0002-1371-5705}\inst{\ref{aff18}}
\and C.~Baccigalupi\orcid{0000-0002-8211-1630}\inst{\ref{aff19},\ref{aff2},\ref{aff20},\ref{aff21}}
\and M.~Baldi\orcid{0000-0003-4145-1943}\inst{\ref{aff22},\ref{aff17},\ref{aff23}}
\and A.~Balestra\orcid{0000-0002-6967-261X}\inst{\ref{aff4}}
\and S.~Bardelli\orcid{0000-0002-8900-0298}\inst{\ref{aff17}}
\and P.~Battaglia\orcid{0000-0002-7337-5909}\inst{\ref{aff17}}
\and A.~N.~Belikov\inst{\ref{aff24},\ref{aff25}}
\and A.~Biviano\orcid{0000-0002-0857-0732}\inst{\ref{aff2},\ref{aff19}}
\and M.~Bolzonella\orcid{0000-0003-3278-4607}\inst{\ref{aff17}}
\and E.~Branchini\orcid{0000-0002-0808-6908}\inst{\ref{aff26},\ref{aff27},\ref{aff16}}
\and M.~Brescia\orcid{0000-0001-9506-5680}\inst{\ref{aff28},\ref{aff8}}
\and J.~Brinchmann\orcid{0000-0003-4359-8797}\inst{\ref{aff29},\ref{aff30},\ref{aff31}}
\and S.~Camera\orcid{0000-0003-3399-3574}\inst{\ref{aff32},\ref{aff33},\ref{aff34}}
\and G.~Ca\~nas-Herrera\orcid{0000-0003-2796-2149}\inst{\ref{aff35}}
\and V.~Capobianco\orcid{0000-0002-3309-7692}\inst{\ref{aff34}}
\and C.~Carbone\orcid{0000-0003-0125-3563}\inst{\ref{aff36}}
\and F.~J.~Castander\orcid{0000-0001-7316-4573}\inst{\ref{aff37},\ref{aff38}}
\and G.~Castignani\orcid{0000-0001-6831-0687}\inst{\ref{aff17}}
\and S.~Cavuoti\orcid{0000-0002-3787-4196}\inst{\ref{aff8},\ref{aff39}}
\and K.~C.~Chambers\orcid{0000-0001-6965-7789}\inst{\ref{aff40}}
\and A.~Cimatti\inst{\ref{aff41}}
\and C.~Colodro-Conde\inst{\ref{aff42}}
\and G.~Congedo\orcid{0000-0003-2508-0046}\inst{\ref{aff43}}
\and C.~J.~Conselice\orcid{0000-0003-1949-7638}\inst{\ref{aff10}}
\and L.~Conversi\orcid{0000-0002-6710-8476}\inst{\ref{aff44},\ref{aff15}}
\and Y.~Copin\orcid{0000-0002-5317-7518}\inst{\ref{aff45}}
\and A.~Costille\inst{\ref{aff46}}
\and F.~Courbin\orcid{0000-0003-0758-6510}\inst{\ref{aff47},\ref{aff48},\ref{aff49}}
\and H.~M.~Courtois\orcid{0000-0003-0509-1776}\inst{\ref{aff50}}
\and M.~Cropper\orcid{0000-0003-4571-9468}\inst{\ref{aff51}}
\and A.~Da~Silva\orcid{0000-0002-6385-1609}\inst{\ref{aff52},\ref{aff53}}
\and R.~da~Silva\orcid{0000-0003-4788-677X}\inst{\ref{aff5},\ref{aff54}}
\and H.~Degaudenzi\orcid{0000-0002-5887-6799}\inst{\ref{aff55}}
\and G.~De~Lucia\orcid{0000-0002-6220-9104}\inst{\ref{aff2}}
\and F.~Dubath\orcid{0000-0002-6533-2810}\inst{\ref{aff55}}
\and X.~Dupac\inst{\ref{aff15}}
\and S.~Dusini\orcid{0000-0002-1128-0664}\inst{\ref{aff56}}
\and S.~Escoffier\orcid{0000-0002-2847-7498}\inst{\ref{aff57}}
\and M.~Fabricius\orcid{0000-0002-7025-6058}\inst{\ref{aff3},\ref{aff1}}
\and M.~Farina\orcid{0000-0002-3089-7846}\inst{\ref{aff58}}
\and S.~Farrens\orcid{0000-0002-9594-9387}\inst{\ref{aff18}}
\and F.~Faustini\orcid{0000-0001-6274-5145}\inst{\ref{aff5}}
\and S.~Ferriol\inst{\ref{aff45}}
\and F.~Finelli\orcid{0000-0002-6694-3269}\inst{\ref{aff17},\ref{aff59}}
\and S.~Fotopoulou\orcid{0000-0002-9686-254X}\inst{\ref{aff60}}
\and N.~Fourmanoit\orcid{0009-0005-6816-6925}\inst{\ref{aff57}}
\and M.~Frailis\orcid{0000-0002-7400-2135}\inst{\ref{aff2}}
\and E.~Franceschi\orcid{0000-0002-0585-6591}\inst{\ref{aff17}}
\and M.~Fumana\orcid{0000-0001-6787-5950}\inst{\ref{aff36}}
\and L.~Gabarra\orcid{0000-0002-8486-8856}\inst{\ref{aff61}}
\and K.~George\orcid{0000-0002-1734-8455}\inst{\ref{aff13}}
\and B.~Gillis\orcid{0000-0002-4478-1270}\inst{\ref{aff43}}
\and C.~Giocoli\orcid{0000-0002-9590-7961}\inst{\ref{aff17},\ref{aff23}}
\and B.~R.~Granett\orcid{0000-0003-2694-9284}\inst{\ref{aff16}}
\and A.~Grazian\orcid{0000-0002-5688-0663}\inst{\ref{aff4}}
\and F.~Grupp\inst{\ref{aff3},\ref{aff1}}
\and L.~Guzzo\orcid{0000-0001-8264-5192}\inst{\ref{aff62},\ref{aff16},\ref{aff63}}
\and S.~Gwyn\orcid{0000-0001-8221-8406}\inst{\ref{aff64}}
\and W.~G.~Hartley\orcid{0000-0002-4563-5101}\inst{\ref{aff55}}
\and S.~V.~H.~Haugan\orcid{0000-0001-9648-7260}\inst{\ref{aff65}}
\and S.~Hemmati\orcid{0000-0003-2226-5395}\inst{\ref{aff66}}
\and H.~Hoekstra\orcid{0000-0002-0641-3231}\inst{\ref{aff35}}
\and W.~Holmes\orcid{0009-0007-8554-4646}\inst{\ref{aff67}}
\and I.~M.~Hook\orcid{0000-0002-2960-978X}\inst{\ref{aff68}}
\and F.~Hormuth\inst{\ref{aff69}}
\and A.~Hornstrup\orcid{0000-0002-3363-0936}\inst{\ref{aff70},\ref{aff71}}
\and P.~Hudelot\inst{\ref{aff72}}
\and M.~Huertas-Company\orcid{0000-0002-1416-8483}\inst{\ref{aff42},\ref{aff73},\ref{aff74}}
\and K.~Jahnke\orcid{0000-0003-3804-2137}\inst{\ref{aff75}}
\and M.~Jhabvala\inst{\ref{aff76}}
\and S.~Kermiche\orcid{0000-0002-0302-5735}\inst{\ref{aff57}}
\and A.~Kiessling\orcid{0000-0002-2590-1273}\inst{\ref{aff67}}
\and M.~Kilbinger\orcid{0000-0001-9513-7138}\inst{\ref{aff18}}
\and R.~Kohley\inst{\ref{aff15}}
\and B.~Kubik\orcid{0009-0006-5823-4880}\inst{\ref{aff45}}
\and M.~Kunz\orcid{0000-0002-3052-7394}\inst{\ref{aff77}}
\and H.~Kurki-Suonio\orcid{0000-0002-4618-3063}\inst{\ref{aff78},\ref{aff79}}
\and R.~Laureijs\inst{\ref{aff24}}
\and A.~M.~C.~Le~Brun\orcid{0000-0002-0936-4594}\inst{\ref{aff80}}
\and V.~Le~Brun\orcid{0000-0002-5027-1939}\inst{\ref{aff46}}
\and S.~Ligori\orcid{0000-0003-4172-4606}\inst{\ref{aff34}}
\and P.~B.~Lilje\orcid{0000-0003-4324-7794}\inst{\ref{aff65}}
\and V.~Lindholm\orcid{0000-0003-2317-5471}\inst{\ref{aff78},\ref{aff79}}
\and I.~Lloro\orcid{0000-0001-5966-1434}\inst{\ref{aff81}}
\and M.~Magliocchetti\orcid{0000-0001-9158-4838}\inst{\ref{aff58}}
\and G.~Mainetti\orcid{0000-0003-2384-2377}\inst{\ref{aff82}}
\and S.~Marcin\inst{\ref{aff83}}
\and O.~Marggraf\orcid{0000-0001-7242-3852}\inst{\ref{aff84}}
\and M.~Martinelli\orcid{0000-0002-6943-7732}\inst{\ref{aff5},\ref{aff85}}
\and N.~Martinet\orcid{0000-0003-2786-7790}\inst{\ref{aff46}}
\and F.~Marulli\orcid{0000-0002-8850-0303}\inst{\ref{aff86},\ref{aff17},\ref{aff23}}
\and R.~J.~Massey\orcid{0000-0002-6085-3780}\inst{\ref{aff87}}
\and N.~Mauri\orcid{0000-0001-8196-1548}\inst{\ref{aff41},\ref{aff23}}
\and S.~Maurogordato\inst{\ref{aff88}}
\and H.~J.~McCracken\orcid{0000-0002-9489-7765}\inst{\ref{aff72}}
\and C.~J.~R.~McPartland\orcid{0000-0003-0639-025X}\inst{\ref{aff71},\ref{aff89}}
\and E.~Medinaceli\orcid{0000-0002-4040-7783}\inst{\ref{aff17}}
\and S.~Mei\orcid{0000-0002-2849-559X}\inst{\ref{aff90},\ref{aff91}}
\and M.~Melchior\inst{\ref{aff92}}
\and M.~Meneghetti\orcid{0000-0003-1225-7084}\inst{\ref{aff17},\ref{aff23}}
\and G.~Meylan\orcid{0000-0001-6503-0209}\inst{\ref{aff93}}
\and P.~Monaco\orcid{0000-0003-2083-7564}\inst{\ref{aff94},\ref{aff2},\ref{aff20},\ref{aff19}}
\and A.~Mora\orcid{0000-0002-1922-8529}\inst{\ref{aff95}}
\and M.~Moresco\orcid{0000-0002-7616-7136}\inst{\ref{aff86},\ref{aff17}}
\and C.~Moretti\orcid{0000-0003-3314-8936}\inst{\ref{aff2},\ref{aff19},\ref{aff20}}
\and L.~Moscardini\orcid{0000-0002-3473-6716}\inst{\ref{aff86},\ref{aff17},\ref{aff23}}
\and E.~Munari\orcid{0000-0002-1751-5946}\inst{\ref{aff2},\ref{aff19}}
\and R.~Nakajima\orcid{0009-0009-1213-7040}\inst{\ref{aff84}}
\and C.~Neissner\orcid{0000-0001-8524-4968}\inst{\ref{aff7},\ref{aff96}}
\and R.~C.~Nichol\orcid{0000-0003-0939-6518}\inst{\ref{aff97}}
\and S.-M.~Niemi\orcid{0009-0005-0247-0086}\inst{\ref{aff98}}
\and C.~Padilla\orcid{0000-0001-7951-0166}\inst{\ref{aff7}}
\and S.~Paltani\orcid{0000-0002-8108-9179}\inst{\ref{aff55}}
\and F.~Pasian\orcid{0000-0002-4869-3227}\inst{\ref{aff2}}
\and W.~J.~Percival\orcid{0000-0002-0644-5727}\inst{\ref{aff99},\ref{aff100},\ref{aff101}}
\and V.~Pettorino\orcid{0000-0002-4203-9320}\inst{\ref{aff98}}
\and A.~Pezzotta\orcid{0000-0003-0726-2268}\inst{\ref{aff16}}
\and S.~Pires\orcid{0000-0002-0249-2104}\inst{\ref{aff18}}
\and G.~Polenta\orcid{0000-0003-4067-9196}\inst{\ref{aff54}}
\and M.~Poncet\inst{\ref{aff102}}
\and L.~A.~Popa\inst{\ref{aff103}}
\and L.~Pozzetti\orcid{0000-0001-7085-0412}\inst{\ref{aff17}}
\and F.~Raison\orcid{0000-0002-7819-6918}\inst{\ref{aff3}}
\and A.~Refregier\inst{\ref{aff104}}
\and A.~Renzi\orcid{0000-0001-9856-1970}\inst{\ref{aff105},\ref{aff56},\ref{aff17}}
\and J.~Rhodes\orcid{0000-0002-4485-8549}\inst{\ref{aff67}}
\and G.~Riccio\inst{\ref{aff8}}
\and I.~Risso\orcid{0000-0003-2525-7761}\inst{\ref{aff26},\ref{aff27},\ref{aff16}}
\and F.~Rizzo\orcid{0000-0002-9407-585X}\inst{\ref{aff2}}
\and M.~Roncarelli\orcid{0000-0001-9587-7822}\inst{\ref{aff17}}
\and C.~Rosset\orcid{0000-0003-0286-2192}\inst{\ref{aff90}}
\and B.~Rusholme\orcid{0000-0001-7648-4142}\inst{\ref{aff66}}
\and R.~Saglia\orcid{0000-0003-0378-7032}\inst{\ref{aff1},\ref{aff3}}
\and Z.~Sakr\orcid{0000-0002-4823-3757}\inst{\ref{aff106},\ref{aff12},\ref{aff107}}
\and A.~G.~S\'anchez\orcid{0000-0003-1198-831X}\inst{\ref{aff3}}
\and D.~Sapone\orcid{0000-0001-7089-4503}\inst{\ref{aff108}}
\and B.~Sartoris\orcid{0000-0003-1337-5269}\inst{\ref{aff88}}
\and M.~Schirmer\orcid{0000-0003-2568-9994}\inst{\ref{aff75}}
\and P.~Schneider\orcid{0000-0001-8561-2679}\inst{\ref{aff84}}
\and M.~Scodeggio\inst{\ref{aff36}}
\and A.~Secroun\orcid{0000-0003-0505-3710}\inst{\ref{aff57}}
\and E.~Sihvola\orcid{0000-0003-1804-7715}\inst{\ref{aff78},\ref{aff79}}
\and C.~Sirignano\orcid{0000-0002-0995-7146}\inst{\ref{aff105},\ref{aff56}}
\and G.~Sirri\orcid{0000-0003-2626-2853}\inst{\ref{aff23}}
\and L.~Stanco\orcid{0000-0002-9706-5104}\inst{\ref{aff56}}
\and P.~Tallada-Cresp\'{i}\orcid{0000-0002-1336-8328}\inst{\ref{aff109},\ref{aff96}}
\and A.~N.~Taylor\inst{\ref{aff43}}
\and H.~I.~Teplitz\orcid{0000-0002-7064-5424}\inst{\ref{aff110}}
\and I.~Tereno\orcid{0000-0002-4537-6218}\inst{\ref{aff52},\ref{aff53}}
\and N.~Tessore\orcid{0000-0002-9696-7931}\inst{\ref{aff51}}
\and S.~Toft\orcid{0000-0003-3631-7176}\inst{\ref{aff111},\ref{aff89}}
\and R.~Toledo-Moreo\orcid{0000-0002-2997-4859}\inst{\ref{aff112},\ref{aff113}}
\and F.~Torradeflot\orcid{0000-0003-1160-1517}\inst{\ref{aff96},\ref{aff109}}
\and A.~Tsyganov\inst{\ref{aff114}}
\and I.~Tutusaus\orcid{0000-0002-3199-0399}\inst{\ref{aff37},\ref{aff38},\ref{aff12}}
\and J.~Valiviita\orcid{0000-0001-6225-3693}\inst{\ref{aff78},\ref{aff79}}
\and G.~Verdoes~Kleijn\orcid{0000-0001-5803-2580}\inst{\ref{aff24}}
\and A.~Veropalumbo\orcid{0000-0003-2387-1194}\inst{\ref{aff16},\ref{aff27},\ref{aff26}}
\and Y.~Wang\orcid{0000-0002-4749-2984}\inst{\ref{aff66}}
\and J.~Weller\orcid{0000-0002-8282-2010}\inst{\ref{aff1},\ref{aff3}}
\and A.~Zacchei\orcid{0000-0003-0396-1192}\inst{\ref{aff2},\ref{aff19}}
\and G.~Zamorani\orcid{0000-0002-2318-301X}\inst{\ref{aff17}}
\and F.~M.~Zerbi\orcid{0000-0002-9996-973X}\inst{\ref{aff16}}
\and E.~Zucca\orcid{0000-0002-5845-8132}\inst{\ref{aff17}}
\and I.~T.~Andika\orcid{0000-0001-6102-9526}\inst{\ref{aff1}}
\and M.~Ballardini\orcid{0000-0003-4481-3559}\inst{\ref{aff115},\ref{aff116},\ref{aff17}}
\and C.~Benoist\orcid{0009-0003-0880-6295}\inst{\ref{aff88}}
\and P.~Bergamini\orcid{0000-0003-1383-9414}\inst{\ref{aff17}}
\and S.~Borgani\orcid{0000-0001-6151-6439}\inst{\ref{aff94},\ref{aff19},\ref{aff2},\ref{aff20},\ref{aff117}}
\and E.~Bozzo\orcid{0000-0002-8201-1525}\inst{\ref{aff55}}
\and C.~Burigana\orcid{0000-0002-3005-5796}\inst{\ref{aff118},\ref{aff59}}
\and M.~Calabrese\orcid{0000-0002-2637-2422}\inst{\ref{aff119},\ref{aff36}}
\and A.~Cappi\inst{\ref{aff88},\ref{aff17}}
\and T.~Castro\orcid{0000-0002-6292-3228}\inst{\ref{aff120},\ref{aff2},\ref{aff19}}
\and D.~Di~Ferdinando\inst{\ref{aff23}}
\and J.~A.~Escartin~Vigo\inst{\ref{aff3}}
\and G.~Fabbian\orcid{0000-0002-3255-4695}\inst{\ref{aff6}}
\and J.~Garc\'ia-Bellido\orcid{0000-0002-9370-8360}\inst{\ref{aff106}}
\and C.~M.~Gutierrez\orcid{0000-0001-7854-783X}\inst{\ref{aff42},\ref{aff121}}
\and A.~Hall\orcid{0000-0002-3139-8651}\inst{\ref{aff43}}
\and Y.~Kang\orcid{0009-0000-8588-7250}\inst{\ref{aff122}}
\and A.~Loureiro\orcid{0000-0002-4371-0876}\inst{\ref{aff123},\ref{aff124}}
\and J.~Macias-Perez\orcid{0000-0002-5385-2763}\inst{\ref{aff125}}
\and C.~Mancini\orcid{0000-0002-4297-0561}\inst{\ref{aff36}}
\and R.~Maoli\orcid{0000-0002-6065-3025}\inst{\ref{aff126},\ref{aff5}}
\and R.~B.~Metcalf\orcid{0000-0003-3167-2574}\inst{\ref{aff86},\ref{aff17}}
\and M.~P\"ontinen\orcid{0000-0001-5442-2530}\inst{\ref{aff78}}
\and E.~Sarpa\orcid{0000-0002-1256-655X}\inst{\ref{aff2}}
\and M.~Schefer\inst{\ref{aff55}}
\and V.~Scottez\orcid{0009-0008-3864-940X}\inst{\ref{aff127},\ref{aff128}}
\and M.~Sereno\orcid{0000-0003-0302-0325}\inst{\ref{aff17},\ref{aff23}}
\and K.~Tanidis\orcid{0000-0001-9843-5130}\inst{\ref{aff129}}
\and M.~Tenti\orcid{0000-0002-4254-5901}\inst{\ref{aff23}}
\and M.~Tucci\inst{\ref{aff55}}
\and J.~R.~Weaver\orcid{0000-0003-1614-196X}\inst{\ref{aff130}}
\and M.~Wiesmann\orcid{0009-0000-8199-5860}\inst{\ref{aff65}}
\and J.~A.~Acevedo~Barroso\orcid{0000-0002-9654-1711}\inst{\ref{aff67}}
\and Y.~Akrami\orcid{0000-0002-2407-7956}\inst{\ref{aff106},\ref{aff131}}
\and G.~Angora\orcid{0000-0002-0316-6562}\inst{\ref{aff8},\ref{aff115}}
\and S.~Anselmi\orcid{0000-0002-3579-9583}\inst{\ref{aff56},\ref{aff105},\ref{aff132}}
\and M.~Archidiacono\orcid{0000-0003-4952-9012}\inst{\ref{aff62},\ref{aff63}}
\and G.~Aric\`o\orcid{0000-0002-2802-2928}\inst{\ref{aff23}}
\and F.~Atrio-Barandela\orcid{0000-0002-2130-2513}\inst{\ref{aff133}}
\and M.~Aubert\orcid{0009-0002-7667-8814}\inst{\ref{aff45}}
\and M.~Baes\orcid{0000-0002-3930-2757}\inst{\ref{aff134}}
\and L.~Bazzanini\orcid{0000-0003-0727-0137}\inst{\ref{aff115},\ref{aff17}}
\and J.~R.~Bermejo-Climent\inst{\ref{aff135},\ref{aff136}}
\and D.~Bertacca\orcid{0000-0002-2490-7139}\inst{\ref{aff105},\ref{aff4},\ref{aff56}}
\and M.~Bethermin\orcid{0000-0002-3915-2015}\inst{\ref{aff137}}
\and F.~Beutler\orcid{0000-0003-0467-5438}\inst{\ref{aff43}}
\and M.~Bonici\orcid{0000-0002-8430-126X}\inst{\ref{aff99},\ref{aff36}}
\and F.~Bouch\`e\orcid{0000-0002-4663-1786}\inst{\ref{aff138},\ref{aff39}}
\and M.~L.~Brown\orcid{0000-0002-0370-8077}\inst{\ref{aff10}}
\and S.~Bruton\orcid{0000-0002-6503-5218}\inst{\ref{aff139}}
\and F.~Buitrago\orcid{0000-0002-2861-9812}\inst{\ref{aff140},\ref{aff141},\ref{aff142}}
\and A.~Calabro\orcid{0000-0003-2536-1614}\inst{\ref{aff5}}
\and B.~Camacho~Quevedo\orcid{0000-0002-8789-4232}\inst{\ref{aff19},\ref{aff21},\ref{aff2}}
\and P.~Carrilho\orcid{0000-0003-1339-0194}\inst{\ref{aff143}}
\and C.~S.~Carvalho\inst{\ref{aff142}}
\and F.~Cogato\orcid{0000-0003-4632-6113}\inst{\ref{aff17}}
\and T.~E.~Collett\orcid{0000-0001-5564-3140}\inst{\ref{aff144}}
\and S.~Conseil\orcid{0000-0002-3657-4191}\inst{\ref{aff145},\ref{aff45}}
\and A.~R.~Cooray\orcid{0000-0002-3892-0190}\inst{\ref{aff146}}
\and P.~Corcho-Caballero\orcid{0000-0001-6327-7080}\inst{\ref{aff24}}
\and M.~Correnti\orcid{0000-0001-6464-3257}\inst{\ref{aff5},\ref{aff54}}
\and B.~Csizi\orcid{0000-0003-3227-6581}\inst{\ref{aff122}}
\and O.~Cucciati\orcid{0000-0002-9336-7551}\inst{\ref{aff17}}
\and S.~Davini\orcid{0000-0003-3269-1718}\inst{\ref{aff27}}
\and T.~de~Boer\orcid{0000-0001-5486-2747}\inst{\ref{aff40}}
\and F.~De~Paolis\orcid{0000-0001-6460-7563}\inst{\ref{aff147},\ref{aff148},\ref{aff149}}
\and G.~Desprez\orcid{0000-0001-8325-1742}\inst{\ref{aff24}}
\and A.~D\'iaz-S\'anchez\orcid{0000-0003-0748-4768}\inst{\ref{aff150}}
\and S.~Di~Domizio\orcid{0000-0003-2863-5895}\inst{\ref{aff26},\ref{aff27}}
\and J.~M.~Diego\orcid{0000-0001-9065-3926}\inst{\ref{aff151}}
\and P.-A.~Duc\orcid{0000-0003-3343-6284}\inst{\ref{aff137}}
\and V.~Duret\orcid{0009-0009-0383-4960}\inst{\ref{aff57}}
\and M.~Y.~Elkhashab\orcid{0000-0001-9306-2603}\inst{\ref{aff2},\ref{aff20},\ref{aff94},\ref{aff19}}
\and A.~Enia\orcid{0000-0002-0200-2857}\inst{\ref{aff152},\ref{aff17}}
\and A.~Farina\orcid{0009-0000-3420-929X}\inst{\ref{aff16},\ref{aff27}}
\and A.~M.~N.~Ferguson\orcid{0000-0001-7934-1278}\inst{\ref{aff43}}
\and A.~Finoguenov\orcid{0000-0002-4606-5403}\inst{\ref{aff78}}
\and A.~Franco\orcid{0000-0002-4761-366X}\inst{\ref{aff147},\ref{aff148},\ref{aff149}}
\and A.~Fumagalli\orcid{0009-0004-0300-2535}\inst{\ref{aff2}}
\and Y.~Fu\orcid{0000-0002-0759-0504}\inst{\ref{aff35},\ref{aff24}}
\and K.~Ganga\orcid{0000-0001-8159-8208}\inst{\ref{aff90}}
\and R.~Gavazzi\orcid{0000-0002-5540-6935}\inst{\ref{aff46},\ref{aff72}}
\and E.~Gaztanaga\orcid{0000-0001-9632-0815}\inst{\ref{aff37},\ref{aff38},\ref{aff144}}
\and Z.~Ghaffari\orcid{0000-0002-6467-8078}\inst{\ref{aff2},\ref{aff19}}
\and F.~Giacomini\orcid{0000-0002-3129-2814}\inst{\ref{aff23}}
\and F.~Gianotti\orcid{0000-0003-4666-119X}\inst{\ref{aff17}}
\and G.~Gozaliasl\orcid{0000-0002-0236-919X}\inst{\ref{aff153},\ref{aff78}}
\and A.~Gruppuso\orcid{0000-0001-9272-5292}\inst{\ref{aff17},\ref{aff23}}
\and M.~Guidi\orcid{0000-0001-9408-1101}\inst{\ref{aff22},\ref{aff17}}
\and N.~A.~Hatch\orcid{0000-0001-5600-0534}\inst{\ref{aff154}}
\and C.~Hern\'andez-Monteagudo\orcid{0000-0001-5471-9166}\inst{\ref{aff121},\ref{aff42}}
\and T.~Heuberger\orcid{0009-0009-3474-8467}\inst{\ref{aff83}}
\and H.~Hildebrandt\orcid{0000-0002-9814-3338}\inst{\ref{aff155}}
\and J.~Hjorth\orcid{0000-0002-4571-2306}\inst{\ref{aff156}}
\and H.~Jansen\orcid{0009-0002-1332-7742}\inst{\ref{aff122}}
\and M.~Jauzac\orcid{0000-0003-1974-8732}\inst{\ref{aff157},\ref{aff87},\ref{aff158},\ref{aff159},\ref{aff12}}
\and J.~J.~E.~Kajava\orcid{0000-0002-3010-8333}\inst{\ref{aff160},\ref{aff161},\ref{aff162}}
\and V.~Kansal\orcid{0000-0002-4008-6078}\inst{\ref{aff163},\ref{aff164}}
\and D.~Karagiannis\orcid{0000-0002-4927-0816}\inst{\ref{aff165},\ref{aff115},\ref{aff166}}
\and J.~Kim\orcid{0000-0003-2776-2761}\inst{\ref{aff61}}
\and C.~C.~Kirkpatrick\inst{\ref{aff167}}
\and A.~Kov\'acs\orcid{0000-0002-5825-579X}\inst{\ref{aff136},\ref{aff135}}
\and I.~Kova{\v{c}}i{\'{c}}\orcid{0000-0001-6751-3263}\inst{\ref{aff134}}
\and K.~Koyama\orcid{0000-0001-6727-6915}\inst{\ref{aff144}}
\and S.~Kruk\orcid{0000-0001-8010-8879}\inst{\ref{aff15}}
\and M.~C.~Lam\orcid{0000-0002-9347-2298}\inst{\ref{aff43}}
\and M.~Lattanzi\orcid{0000-0003-1059-2532}\inst{\ref{aff116}}
\and S.~Lee\orcid{0000-0002-8289-740X}\inst{\ref{aff168},\ref{aff67}}
\and L.~Legrand\orcid{0000-0003-0610-5252}\inst{\ref{aff169},\ref{aff6}}
\and M.~Lembo\orcid{0000-0002-5271-5070}\inst{\ref{aff72}}
\and G.~Leroy\orcid{0009-0004-2523-4425}\inst{\ref{aff157},\ref{aff87}}
\and J.~Lesgourgues\orcid{0000-0001-7627-353X}\inst{\ref{aff170}}
\and T.~I.~Liaudat\orcid{0000-0002-9104-314X}\inst{\ref{aff171}}
\and L.~Linke\orcid{0000-0002-2622-8113}\inst{\ref{aff122}}
\and S.~J.~Liu\orcid{0000-0001-7680-2139}\inst{\ref{aff58}}
\and G.~Maggio\orcid{0000-0003-4020-4836}\inst{\ref{aff2}}
\and E.~A.~Magnier\orcid{0000-0002-7965-2815}\inst{\ref{aff40}}
\and F.~Mannucci\orcid{0000-0002-4803-2381}\inst{\ref{aff9}}
\and F.~R.~Marleau\orcid{0000-0002-1442-2947}\inst{\ref{aff122}}
\and C.~J.~A.~P.~Martins\orcid{0000-0002-4886-9261}\inst{\ref{aff172},\ref{aff29}}
\and M.~Miluzio\inst{\ref{aff15},\ref{aff173}}
\and G.~Morgante\inst{\ref{aff17}}
\and S.~Nadathur\orcid{0000-0001-9070-3102}\inst{\ref{aff144}}
\and K.~Naidoo\orcid{0000-0002-9182-1802}\inst{\ref{aff144},\ref{aff75}}
\and A.~Navarro-Alsina\orcid{0000-0002-3173-2592}\inst{\ref{aff84}}
\and A.~Nersesian\orcid{0000-0001-6843-409X}\inst{\ref{aff174},\ref{aff134}}
\and S.~Nesseris\orcid{0000-0002-0567-0324}\inst{\ref{aff106}}
\and L.~Nicastro\orcid{0000-0001-8534-6788}\inst{\ref{aff17}}
\and F.~Oppizzi\orcid{0000-0003-3904-8370}\inst{\ref{aff27},\ref{aff56},\ref{aff105}}
\and F.~Pace\orcid{0000-0001-8039-0480}\inst{\ref{aff32},\ref{aff33},\ref{aff34},\ref{aff53}}
\and L.~Pagano\orcid{0000-0003-1820-5998}\inst{\ref{aff115},\ref{aff116}}
\and D.~Paoletti\orcid{0000-0003-4761-6147}\inst{\ref{aff17},\ref{aff59}}
\and F.~Passalacqua\orcid{0000-0002-8606-4093}\inst{\ref{aff56}}
\and K.~Paterson\orcid{0000-0001-8340-3486}\inst{\ref{aff75}}
\and L.~Patrizii\inst{\ref{aff23}}
\and C.~Pattison\orcid{0000-0003-3272-2617}\inst{\ref{aff144}}
\and R.~Paviot\orcid{0009-0002-8108-3460}\inst{\ref{aff18},\ref{aff102}}
\and A.~Pisani\orcid{0000-0002-6146-4437}\inst{\ref{aff57}}
\and L.~Pizzuti\orcid{0000-0001-5654-7580}\inst{\ref{aff175}}
\and J.~J.~Popp\orcid{0000-0002-3724-1727}\inst{\ref{aff176}}
\and D.~Potter\orcid{0000-0002-0757-5195}\inst{\ref{aff177}}
\and M.~Poulain\orcid{0000-0002-7664-4510}\inst{\ref{aff178}}
\and G.~W.~Pratt\inst{\ref{aff18}}
\and S.~Quai\orcid{0000-0002-0449-8163}\inst{\ref{aff86},\ref{aff17}}
\and M.~Radovich\orcid{0000-0002-3585-866X}\inst{\ref{aff4}}
\and F.~Ricci\orcid{0000-0001-5742-5980}\inst{\ref{aff179},\ref{aff5}}
\and G.~Rodighiero\orcid{0000-0002-9415-2296}\inst{\ref{aff105},\ref{aff4}}
\and W.~Roster\orcid{0000-0002-9149-6528}\inst{\ref{aff3}}
\and M.~Sahl\'en\orcid{0000-0003-0973-4804}\inst{\ref{aff180}}
\and D.~B.~Sanders\orcid{0000-0002-1233-9998}\inst{\ref{aff40}}
\and J.~Schaye\orcid{0000-0002-0668-5560}\inst{\ref{aff35}}
\and A.~Schneider\orcid{0000-0001-7055-8104}\inst{\ref{aff177}}
\and D.~Sciotti\orcid{0009-0008-4519-2620}\inst{\ref{aff5},\ref{aff85}}
\and E.~Sellentin\orcid{0009-0002-2655-3458}\inst{\ref{aff181},\ref{aff35}}
\and S.~Serjeant\orcid{0000-0002-0517-7943}\inst{\ref{aff176}}
\and L.~C.~Smith\orcid{0000-0002-3259-2771}\inst{\ref{aff182}}
\and F.~Soldano\inst{\ref{aff72}}
\and I.~Szapudi\orcid{0000-0003-2274-0301}\inst{\ref{aff40}}
\and M.~Talia\orcid{0000-0003-4352-2063}\inst{\ref{aff86},\ref{aff17}}
\and F.~Tarsitano\orcid{0000-0002-5919-0238}\inst{\ref{aff183},\ref{aff104}}
\and G.~Testera\orcid{0000-0003-2970-766X}\inst{\ref{aff27}}
\and M.~Tewes\orcid{0000-0002-1155-8689}\inst{\ref{aff84}}
\and R.~Teyssier\orcid{0000-0001-7689-0933}\inst{\ref{aff184}}
\and S.~Tosi\orcid{0000-0002-7275-9193}\inst{\ref{aff26},\ref{aff16},\ref{aff27}}
\and A.~Troja\orcid{0000-0003-0239-4595}\inst{\ref{aff2}}
\and C.~Uhlemann\orcid{0000-0001-7831-1579}\inst{\ref{aff185},\ref{aff186}}
\and M.~Urbano\orcid{0000-0001-5640-0650}\inst{\ref{aff154},\ref{aff137}}
\and C.~Valieri\inst{\ref{aff23}}
\and A.~Venhola\orcid{0000-0001-6071-4564}\inst{\ref{aff178}}
\and D.~Vergani\orcid{0000-0003-0898-2216}\inst{\ref{aff17}}
\and P.~Vielzeuf\orcid{0000-0003-2035-9339}\inst{\ref{aff57}}
\and A.~Viitanen\orcid{0000-0001-9383-786X}\inst{\ref{aff55},\ref{aff167},\ref{aff5}}
\and E.~Vilenius\orcid{0000-0002-6184-7681}\inst{\ref{aff51}}
\and S.~Vinciguerra\orcid{0009-0005-4018-3184}\inst{\ref{aff46}}
\and M.~von~Wietersheim-Kramsta\orcid{0000-0003-4986-5091}\inst{\ref{aff87},\ref{aff157}}
\and N.~A.~Walton\orcid{0000-0003-3983-8778}\inst{\ref{aff182}}
\and L.~Wang\orcid{0000-0002-6736-9158}\inst{\ref{aff187},\ref{aff24}}
\and A.~H.~Wright\orcid{0000-0001-7363-7932}\inst{\ref{aff155}}
\and H.~W.~Yeung\orcid{0000-0002-4993-9014}\inst{\ref{aff43}}
\and D.~Scott\orcid{0000-0002-6878-9840}\inst{\ref{aff188}}}
										   
\institute{Universit\"ats-Sternwarte M\"unchen, Fakult\"at f\"ur Physik, Ludwig-Maximilians-Universit\"at M\"unchen, Scheinerstr.~1, 81679 M\"unchen, Germany\label{aff1}
\and
INAF-Osservatorio Astronomico di Trieste, Via G. B. Tiepolo 11, 34143 Trieste, Italy\label{aff2}
\and
Max Planck Institute for Extraterrestrial Physics, Giessenbachstr. 1, 85748 Garching, Germany\label{aff3}
\and
INAF-Osservatorio Astronomico di Padova, Via dell'Osservatorio 5, 35122 Padova, Italy\label{aff4}
\and
INAF-Osservatorio Astronomico di Roma, Via Frascati 33, 00078 Monteporzio Catone, Italy\label{aff5}
\and
Universit\'e Paris-Saclay, CNRS, Institut d'astrophysique spatiale, 91405, Orsay, France\label{aff6}
\and
Institut de F\'{i}sica d'Altes Energies (IFAE), The Barcelona Institute of Science and Technology, Campus UAB, 08193 Bellaterra (Barcelona), Spain\label{aff7}
\and
INAF-Osservatorio Astronomico di Capodimonte, Via Moiariello 16, 80131 Napoli, Italy\label{aff8}
\and
INAF-Osservatorio Astrofisico di Arcetri, Largo E. Fermi 5, 50125, Firenze, Italy\label{aff9}
\and
Jodrell Bank Centre for Astrophysics, Department of Physics and Astronomy, University of Manchester, Oxford Road, Manchester M13 9PL, UK\label{aff10}
\and
Observatorio Nacional, Rua General Jose Cristino, 77-Bairro Imperial de Sao Cristovao, Rio de Janeiro, 20921-400, Brazil\label{aff11}
\and
Institut de Recherche en Astrophysique et Plan\'etologie (IRAP), Universit\'e de Toulouse, CNRS, UPS, CNES, 14 Av. Edouard Belin, 31400 Toulouse, France\label{aff12}
\and
University Observatory, LMU Faculty of Physics, Scheinerstr.~1, 81679 Munich, Germany\label{aff13}
\and
Astronomisches Rechen-Institut, Zentrum f\"ur Astronomie der Universit\"at Heidelberg, M\"onchhofstr. 12-14, 69120 Heidelberg, Germany\label{aff14}
\and
ESAC/ESA, Camino Bajo del Castillo, s/n., Urb. Villafranca del Castillo, 28692 Villanueva de la Ca\~nada, Madrid, Spain\label{aff15}
\and
INAF-Osservatorio Astronomico di Brera, Via Brera 28, 20122 Milano, Italy\label{aff16}
\and
INAF-Osservatorio di Astrofisica e Scienza dello Spazio di Bologna, Via Piero Gobetti 93/3, 40129 Bologna, Italy\label{aff17}
\and
Universit\'e Paris-Saclay, Universit\'e Paris Cit\'e, CEA, CNRS, AIM, 91191, Gif-sur-Yvette, France\label{aff18}
\and
IFPU, Institute for Fundamental Physics of the Universe, via Beirut 2, 34151 Trieste, Italy\label{aff19}
\and
INFN, Sezione di Trieste, Via Valerio 2, 34127 Trieste TS, Italy\label{aff20}
\and
SISSA, International School for Advanced Studies, Via Bonomea 265, 34136 Trieste TS, Italy\label{aff21}
\and
Dipartimento di Fisica e Astronomia, Universit\`a di Bologna, Via Gobetti 93/2, 40129 Bologna, Italy\label{aff22}
\and
INFN-Sezione di Bologna, Viale Berti Pichat 6/2, 40127 Bologna, Italy\label{aff23}
\and
Kapteyn Astronomical Institute, University of Groningen, PO Box 800, 9700 AV Groningen, The Netherlands\label{aff24}
\and
ATG Europe BV, Huygensstraat 34, 2201 DK Noordwijk, The Netherlands\label{aff25}
\and
Dipartimento di Fisica, Universit\`a di Genova, Via Dodecaneso 33, 16146, Genova, Italy\label{aff26}
\and
INFN-Sezione di Genova, Via Dodecaneso 33, 16146, Genova, Italy\label{aff27}
\and
Department of Physics "E. Pancini", University Federico II, Via Cinthia 6, 80126, Napoli, Italy\label{aff28}
\and
Instituto de Astrof\'isica e Ci\^encias do Espa\c{c}o, Universidade do Porto, CAUP, Rua das Estrelas, PT4150-762 Porto, Portugal\label{aff29}
\and
Faculdade de Ci\^encias da Universidade do Porto, Rua do Campo de Alegre, 4150-007 Porto, Portugal\label{aff30}
\and
European Southern Observatory, Karl-Schwarzschild-Str.~2, 85748 Garching, Germany\label{aff31}
\and
Dipartimento di Fisica, Universit\`a degli Studi di Torino, Via P. Giuria 1, 10125 Torino, Italy\label{aff32}
\and
INFN-Sezione di Torino, Via P. Giuria 1, 10125 Torino, Italy\label{aff33}
\and
INAF-Osservatorio Astrofisico di Torino, Via Osservatorio 20, 10025 Pino Torinese (TO), Italy\label{aff34}
\and
Leiden Observatory, Leiden University, Einsteinweg 55, 2333 CC Leiden, The Netherlands\label{aff35}
\and
INAF-IASF Milano, Via Alfonso Corti 12, 20133 Milano, Italy\label{aff36}
\and
Institute of Space Sciences (ICE, CSIC), Campus UAB, Carrer de Can Magrans, s/n, 08193 Barcelona, Spain\label{aff37}
\and
Institut d'Estudis Espacials de Catalunya (IEEC),  Edifici RDIT, Campus UPC, 08860 Castelldefels, Barcelona, Spain\label{aff38}
\and
INFN -- Sezione di Napoli, Via Cinthia 6, 80126, Napoli, Italy\label{aff39}
\and
Institute for Astronomy, University of Hawaii, 2680 Woodlawn Drive, Honolulu, HI 96822, USA\label{aff40}
\and
Dipartimento di Fisica e Astronomia "Augusto Righi" - Alma Mater Studiorum Universit\`a di Bologna, Viale Berti Pichat 6/2, 40127 Bologna, Italy\label{aff41}
\and
Instituto de Astrof\'{\i}sica de Canarias, E-38205 La Laguna, Tenerife, Spain\label{aff42}
\and
Institute for Astronomy, University of Edinburgh, Royal Observatory, Blackford Hill, Edinburgh EH9 3HJ, UK\label{aff43}
\and
European Space Agency/ESRIN, Largo Galileo Galilei 1, 00044 Frascati, Roma, Italy\label{aff44}
\and
Universit\'e Lyon 1, CNRS/IN2P3, IP2I Lyon, Villeurbanne, France\label{aff45}
\and
Aix-Marseille Universit\'e, CNRS, CNES, LAM, Marseille, France\label{aff46}
\and
Institut de Ci\`{e}ncies del Cosmos (ICCUB), Universitat de Barcelona (IEEC-UB), Mart\'{i} i Franqu\`{e}s 1, 08028 Barcelona, Spain\label{aff47}
\and
Instituci\'o Catalana de Recerca i Estudis Avan\c{c}ats (ICREA), Passeig de Llu\'{\i}s Companys 23, 08010 Barcelona, Spain\label{aff48}
\and
Institut de Ciencies de l'Espai (IEEC-CSIC), Campus UAB, Carrer de Can Magrans, s/n Cerdanyola del Vall\'es, 08193 Barcelona, Spain\label{aff49}
\and
Universit\'e Lyon 1, CNRS/IN2P3, IUF, IP2I Lyon, Villeurbanne, France \label{aff50}
\and
Mullard Space Science Laboratory, University College London, Holmbury St Mary, Dorking, Surrey RH5 6NT, UK\label{aff51}
\and
Departamento de F\'isica, Faculdade de Ci\^encias, Universidade de Lisboa, Edif\'icio C8, Campo Grande, PT1749-016 Lisboa, Portugal\label{aff52}
\and
Instituto de Astrof\'isica e Ci\^encias do Espa\c{c}o, Faculdade de Ci\^encias, Universidade de Lisboa, Campo Grande, 1749-016 Lisboa, Portugal\label{aff53}
\and
Space Science Data Center, Italian Space Agency, via del Politecnico snc, 00133 Roma, Italy\label{aff54}
\and
Department of Astronomy, University of Geneva, ch. d'Ecogia 16, 1290 Versoix, Switzerland\label{aff55}
\and
INFN-Padova, Via Marzolo 8, 35131 Padova, Italy\label{aff56}
\and
Aix-Marseille Universit\'e, CNRS/IN2P3, CPPM, Marseille, France\label{aff57}
\and
INAF-Istituto di Astrofisica e Planetologia Spaziali, via del Fosso del Cavaliere, 100, 00100 Roma, Italy\label{aff58}
\and
INFN-Bologna, Via Irnerio 46, 40126 Bologna, Italy\label{aff59}
\and
School of Physics, HH Wills Physics Laboratory, University of Bristol, Tyndall Avenue, Bristol, BS8 1TL, UK\label{aff60}
\and
Department of Physics, University of Oxford, Keble Road, Oxford OX1 3RH, UK\label{aff61}
\and
Dipartimento di Fisica "Aldo Pontremoli", Universit\`a degli Studi di Milano, Via Celoria 16, 20133 Milano, Italy\label{aff62}
\and
INFN-Sezione di Milano, Via Celoria 16, 20133 Milano, Italy\label{aff63}
\and
National Research Council, Herzberg Astronomy and Astrophysics Research Centre, 5071 W. Saanich Rd. Victoria, BC, V9E 2E7, Canada\label{aff64}
\and
Institute of Theoretical Astrophysics, University of Oslo, P.O. Box 1029 Blindern, 0315 Oslo, Norway\label{aff65}
\and
Caltech/IPAC, 1200 E. California Blvd., Pasadena, CA 91125, USA\label{aff66}
\and
Jet Propulsion Laboratory, California Institute of Technology, 4800 Oak Grove Drive, Pasadena, CA, 91109, USA\label{aff67}
\and
School of Physics and Astronomy, Lancaster University, Lancaster, LA1 4YB, UK\label{aff68}
\and
Felix Hormuth Engineering, Goethestr. 17, 69181 Leimen, Germany\label{aff69}
\and
Technical University of Denmark, Elektrovej 327, 2800 Kgs. Lyngby, Denmark\label{aff70}
\and
Cosmic Dawn Center (DAWN), Denmark\label{aff71}
\and
Institut d'Astrophysique de Paris, UMR 7095, CNRS, and Sorbonne Universit\'e, 98 bis boulevard Arago, 75014 Paris, France\label{aff72}
\and
Universit\'e PSL, Observatoire de Paris, Sorbonne Universit\'e, CNRS, LERMA, 75014, Paris, France\label{aff73}
\and
Universit\'e Paris-Cit\'e, 5 Rue Thomas Mann, 75013, Paris, France\label{aff74}
\and
Max-Planck-Institut f\"ur Astronomie, K\"onigstuhl 17, 69117 Heidelberg, Germany\label{aff75}
\and
NASA Goddard Space Flight Center, Greenbelt, MD 20771, USA\label{aff76}
\and
Universit\'e de Gen\`eve, D\'epartement de Physique Th\'eorique and Centre for Astroparticle Physics, 24 quai Ernest-Ansermet, CH-1211 Gen\`eve 4, Switzerland\label{aff77}
\and
Department of Physics, P.O. Box 64, University of Helsinki, 00014 Helsinki, Finland\label{aff78}
\and
Helsinki Institute of Physics, Gustaf H{\"a}llstr{\"o}min katu 2, University of Helsinki, 00014 Helsinki, Finland\label{aff79}
\and
Laboratoire d'etude de l'Univers et des phenomenes eXtremes, Observatoire de Paris, Universit\'e PSL, Sorbonne Universit\'e, CNRS, 92190 Meudon, France\label{aff80}
\and
SKAO, Jodrell Bank, Lower Withington, Macclesfield SK11 9FT, UK\label{aff81}
\and
Centre de Calcul de l'IN2P3/CNRS, 21 avenue Pierre de Coubertin 69627 Villeurbanne Cedex, France\label{aff82}
\and
University of Applied Sciences and Arts of Northwestern Switzerland, School of Computer Science, 5210 Windisch, Switzerland\label{aff83}
\and
Universit\"at Bonn, Argelander-Institut f\"ur Astronomie, Auf dem H\"ugel 71, 53121 Bonn, Germany\label{aff84}
\and
INFN-Sezione di Roma, Piazzale Aldo Moro, 2 - c/o Dipartimento di Fisica, Edificio G. Marconi, 00185 Roma, Italy\label{aff85}
\and
Dipartimento di Fisica e Astronomia "Augusto Righi" - Alma Mater Studiorum Universit\`a di Bologna, via Piero Gobetti 93/2, 40129 Bologna, Italy\label{aff86}
\and
Department of Physics, Institute for Computational Cosmology, Durham University, South Road, Durham, DH1 3LE, UK\label{aff87}
\and
Universit\'e C\^{o}te d'Azur, Observatoire de la C\^{o}te d'Azur, CNRS, Laboratoire Lagrange, Bd de l'Observatoire, CS 34229, 06304 Nice cedex 4, France\label{aff88}
\and
Niels Bohr Institute, University of Copenhagen, Jagtvej 128, 2200 Copenhagen, Denmark\label{aff89}
\and
Universit\'e Paris Cit\'e, CNRS, Astroparticule et Cosmologie, 75013 Paris, France\label{aff90}
\and
CNRS-UCB International Research Laboratory, Centre Pierre Bin\'etruy, IRL2007, CPB-IN2P3, Berkeley, USA\label{aff91}
\and
University of Applied Sciences and Arts of Northwestern Switzerland, School of Engineering, 5210 Windisch, Switzerland\label{aff92}
\and
Institute of Physics, Laboratory of Astrophysics, Ecole Polytechnique F\'ed\'erale de Lausanne (EPFL), Observatoire de Sauverny, 1290 Versoix, Switzerland\label{aff93}
\and
Dipartimento di Fisica - Sezione di Astronomia, Universit\`a di Trieste, Via Tiepolo 11, 34131 Trieste, Italy\label{aff94}
\and
Telespazio UK S.L. for European Space Agency (ESA), Camino bajo del Castillo, s/n, Urbanizacion Villafranca del Castillo, Villanueva de la Ca\~nada, 28692 Madrid, Spain\label{aff95}
\and
Port d'Informaci\'{o} Cient\'{i}fica, Campus UAB, C. Albareda s/n, 08193 Bellaterra (Barcelona), Spain\label{aff96}
\and
School of Mathematics and Physics, University of Surrey, Guildford, Surrey, GU2 7XH, UK\label{aff97}
\and
European Space Agency/ESTEC, Keplerlaan 1, 2201 AZ Noordwijk, The Netherlands\label{aff98}
\and
Waterloo Centre for Astrophysics, University of Waterloo, Waterloo, Ontario N2L 3G1, Canada\label{aff99}
\and
Department of Physics and Astronomy, University of Waterloo, Waterloo, Ontario N2L 3G1, Canada\label{aff100}
\and
Perimeter Institute for Theoretical Physics, Waterloo, Ontario N2L 2Y5, Canada\label{aff101}
\and
Centre National d'Etudes Spatiales -- Centre spatial de Toulouse, 18 avenue Edouard Belin, 31401 Toulouse Cedex 9, France\label{aff102}
\and
Institute of Space Science, Str. Atomistilor, nr. 409 M\u{a}gurele, Ilfov, 077125, Romania\label{aff103}
\and
Institute for Particle Physics and Astrophysics, Dept. of Physics, ETH Zurich, Wolfgang-Pauli-Strasse 27, 8093 Zurich, Switzerland\label{aff104}
\and
Dipartimento di Fisica e Astronomia "G. Galilei", Universit\`a di Padova, Via Marzolo 8, 35131 Padova, Italy\label{aff105}
\and
Instituto de F\'isica Te\'orica UAM-CSIC, Campus de Cantoblanco, 28049 Madrid, Spain\label{aff106}
\and
Universit\'e St Joseph; Faculty of Sciences, Beirut, Lebanon\label{aff107}
\and
Departamento de F\'isica, FCFM, Universidad de Chile, Blanco Encalada 2008, Santiago, Chile\label{aff108}
\and
Centro de Investigaciones Energ\'eticas, Medioambientales y Tecnol\'ogicas (CIEMAT), Avenida Complutense 40, 28040 Madrid, Spain\label{aff109}
\and
Infrared Processing and Analysis Center, California Institute of Technology, Pasadena, CA 91125, USA\label{aff110}
\and
Cosmic Dawn Center (DAWN)\label{aff111}
\and
Universidad Polit\'ecnica de Cartagena, Departamento de Electr\'onica y Tecnolog\'ia de Computadoras,  Plaza del Hospital 1, 30202 Cartagena, Spain\label{aff112}
\and
European University of Technology EUt+, European Union\label{aff113}
\and
Centre for Information Technology, University of Groningen, P.O. Box 11044, 9700 CA Groningen, The Netherlands\label{aff114}
\and
Dipartimento di Fisica e Scienze della Terra, Universit\`a degli Studi di Ferrara, Via Giuseppe Saragat 1, 44122 Ferrara, Italy\label{aff115}
\and
Istituto Nazionale di Fisica Nucleare, Sezione di Ferrara, Via Giuseppe Saragat 1, 44122 Ferrara, Italy\label{aff116}
\and
ICSC - Centro Nazionale di Ricerca in High Performance Computing, Big Data e Quantum Computing, Via Magnanelli 2, Bologna, Italy\label{aff117}
\and
INAF, Istituto di Radioastronomia, Via Piero Gobetti 101, 40129 Bologna, Italy\label{aff118}
\and
Astronomical Observatory of the Autonomous Region of the Aosta Valley (OAVdA), Loc. Lignan 39, I-11020, Nus (Aosta Valley), Italy\label{aff119}
\and
Department of Mathematical Physics, Institute of Physics, University of S\~ao Paulo, R. do Mat\~ao 1371, 05508-090, S\~ao Paulo, SP, Brazil\label{aff120}
\and
Universidad de La Laguna, Dpto. Astrof\'\i sica, E-38206 La Laguna, Tenerife, Spain\label{aff121}
\and
Universit\"at Innsbruck, Institut f\"ur Astro- und Teilchenphysik, Technikerstr. 25/8, 6020 Innsbruck, Austria\label{aff122}
\and
Oskar Klein Centre for Cosmoparticle Physics, Department of Physics, Stockholm University, Stockholm, SE-106 91, Sweden\label{aff123}
\and
Astrophysics Group, Blackett Laboratory, Imperial College London, London SW7 2AZ, UK\label{aff124}
\and
Univ. Grenoble Alpes, CNRS, Grenoble INP, LPSC-IN2P3, 53, Avenue des Martyrs, 38000 Grenoble, France\label{aff125}
\and
Dipartimento di Fisica, Sapienza Universit\`a di Roma, Piazzale Aldo Moro 2, 00185 Roma, Italy\label{aff126}
\and
Institut d'Astrophysique de Paris, 98bis Boulevard Arago, 75014, Paris, France\label{aff127}
\and
ICL, Junia, Universit\'e Catholique de Lille, LITL, 59000 Lille, France\label{aff128}
\and
Center for Astrophysics and Cosmology, University of Nova Gorica, Nova Gorica, Slovenia\label{aff129}
\and
MIT Kavli Institute for Astrophysics and Space Research, Massachusetts Institute of Technology, Cambridge, MA 02139, USA\label{aff130}
\and
CERCA/ISO, Department of Physics, Case Western Reserve University, 10900 Euclid Avenue, Cleveland, OH 44106, USA\label{aff131}
\and
Laboratoire Univers et Th\'eorie, Observatoire de Paris, Universit\'e PSL, Universit\'e Paris Cit\'e, CNRS, 92190 Meudon, France\label{aff132}
\and
Departamento de F{\'\i}sica Fundamental. Universidad de Salamanca. Plaza de la Merced s/n. 37008 Salamanca, Spain\label{aff133}
\and
Universiteit Gent, Department of Physics and Astronomy, Proeftuinstraat 86 N3, 9000 Ghent, Belgium
\label{aff134}
\and
Konkoly Observatory, HUN-REN CSFK, MTA Centre of Excellence, Budapest, Konkoly Thege Mikl\'os {\'u}t 15-17. H-1121, Hungary\label{aff135}
\and
MTA-CSFK Lend\"ulet Large-Scale Structure Research Group, Konkoly-Thege Mikl\'os \'ut 15-17, H-1121 Budapest, Hungary\label{aff136}
\and
Universit\'e de Strasbourg, CNRS, Observatoire astronomique de Strasbourg, UMR 7550, 67000 Strasbourg, France\label{aff137}
\and
Scuola Superiore Meridionale, Via Mezzocannone 4, 80138, Napoli, Italy\label{aff138}
\and
California Institute of Technology, 1200 E California Blvd, Pasadena, CA 91125, USA\label{aff139}
\and
Departamento de F\'{i}sica Te\'{o}rica, At\'{o}mica y \'{O}ptica, Universidad de Valladolid, 47011 Valladolid, Spain\label{aff140}
\and
Laboratory for Disruptive Interdisciplinary Science (LaDIS), Universidad de Valladolid, 47011 Valladolid, Spain\label{aff141}
\and
Instituto de Astrof\'isica e Ci\^encias do Espa\c{c}o, Faculdade de Ci\^encias, Universidade de Lisboa, Tapada da Ajuda, 1349-018 Lisboa, Portugal\label{aff142}
\and
Department of Physics, Astronomy and Mathematics, University of Hertfordshire, College Lane, Hatfield AL10 9AB, UK\label{aff143}
\and
Institute of Cosmology and Gravitation, University of Portsmouth, Portsmouth PO1 3FX, UK\label{aff144}
\and
Centre de Recherche Astrophysique de Lyon, UMR5574, CNRS, Universit\'e Claude Bernard Lyon 1, ENS de Lyon, 69230, Saint-Genis-Laval, France\label{aff145}
\and
Department of Physics \& Astronomy, University of California Irvine, Irvine CA 92697, USA\label{aff146}
\and
Department of Mathematics and Physics E. De Giorgi, University of Salento, Via per Arnesano, CP-I93, 73100, Lecce, Italy\label{aff147}
\and
INFN, Sezione di Lecce, Via per Arnesano, CP-193, 73100, Lecce, Italy\label{aff148}
\and
INAF-Sezione di Lecce, c/o Dipartimento Matematica e Fisica, Via per Arnesano, 73100, Lecce, Italy\label{aff149}
\and
Departamento F\'isica Aplicada, Universidad Polit\'ecnica de Cartagena, Campus Muralla del Mar, 30202 Cartagena, Murcia, Spain\label{aff150}
\and
Instituto de F\'isica de Cantabria, Edificio Juan Jord\'a, Avenida de los Castros, 39005 Santander, Spain\label{aff151}
\and
CEA Saclay, DFR/IRFU, Service d'Astrophysique, Bat. 709, 91191 Gif-sur-Yvette, France\label{aff152}
\and
Department of Computer Science, Aalto University, PO Box 15400, Espoo, FI-00 076, Finland\label{aff153}
\and
School of Physics and Astronomy, University of Nottingham, University Park, Nottingham NG7 2RD, UK\label{aff154}
\and
Ruhr University Bochum, Faculty of Physics and Astronomy, Astronomical Institute (AIRUB), German Centre for Cosmological Lensing (GCCL), 44780 Bochum, Germany\label{aff155}
\and
DARK, Niels Bohr Institute, University of Copenhagen, Jagtvej 155, 2200 Copenhagen, Denmark\label{aff156}
\and
Department of Physics, Centre for Extragalactic Astronomy, Durham University, South Road, Durham, DH1 3LE, UK\label{aff157}
\and
Astrophysics Research Centre, University of KwaZulu-Natal, Westville Campus, Durban 4041, South Africa\label{aff158}
\and
School of Mathematics, Statistics \& Computer Science, University of KwaZulu-Natal, Westville Campus, Durban 4041, South Africa\label{aff159}
\and
Department of Physics and Astronomy, Vesilinnantie 5, University of Turku, 20014 Turku, Finland\label{aff160}
\and
Finnish Centre for Astronomy with ESO (FINCA), Quantum, Vesilinnantie 5, University of Turku, 20014 Turku, Finland\label{aff161}
\and
Serco for European Space Agency (ESA), Camino bajo del Castillo, s/n, Urbanizacion Villafranca del Castillo, Villanueva de la Ca\~nada, 28692 Madrid, Spain\label{aff162}
\and
ARC Centre of Excellence for Dark Matter Particle Physics, Melbourne, Australia\label{aff163}
\and
Centre for Astrophysics \& Supercomputing, Swinburne University of Technology,  Hawthorn, Victoria 3122, Australia\label{aff164}
\and
Van Swinderen Institute, University of Groningen, Nijenborgh 3, 9747 AG Groningen, The Netherlands\label{aff165}
\and
Department of Physics and Astronomy, University of the Western Cape, Bellville, Cape Town, 7535, South Africa\label{aff166}
\and
Department of Physics and Helsinki Institute of Physics, Gustaf H\"allstr\"omin katu 2, University of Helsinki, 00014 Helsinki, Finland\label{aff167}
\and
Ohio University, Physics \& Astronomy Department,1 Ohio University, Athens, OH 45701, USA\label{aff168}
\and
Brazilian Center for Research in Physics (CBPF), Dr. Xavier Sigaud st. 150, zip 22290-180, Rio de Janeiro, RJ, Brazil\label{aff169}
\and
Institute for Theoretical Particle Physics and Cosmology (TTK), RWTH Aachen University, 52056 Aachen, Germany\label{aff170}
\and
IRFU, CEA, Universit\'e Paris-Saclay 91191 Gif-sur-Yvette Cedex, France\label{aff171}
\and
Centro de Astrof\'{\i}sica da Universidade do Porto, Rua das Estrelas, 4150-762 Porto, Portugal\label{aff172}
\and
HE Space for European Space Agency (ESA), Camino bajo del Castillo, s/n, Urbanizacion Villafranca del Castillo, Villanueva de la Ca\~nada, 28692 Madrid, Spain\label{aff173}
\and
STAR Institute, University of Li{\`e}ge, Quartier Agora, All\'ee du six Ao\^ut 19c, 4000 Li\`ege, Belgium\label{aff174}
\and
Dipartimento di Fisica ``G. Occhialini", Universit\`a degli Studi di Milano Bicocca, Piazza della Scienza 3, 20126 Milano, Italy\label{aff175}
\and
School of Physical Sciences, The Open University, Milton Keynes, MK7 6AA, UK\label{aff176}
\and
Department of Astrophysics, University of Zurich, Winterthurerstrasse 190, 8057 Zurich, Switzerland\label{aff177}
\and
Space physics and astronomy research unit, University of Oulu, Pentti Kaiteran katu 1, FI-90014 Oulu, Finland\label{aff178}
\and
Department of Mathematics and Physics, Roma Tre University, Via della Vasca Navale 84, 00146 Rome, Italy\label{aff179}
\and
Theoretical astrophysics, Department of Physics and Astronomy, Uppsala University, Box 524, 751 20 Uppsala, Sweden\label{aff180}
\and
Mathematical Institute, University of Leiden, Einsteinweg 55, 2333 CA Leiden, The Netherlands\label{aff181}
\and
Institute of Astronomy, University of Cambridge, Madingley Road, Cambridge CB3 0HA, UK\label{aff182}
\and
Kobayashi-Maskawa Institute for the Origin of Particles and the Universe, Nagoya University, Chikusa-ku, Nagoya, 464-8602, Japan\label{aff183}
\and
Department of Astrophysical Sciences, Peyton Hall, Princeton University, Princeton, NJ 08544, USA\label{aff184}
\and
Fakult\"at f\"ur Physik, Universit\"at Bielefeld, Postfach 100131, 33501 Bielefeld, Germany\label{aff185}
\and
School of Mathematics, Statistics and Physics, Newcastle University, Herschel Building, Newcastle-upon-Tyne, NE1 7RU, UK\label{aff186}
\and
SRON Netherlands Institute for Space Research, Landleven 12, 9747 AD, Groningen, The Netherlands\label{aff187}
\and
Department of Physics and Astronomy, University of British Columbia, Vancouver, BC V6T 1Z1, Canada\label{aff188}}

\date{\today}
\abstract{
The ESA satellite mission \Euclid was launched in July 2023 and started its regular observing mode in February 2024. 
The data from the \Euclid instruments Visible Camera (VIS) and the Near IR Spectrometer and Photometer (NISP) are complemented by ground-based imaging data in the $u$, $g$ $r$, $i$, and $z$ bands to enable the determination of photometric redshifts and other galaxy properties. This paper describes the MERge processing function (MER PF) that has been developed and run for the Euclid Data Release 1 (DR1). The aim of the MER PF is to detect objects in the \Euclid imaging data, measure their properties, and combine them into a single multi-wavelength catalogue. In this paper we illustrate the design of the MER PF and introduce the different photometric methods applied to the data, which are Kron total flux, aperture photometry on Point Spread Functions (PSF) -matched images, template fitting photometry, and S\'ersic fitting photometry. We also present the various morphological measurements and quality flags that improve the scientific content of the output catalogues and better characterize the detected sources. We illustrate the processing and measurements that help to validate the MER PF results, which is an essential step to qualify the output for the public DR1 release. We discuss the global assessment of the photometric homogeneity and provide an internal comparison of our processing results. In addition, we give an extensive overview of the DR1 output data and the various processing scenarios.
}
%
%
%
%
   \titlerunning{The \Euclid MERge DR1 processing}
   \authorrunning{Euclid Collaboration: M.\ K\"ummel et al.}
   
   \maketitle
   \nolinenumbers
%
%
%
%
\section{\label{sc:intro}Introduction}
The cosmology and fundamental physics mission \Euclid \citep{EuclidSkyOverview} was successfully launched in July 2023. \Euclid aims at collecting data in the $14\,000\,\text{deg}^2$ large Euclid Wide Survey (EWS) and the approximately $63\,\text{deg}^2$ large Euclid Deep Survey \citep[EDS;][]{Scaramella-EP1}. In additon to the VIS and NIR data directly from \Euclid, the project also includes data from ground-based photometric surveys such as DES \citep{Abbott2021}, CFIS \citep{Ibata2017}, Pan-STARRS \citep{Magnier2020}, WHIGS and WHISHES \citep[see][]{2018PASJ...70S...4A} to enable photometric redshift determination and determine the SED of the PSF to be used for the shape measurements \citep{2018MNRAS.477.3433E}.
\begin{figure*}
    \begin{minipage}[c]{0.7\textwidth}
        \centering
        \includegraphics[width=0.9\textwidth]{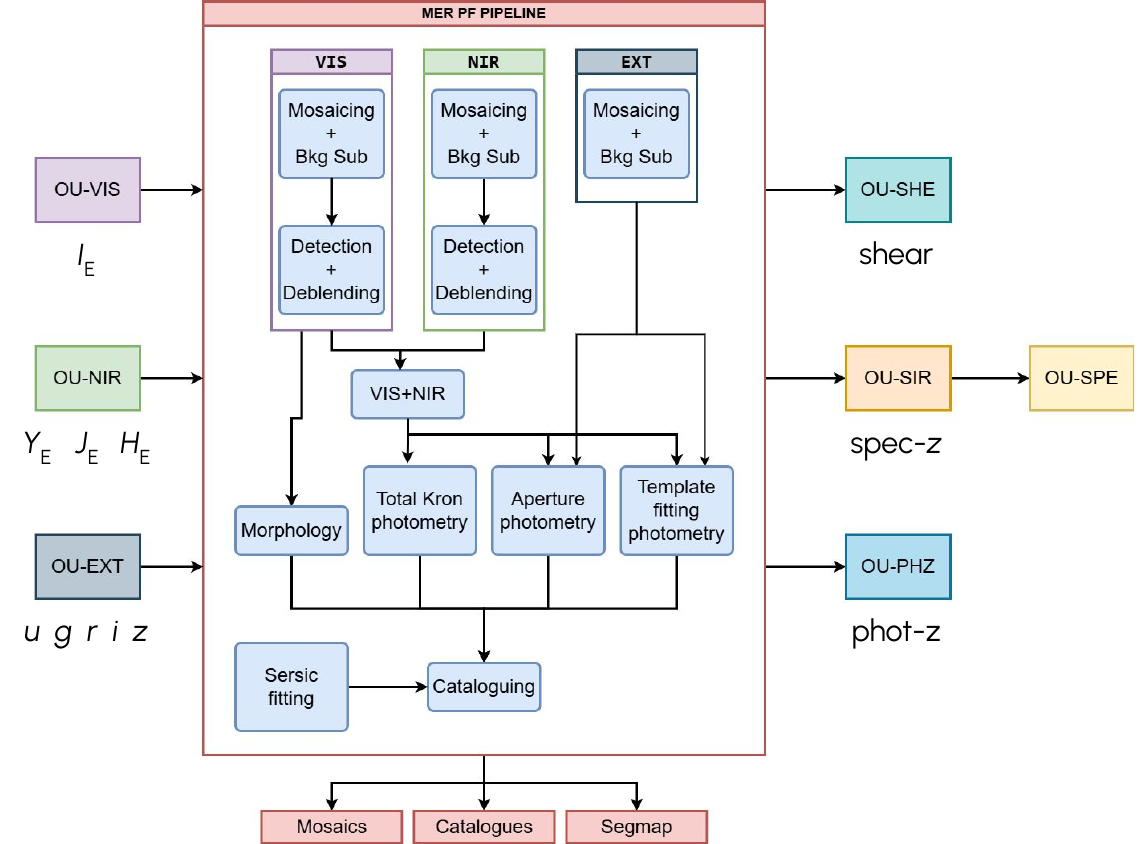}
    \end{minipage}
    \begin{minipage}[c]{0.3\textwidth}
        \caption{Simplified workflow of the OU-MER pipeline. The pipeline is fed with a list of input products coming from OU-VIS (VIS band \IE), OU-NIR (NISP-P bands \YE, \JE, and \HE), and OU-EXT (external bands). The pipeline performs a mosaicing of input images and generates the source list from a combination of detection and optimal deblending procedures.
        The pipeline computes photometry and morphology measurements and collects all this information in a set of output catalogues. Those catalogues, together with the reference images, represent the output of the MER-PF pipeline and constitute one of the main inputs to OU-PHZ (photo-$z$ measurements), OU-SHE (shear measurements) and OU-SIR (spectroscopic $z$ measurements).}
    \end{minipage}
    \label{fig:pipeline_workflow}
\end{figure*}

The processing of the very large volume of \Euclid data is, in the context of the \Euclid Science Ground Segment (SGS), organized into a set of processing functions (PFs), with each PF developed and maintained by a corresponding Organisation Unit (OU). The Euclid Consortium (EC) is committed to release fully processed data at pre-defined dates, and this paper describes the MERge PF that contributed to the core products of Euclid Data Release 1 (DR1). A full description of the area selection, processing organization, and processing details of Euclid DR1 are provided in \citet{DR1-TP001}. After Euclid Quick Release Q1 \citep{Q1-TP001} 
DR1 now comprises $\sim 1900\,{\rm deg}^2$ of EWS and EDS in all three sub-fields down to different depth.

The main output products of the MER PF are coadded images and a final catalogue with the photometric and morphological properties of all detected sources. As input, the MER PF needs VIS and NISP calibrated frames, processed by OU-VIS \citep{DR1-TP002} and OU-NIR \citep{DR1-TP003}, respectively, and stacked frames from selected external surveys as processed by OU-EXT \citep{DR1-TP006}, together with related ancillary data and PSF models. Object detection is done on both the VIS and NIR (stacking the three NISP bands \JE, \HE, and \YE) imaging data. A multi-wavelength flux determination is performed on detected objects with different photometric techniques. The final product of the MER PF is a catalogue of sources with unique identification numbers, multi-wavelength photometric measurements, and ancillary information (morphological estimates, quality flags, etc.).

Within the \Euclid SGS the output of the MER PF is fed to the pipelines for photometric redshift computation \citep{DR1-TP011} and spectra extraction \citep{DR1-TP023}. The object catalogues are used as an input for the shear estimation which is performed by OU-SHE and the subsequent OU-LE3 work on cosmology, which constitutes the core science of \Euclid.

This paper is organized as follows. Section \ref{sect:pipeline} describes the MER pipeline and the processing elements (PEs) that were implemented and utilized for the Euclid DR1 processing. In Sect.\ \ref{sect:validation} we discuss the data validation, Sect.\ \ref{subsubsect:relative_photometry} shows a method to assess the homogeneity of the MER photometry, Sect.\ \ref{sect:psf_discussion} discusses the treatment of the PSF models, and Sect.\ \ref{sect:deepwide} shows a comparison of the properties of the objects measured on EWS and EDS. Section \ref{sect:overview} contains an overview of the MER EWS and EDS data released in DR1. We close the paper with an outlook in Sect.\ \ref{sect:conclusions}.

\section{\label{sect:pipeline}The MER DR1 processing pipeline}
The MER processing pipeline deployed for DR1 was developed and improved from its predecessor used for Q1 \citep{Q1-TP004}. Figure \ref{fig:pipeline_workflow} shows a simplified workflow and places the MER PF in the larger context of the SGS DR1 processing. Compared to Q1, for DR1,    
\begin{itemize}
    \item the MER PF was run on a much larger data volume and also \Euclid data taken under various observing conditions (see Sect.\ \ref{sect:overview});
    \item the MER PF was run on EDS data with much larger exposure times;
    \item all input data (images and PSF models and related ancillary data) for the MER PF have been re-processed by the upstream OUs (VIS, NIR, and EXT) with updated pipelines and calibrations, which results in different outputs from the MER PF, even if run again on the now outdated Q1 version.
\end{itemize}

Before the final DR1 processing, we assured ourselves in many tests that our MER PF is generic enough to provide satisfactory results also for the non-EWS parts of DR1, and consequently the same PF was run on all data.
The following sections discuss elements of the MER pipeline with an emphasis on new products, modifications, and bug-fixes implemented for DR1.

\subsection{\label{subsect:tiling}Tiling for DR1}
\subsubsection{\label{subsubsect:deep_wide}Wide and Deep tiling}
\begin{figure}[t!]
    \centering
    \includegraphics[width=0.5\textwidth]{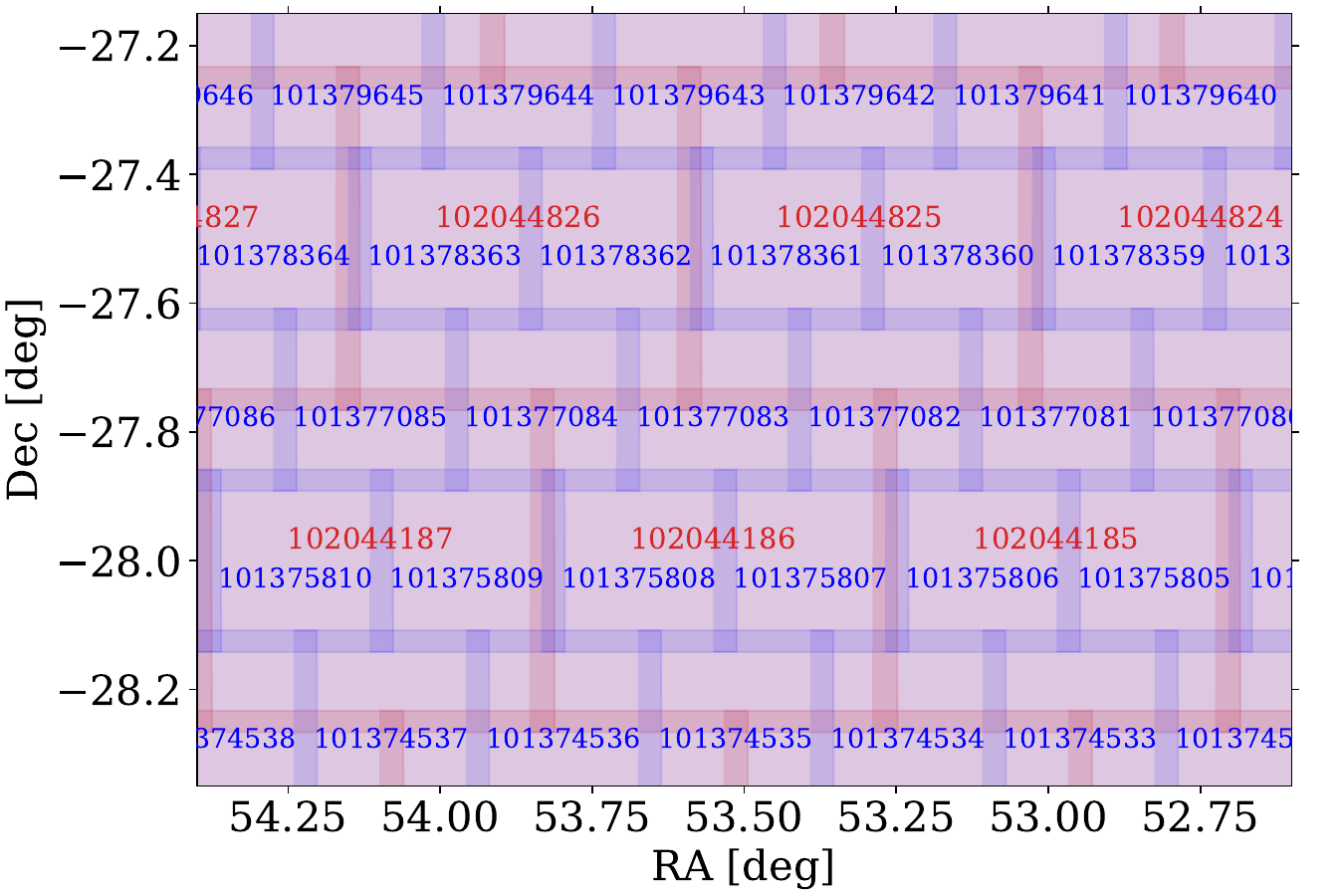}
    \caption{The pattern of the Wide (green) and Deep (blue) tiling at the EDF-F. Euclid DR1 contains both Deep and Wide tiling data of this area.}
    \label{figure:tiles_deep_wide}
\end{figure}
As in Q1, the DR1 pipeline runs in subsets of data that produce the output products for a predefined area on the sky called a `tile' \citep[see][]{2022ASPC..532..329K}, which is independent of both the \Euclid survey plan \citep{Scaramella-EP1} and the observing sequence. For most of the MER DR1, the same tiling solution as in Q1 was used, which means that a tile ID in DR1 covers the identical sky area as in Q1.

During the preparations for the \Euclid mission, it turned out that it would be very difficult to accomplish the MER processing for regions with a very large exposure time, such as Euclid Deep Fields (EDFs) that constitute the EDS or the photometric redshift calibration fields (see Table \ref{tab:field_overview}). The efficient processing of the image co-addition and the PSF processing \citep[see][]{DR1-TP033} requires the usage of local disk space ($\sim50\,$GB for every \Euclid band \IE, \YE, \JE, and \HE for normal wide tile) at the computing node for a fast data transfer, and the enormous incoming data for those fields are too large for the available computing nodes. As a consequence, we have created, in addition to the normal `Wide' tiling solution with an extended tile area of $32'\times32'$, a new, `Deep' tiling solution with
\begin{itemize}
\item an extended tile area of $17'\times17'$ in tangential projection;
\item increments of the tile centres by $15'$, providing an overlap  of $2'$ between adjacent tiles;
\item a core area definition at a \texttt{HEALPix} \citep{2005ApJ...622..759G} resolution of $N_{\text{side}} = 8192$.
\end{itemize}
The Deep tiling solution is independent of the Wide tiling, and has been defined for all sky regions with deep observations such at the EDFs.
Figure \ref{figure:tiles_deep_wide} displays the Deep and Wide tiling for EDF-Fornax (EDF-F). For the EDFs the Euclid DR1 contains MER data for both Deep and Wide tile (single reference observing sequence) processing. Tile indices for Wide start with the numbers `$102$' and indices for Deep with `$101$', which gives a convenient method to discriminate the corresponding output data.
\subsubsection{\label{subsubsect:spec_obj}Special objects}
Given the large coverage of the Euclid DR1 and fixed tile sizes, it is inevitable that some galaxies with sizes larger than the overlap size extend beyond one tile. As indicated in \cite{Q1-TP004} we have made adjustments to the Wide tiling solution to produce undisturbed imaging data for both the further processing (photometry, morphology) within the MER pipeline as well as for further scientific investigations of those objects.

To identify affected tiles, we collected, from publicly available catalogues, all large objects ($\rm size>2'$) located in the \Euclid region of interest \citep[ROI, see][]{Scaramella-EP1} area as `special' objects. This catalogue for the special tiling, consisting of 164\,526 objects, was constructed by aggregating several sources to maximize the inclusion of large objects. The primary foundation is HyperLEDA\footnote{\url{https://leda.univ-lyon1.fr/}} \citep{Makarov+14_Hyperleda}, which serves as the starting point. This is supplemented by the Heraklion Extragalactic CATaloguE\footnote{\url{https://hecate.ia.forth.gr/index.php}} \citep[HECATE;][]{Kovlakas+21_HECATE}, a value-added catalogue of nearby galaxies with robust distance measurements out to nearly 200\,Mpc, the Extragalactic Distance Database (EDD\footnote{\url{https://edd.ifa.hawaii.edu/dfirst.php}}; \citealt{Tully+2009_EDD}), SIMBAD\footnote{\url{https://simbad.cds.unistra.fr/simbad/}} \citep{Wenger+00_simbad}, the Third Reference Catalogue of Bright Galaxies\footnote{\url{http://cdsarc.u-strasbg.fr/viz-bin/Cat?VII/155}} (RC3; \citealt{1991rc3..book.....D}), and the Updated Nearby Galaxy Catalog\footnote{\url{https://vizier.cds.unistra.fr/viz-bin/VizieR?-source=J/AJ/145/101}} \citep{K+13}. Finally, Milky Way globular clusters were added to the sample \citep{Harris96}. Beyond these, while the aggregated catalogue primarily targets galaxies, it also retains other potentially interesting objects originating from HyperLEDA (e.g., star clusters, nebulae, and objects of uncertain nature).We explicitly removed entries from our initial HyperLEDA query classified as fake, absent, or identified as single, double, or triple stars.

For each selected source, we defined a square bounding box with a side length equal to four times its isophotal diameter (depending on the source catalogue either Holmberg $a_{26}$ or de Vaucouleurs $D_{25}$). To ensure proper tiling, we constrained the box sizes to a range between $2'$ and $120'$. For Milky Way globular clusters, the side of the square box is determined by the tidal radius $r_{\rm t}$ from \cite{Harris96}: it is fixed at $30'$ if $3 r_{\rm t} < 30'$ and capped at $120'$ if $3 r_{\rm t} > 120'$; otherwise, it is set to $3 r_{\rm t}$. These conservative dimensions ensure that each tile includes the primary object and its surrounding sky, potentially capturing extended tidal features. No initial (RA, Dec) selection was applied, meaning the catalogue currently includes objects outside the \Euclid ROI and also some potential overlapping boxes. This approach allows for future modifications to the footprint; naturally, only sources falling within the final \Euclid survey area are included in the final tiling.

Enlarging the `normal' Wide tiles to also extend over the special objects \citep[see][]{Q1-TP004} cannot provide solutions for every special object, especially in areas with a high density of special objects or for special objects that require a large bounding box. To define a processing solution also for these special objects that neither fit into the normal-sized wide tiles nor the extended wide tiles, we created a third flavour of `special' tiles. Every special tile is created for one specific special object that can not be covered otherwise. The special tiles have a tile index starting with `$1025$' (no interference with the indices of normal Wide tiles), are fully integrated in the \Euclid processing system and are typically larger than a wide tile. Processing such large amounts of data needs dedicated processing campaigns, and no special tile was processed in DR1.

\subsection{\label{subsect:mosaicing}Mosaic creation}
\begin{figure}
    \centering
    \includegraphics[width=0.48\textwidth]{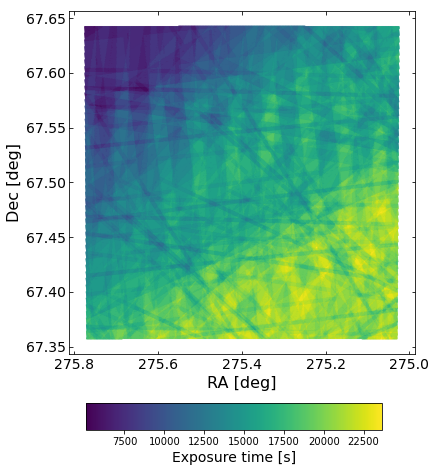}
    \caption{The VIS exposure time map for the EDF-N tile 101837492 computed from the layer file.}
    \label{figure:vis_expt_TILE101837492}
\end{figure}
For the later steps, in order to measure the photometric and morphological object properties, we create a multi-band set of co-aligned images where each pixel represents the same position on the sky with the common pixel scale of $0\overset{\prime\prime}{.}1\,{\rm pixel}^{-1}$ and the dimensions of the Wide and Deep \Euclid tiles (see Sect.~\ref{subsect:tiling}). Using the methods outlined in \cite{Q1-TP004} we coadd the fully calibrated VIS and NIR images and rebin the incoming EXT imaging data to that common pixel scale which is close to the VIS pixel scale \citep{EuclidSkyVIS}.

As a new data product for DR1, which is created during the mosaicing stage, we release for all co-added images in the \Euclid bands \IE, \YE, \JE, and \HE the so called `layer files'. For every VIS or NIR calibrated image contributing to the co-added image, the layer file contains: (i) the exposure time; (ii) the three angles (solar aspect angle, alpha and beta) describing the orientation of \Euclid with respect to the Sun \citep{EuclidSkyOverview}; (iii) the observing date; (iv) the coordinates of the image corners.
The primary purpose to directly collect and associate this metadata with the co-added image is to facilitate the search for systematic effects in the cosmology analysis in OU-LE3.

The layer files also allow reconstructing the exposure time map or the  number of image layers for a tile, which can then be transferred to the objects in the final MER catalogue. As an example, Fig.\  \ref{figure:vis_expt_TILE101837492} shows the exposure time map of the deep EDF-North (EDF-N) tile 101837492 for the co-added \IE image as derived from the corresponding layer file. The exposure time is much larger than the typical $2.3\,\mathrm{ks}$ for the EWS. The tile is positioned in an area with a strong gradient in the exposure time, and the line-like features are created by the various orientation angles of the \Euclid observations covering this tile.

Note that all images are created for a specific tile. Since the tiles are defined independent from the \Euclid observations, there are images that are not completely filled with pixel data, especially at the border of an observed area
(see Sect.\ \ref{subsubsect:border_tiles} and Fig.\ \ref{fig:border_tile}). In the \Euclid survey design, a reference observing sequence (ROS) has six and four exposures in \IE and the NIR bands, respectively \citep{EuclidSkyOverview}. The dithers within a ROS create areas with less exposures, and at the overlap with adjacent ROS there are areas with more exposures. As a consequence, in the \Euclid bands the exposure time and the S/N (and number of layers) has an individual  pattern for each image.

\subsection{\label{subsect:detection}Assembly of the source list}
The source detection in DR1 follows the procedures outlined in \cite{Q1-TP004}. This means we do an independent source detection on VIS and NIR co-added images. The deblending is done independently in VIS and NIR using a dedicated software component based on \texttt{ASTErIsM} \citep{Tramacere2016} which uses two topometric clustering algorithms. The VIS and NIR deblended sources are combined into a single list of objects and a segmentation map, which are the basis for the later measurement steps for photometry and morphology.  The source detection in VIS and NIR is done with \texttt{SourceExtractor++} \citep[\texttt{SE++,}][]{2020ASPC..527..461B, 2022arXiv221202428K} and for convenience we list the most important detection parameters in Table \ref{tab:detection_params}. Note that we use the same detection parameters for the EWS and the EDS data. The merging of VIS and NIR deblended sources is done as described in \cite{Q1-TP004}.
\begin{table}[t]
    \centering
    \caption{\texttt{SE++} parameters used for the VIS and NIR detections in all DR1 data.}
    \begin{tabular}{lll}
       \hline
       \hline
       Parameter & VIS & NIR \\
       \hline
       \small\texttt{segmentation-filter} (Gaussian FWHM [pix])&$2.4$ & $5.0$\\
       \small\texttt{detection-threshold (S/N in a pixel)} & 0.9&  0.4\\
       \small\texttt{detection-minimum-area [pix]} & 10& 10\\
       \small\texttt{core-threshold-value (S/N in a pixels)} & 1.3&  0.55\\
       \small\texttt{core-minimum-area [pix]} & 10& 10\\
       \hline
    \end{tabular}
    \label{tab:detection_params}
\end{table}
\subsection{\label{subsect:psf_calculation} Mapping of the point-spread functions}

As in Q1, the DR1 pipeline propagates the input PSFs calculated by the VIS, NISP, and EXT PFs to the mosaic grid, taking into account the effects of the image stacking process \cite[see][]{Q1-TP004}. 
In Q1 we found some discrepancies between the propagated NIR PSFs and the profiles of \textit{Gaia} stars on the MER mosaics. The propagated NIR PSFs had $1.75\%$ to $3.5\%$ larger full width half maximum (FWHM) values than \textit{Gaia} stars, which indicated some issues with the input NIR PSFs. These differences are now much smaller in DR1, thanks to the introduction of a new method for calculating the PSFs in the NIR PF \citep[][]{DR1-TP005}. The normalization of the PSF through the processing is also handled differently in DR1, which has consequences for essentially all photometry measurements and is extensively discussed in Sect.\ \ref{sect:psf_discussion}.

\subsection{\label{subsect:photometry}Photometry}
The photometric techniques adopted for the DR1 catalogues are the same as those described in \cite{Q1-TP004} and in the photometry cookbook (see link in Sect.\ \ref{sect:conclusions}).
Briefly, for all the objects in the source list, three types of measurements are performed: \newline
(i) Aperture photometry is obtained using a custom version of the code \textsc{A-PHOT} \citep{Merlin2019} on source cutouts that are PSF-matched on-the-fly to the worst resolution on an object-by-object basis, using the convolution kernels described in section 6 of \cite{Q1-TP004}. Fluxes are measured for a set of four apertures with diameters equal to 1, 2, 3, and 4 times the FWHM of the worst PSF (typically an EXT band; the objects specific values are stored in the column \texttt{FWHM} of the final catalog. Pixels assigned to different sources in the segmentation map are excluded in the aperture fluxes of a given object; however, contamination from faint extended wings remains present. Additionally, a total flux in the detection band (\IE or the NIR stack as described in Sect. \ref{subsect:detection}) is estimated by means of a \citet{Kron1980} elliptical aperture, so that the total flux in each band can be obtained by scaling it by the colours computed with the circular apertures (see the Photometry Cookbook for the details of the procedure).\newline
(ii) Template-fitting photometry is performed using a custom version of the code \textsc{T-PHOT} \citep{Merlin2015,Merlin2016}, which uses the morphological information of the detection band (2-dimensional stamps cut from the detection images) as priors which are then degraded to the resolution of the measurement bands, again using the convolution kernels, and used to fit individual regions around each source. \textsc{T-PHOT} is designed to reduce the impact of blending in low-resolution data, and in principle can directly provide total fluxes; however this direct approach is limited by the area assigned to the measured source in the segmentation map and by the extension of the PSF models, so another possibility is to compute colours as in the aperture case (the fluxes from both the measurement image and the corresponding PSF-matched detection image are provided in the catalogue to this aim, again see the Photometry Cookbook).\newline
(iii) For model-fitting photometry we use S\'ersic models that provide the integrated objects flux as a parameter. Further details are provided in Sect.\ \ref{subsubsect:para_morphology}.

We refer the reader to the Q1 paper for more details. The most important difference with respect to Q1 in the photometric algorithms is that now the PSFs are not normalised to unit flux before constructing the convolution kernels, thus taking into account the flux missed by cutting their extended wings. As discussed in Sect.\ \ref{sect:psf_discussion}, this impacts almost all the photometric methods, because PSF-matched images are always involved in some way. Our data validation shows (see Sect.\ \ref{sect:validation}) that this change improved the measurements, especially in the case of the Pan-STARRS $i$ band, which has a very extended PSF larger than \ang{;;5}.

The photometry calibration is validated against {\it Gaia} DR3 \citep{GAIADR3}, see Sect. \ref{sect:validation}. It must be stressed that the photometric methods are built and optimized to perform measurements of faint and extended sources, rather than on bright point-like objects that dominate the {\it Gaia} catalog. For this reason, the performance on stars is sub-optimal. A direct comparison of the MER measured fluxes against external catalogues of extended objects is provided in \citet{DR1-TP010}.
\begin{figure}
    \centering
    \includegraphics[width=0.5\textwidth]{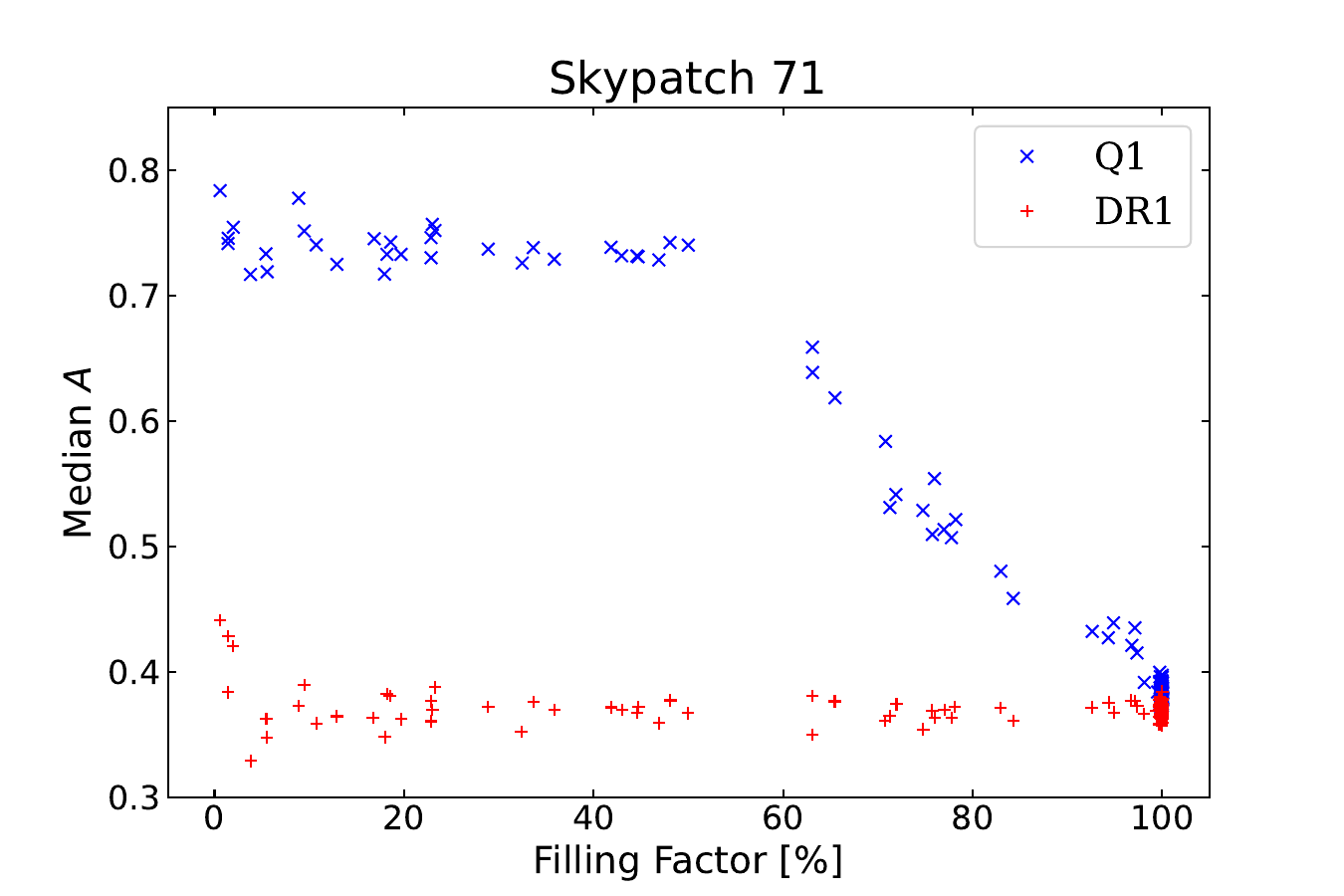}
    \caption{Distribution of the tile median asymmetry $A$ as a function of the tile filling factor in data covering EDF-S. Blue crosses represent the distribution in the Q1 release, while the red crosses show the distribution of the same data as processed for DR1, after the switch from global to local noise evaluation.}
    \label{figure:a_vs_filling}
\end{figure}
\subsection{\label{subsect:morphology} Morphological characterization}
\subsubsection{\label{subsubsect:nonpara_morphology} Non-parametric morphology}
During the further analysis of the non-parametric parameters concentration ($C$), asymmetry ($A$), and smoothness ($S$) \citep[$CAS$,][]{Conselice2014,Tohill2021} published in Q1, we discovered a peculiar correlation between the median value of $A$ within the tile and the tile \verb|FILLING_FACTOR|, which is the fraction of \texttt{HealPix} pixels containing at least one detected object. 
This was caused by a wrong computation of the noise level in the source cutouts: the median and normalized median absolute deviation (NMAD) of the background noise were computed globally, using all the mosaic pixels not covered by sources by taking all the pixels where the segmentation map equals zero. 
For partially covered tiles (see Sect.\ \ref{subsubsect:border_tiles}) this noise computation included also the empty image areas that are filled with values $0.0$. In those tiles, that are marked by their low \verb|FILLING_FACTOR|, the filled in zeros underestimated the global background noise, and thus led to skewed measurements of $A$.
The pipeline has been updated for DR1, and now we compute, for each source, the local noise level within the source cutout, thus mitigating the dependency of the $A$ measurements on the
\verb|FILLING_FACTOR|. Figure \ref{figure:a_vs_filling} shows a comparison of the Q1 and DR1 asymmetry values for the same dataset taken on EDF-South (EDF-S).

At the request of several \Euclid Science Working Groups (SWGs) and also some downstream OUs (SIR, responsible for slitless spectroscopy and SHE, responsible for shear measurements), we added the computation of the \verb|FLUX_RADIUS| as an additional quantity for object morphology in DR1. This quantity is familiar from the \texttt{SExtractor2} \citep{1996A&AS..117..393B} object detection software and measures the radial distance from the object centre containing $50\%$ of the isophotal object flux.
The implementation of \verb|FLUX_RADIUS| within the MER pipeline is inherited from \texttt{SExtractor2}. We define a circular disk of radius $r=6\, a_M$ around the source's centroid, where $a_M$ is the source's semi-major axis reported in the MER final catalogue (column \texttt{SEMIMAJOR\_AXIS}). To be compliant with the \texttt{SExtractor2} implementation, we split the disk in 64 annuli of equal thickness in order to compute a brightness profile of the source. We then interpolate the value of the radius where the flux is 50\% of the total flux within the disk.

\begin{table}[t!]
    \centering
    \caption{\texttt{SourceExtractor++} model fitting flags.}
    \label{tab:sepp_flags}
    \resizebox{\linewidth}{!}{
    \begin{tabular}{ll}
    \hline
    \hline
    Flag bit & Explanation \\
    \hline
    0 & Error flag (from levmar) \\
    1 & Size correction (manual resizing of fitting box) \\
    2 & Outside of the measurement frame (no model fitting run) \\
    3 & Insufficient data (no model fitting run) \\
    4 & Some/all of the model parameters could not be fitted \\
    5 & The fit was done on a downsampled image\\
      & due to exceeding the maximum size \\
    6 & Small gradient, $\mathbf{J}^{T}\mathbf{e}$\\
    7 & Small step, $\Delta\mathbf{p}$\\
    8 & Maximum iterations, \texttt{it$_\mathrm{max}$}, reached\\
    9 & Singular matrix $\mathbf{J}^{T}\mathbf{J}+\mu\mathbf{I}$ (raises error flag)\\
    10 & No further error reduction possible\\
    11 & Small residual, $\lVert\mathbf{e}\rVert_2$\\
    12 & Invalid (NaN/Inf) model values (raises error flag)\\
    \hline
    \end{tabular}
    }
\end{table}
\subsubsection{\label{subsubsect:para_morphology}Parametric morphology}
Similarly as in Q1, for all sources within a \Euclid tile, we fit a two-dimensional S\'ersic profile \citep{1963BAAA....6...41S} in all bands. We refer the reader to \cite{Q1-TP004} for details on the fitting procedure; here, we provide a general description of the fitting code and the changes with respect to Q1. We use the SE++ \citep{2022arXiv221202428K} version 1.0.2 for modeling the S\'ersic parameters. Specifically, we work in no-detection mode, which means we feed SE++ the objects' centres and fitting areas along with initial guesses for the S\'ersic parameters.

After a careful analysis of the Q1 data we extended the maximum S\'ersic index from $5.5$ to $8.0$ to avoid an artificial parameter cut for red galaxies. To reduce the runtime of SE++, we had introduced an upper limit for the size of the fitting area, which is derived from the segmentation area. Using the Q1 data as a guide, in DR1 we have improved this selection (using the detection flag in \texttt{DET\_QUALITY\_FLAG}) such that bright, saturated stars get a limited fitting area, but bright galaxies are still fitted with the full area.

For DR1, we combine the information of the flags from \texttt{SourceExtractor++} (i.e., from the model fitting) and from the fitting engine (i.e., the stopping criterion for the fit) into one flag for VIS+NIR, \verb|SERSIC_VISNIR_FLAGS|, and for EXT data, \verb|SERSIC_EXT_FLAGS|. These flags are in binary format and are explained in Table~\ref{tab:sepp_flags}. Flag values from bits 6\textendash12 represent the stopping criterion for the model fitting and are taken from the fitting engine \texttt{levmar}\footnote{\url{https://users.ics.forth.gr/~lourakis/levmar/}}. We also set a flag in case the size of big objects was manually corrected (bit 1). As in all DR1 photometry, we use non-normalized PSFs in the S\'ersic fits.

\subsubsection{\label{subsubsect:star galaxy separation}Star-galaxy separation}
As in Q1 the rather simple star-galaxy (S/G) classifier uses the quantity \verb|MU_MAX| $-$ \verb|MAG_AUTO| plus a calibration derived from pre-flight \Euclid Scientific Challenge (SC8) to define a point-like probability \verb|POINT_LIKE_PROB| for VIS-detected objects. 
We also define a flag POINT\_LIKE\_FLAG that selects the VIS detected objects with POINT\_LIKE\_PROB greater than 0.96 for which the requested 99\% star purity was reached in SC8. Other use cases with a lower probability threshold value can be useful to have a better completeness of the star sample.

\subsubsection{\label{subsubsect:zoobot}Zoobot morphology}
Similar to Q1, the MER final catalogue includes visual-like morphologies for a subsample of galaxies, which is now applied to both the wide and deep data of DR1. This visual classification uses the deep foundation model which was created by fine-tuning the \texttt{Zoobot} galaxy foundation models as described in \cite{Q1-SP047}. In DR1 we extend the subsample by selecting all galaxies brighter than \IE$=21.0$ or with a segmentation area larger than $700$ pixels. This enlarges the subsample by a factor of $4$ w.r.t. Q1.

\begin{figure}[t!]
\centering
\includegraphics[width=0.45\textwidth]{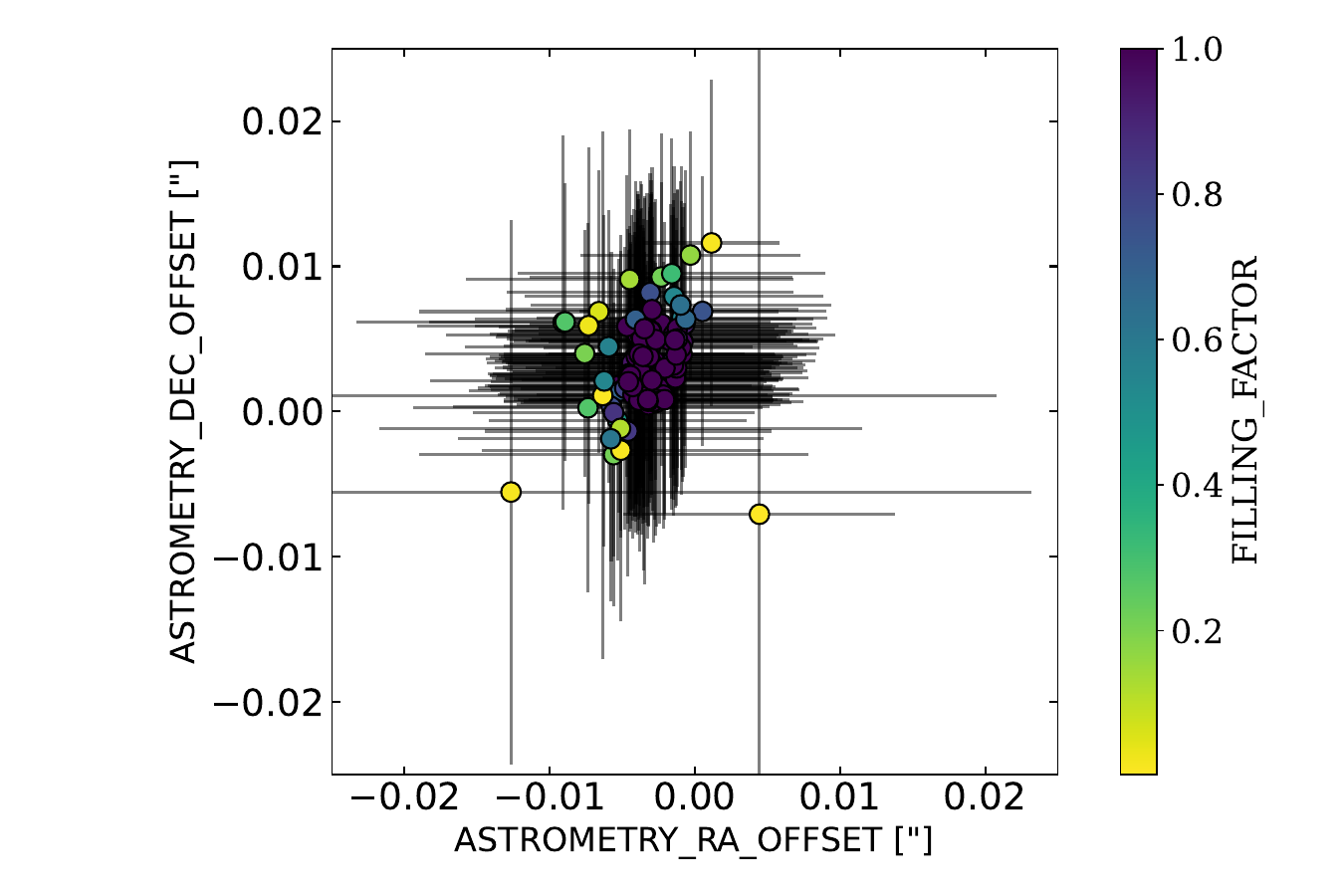}
\caption{The astrometric offsets for all tiles in the EDF-N wide-like processing of patch 64}
\label{fig:astro_validation}
\end{figure}
\subsection{Identification of spurious sources}\label{subsect:spurious_proba}
In DR1 we used the same tool to identify spurious sources as in Q1. The model used to compute the spurious probability is trained on pre-launch simulations, with no DEEP-like training set available. For that reason, this measurement is currently optimised for WIDE processing. Feedback from \Euclid SWGs has shown that a lower limit of $0.5$ in \texttt{SPURIOUS\_PROB} to set the \texttt{SPURIOUS\_FLAG} seems too aggressive. Use cases from those groups suggest using $0.7$ instead. A detailed analysis is currently ongoing in order to improve these measurements in the next releases.

\subsection{Bright star masking}\label{subsect:star_masking}
As in Q1 we have mapped the imprint of bright {\it Gaia} stars onto the coadded \IE, \YE, \JE, and \HE images and marked the objects in the saturated centers and along the diffraction spikes in the column \texttt{DET\_QUALITY\_FLAG}. Depending on the Galactic latitude, the masked area is in the range of about $1-5\%$ for EWS tiles. For deeper data that was observed at several roll angles of \Euclid, specifically EDF-S and EDF-N, the diffraction spikes of each roll angle make an imprint on the stellar mask. This increases the masked area to ranges of $2--10\%$ and $3--15\%$ for EDF-S and EDF-N, respectively. However, a close inspection of this data shows that the diffraction spike signatures have frequently cancelled out, and these numbers rather represent an upper limit.

\subsection{\label{subsect:catalog} The MER catalogues}
As in Q1, the MER output lists are distributed over several files with the main photometry catalogue (\verb|EUC_MER_FINAL-CAT|), the catalogue with the morphology measurements (\verb|EUC_MER_FINAL-MORPH-CAT|), the cutout catalogue (\verb|EUC_MER_FINAL-CUTOUTS-CAT|), and the deep catalogue (\verb|EUC_MER_FINAL-DEEP-CAT|) with the photometry of the bands used in EDFs, calibration and extra fields (see Tab.\ \ref{tab:field_overview}).

\begin{figure}[t!]
\centering
\includegraphics[width=0.45\textwidth]{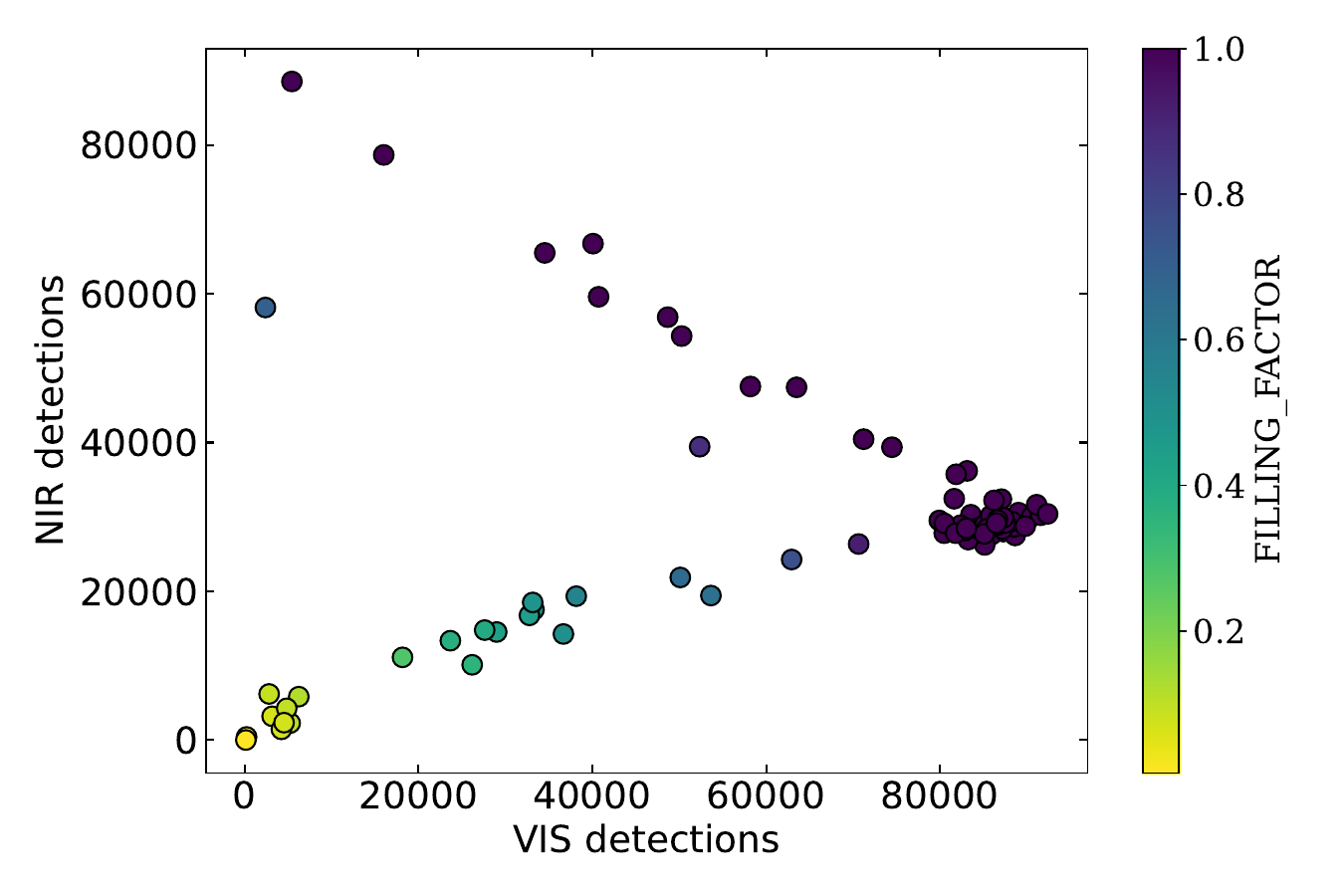}
\caption{The number of VIS and NIR detected objects for all tiles in the wide-like processing of the EDF-N patch 44.}
\label{fig:detection_validation}
\end{figure}
\section{\label{sect:validation}Data validation}
After finishing the processing in the MER PF, the resulting output data is pushed through the validation process. Data validation checks whether the pipeline output is within the specification and requirements and also aims to identify processing issues and data quality problems.
After running a dedicated validation pipeline, we collect and visualize its output, characteristic quantities such as image depths, photometric offsets, or the PSF sizes. The large amount of diverse properties makes an automatic validation with formulas and strict limits not feasible, and we examine the aggregated validation numbers for any inconsistencies and possible problems in the pipeline processing; if necessary there are further investigations in the validation results or even output data for specific tiles. While the actual number of invalidated tiles was small and mostly due to severe processing errors, we were able to identify areas with problematic photometry that could either be reprocessed or taken into account in the further SGS processing.

In the following sections, we present and discuss the aggregated validation results for selected areas on the sky, which are usually single patch\footnote{The \Euclid survey is organized in consecutively, spatially close observations that require only small slew movements and are called patch \citep[see][]{EuclidSkyOverview} observations of} one EDF. While the deep results go through the same validation process, we are using wide survey data as examples since this comprises the majority of the data and show a larger range of observing conditions (see Sect.\ \ref{sect:overview}).
\subsection{\label{subsect:astrometry}Astrometry}
We check the astrometry of our objects by computing the offsets between the catalogue positions and the proper motion-corrected positions of all {\it Gaia} stars in a tile. Figure \ref{fig:astro_validation} shows the median offset and NMAD for all tiles of patch 64. Tiles with a low \verb|FILLING_FACTOR| show relatively large offsets due to the smaller numbers of {\it Gaia} stars and the uneven sampling of residual systematic astrometric errors from the OU-VIS input images. Tiles with a \verb|FILLING_FACTOR| larger than $0.5$ are within an offset $<10\,\mathrm{mas}$ in both right ascension and declination and well within the requirement for the absolute accuracy.
\subsection{\label{subsect:detections}Detections}
In the detection validation, we collect all characteristic numbers associated to object detection, cross-matching, and detection qualification. Figure \ref{fig:detection_validation} shows the number of VIS and NIR detections in the wide-like processed tiles of patch 44, which is in the EDF-N. The points in the lower branch mark the typical ratio between VIS and NIR detections with the absolute numbers scaling with the \verb|FILLING_FACTOR|. The tiles in the upper branch have a large \verb|FILLING_FACTOR| with an unusually large number of NIR detections. This is the result of an incomplete coverage of VIS data for those tiles, which is usually caused by large proton showers coming from the Sun (see section\ \ref{subsubsect:inhomo_tiles} and \citealt{DR1-TP002}). Note that tiles with these signatures are still validated and part of DR1 if the available data matches the quality standards.
\subsection{\label{subsect:pointlike}Point-like probability}
\begin{figure}[t!]
\centering
\includegraphics[width=0.45\textwidth]{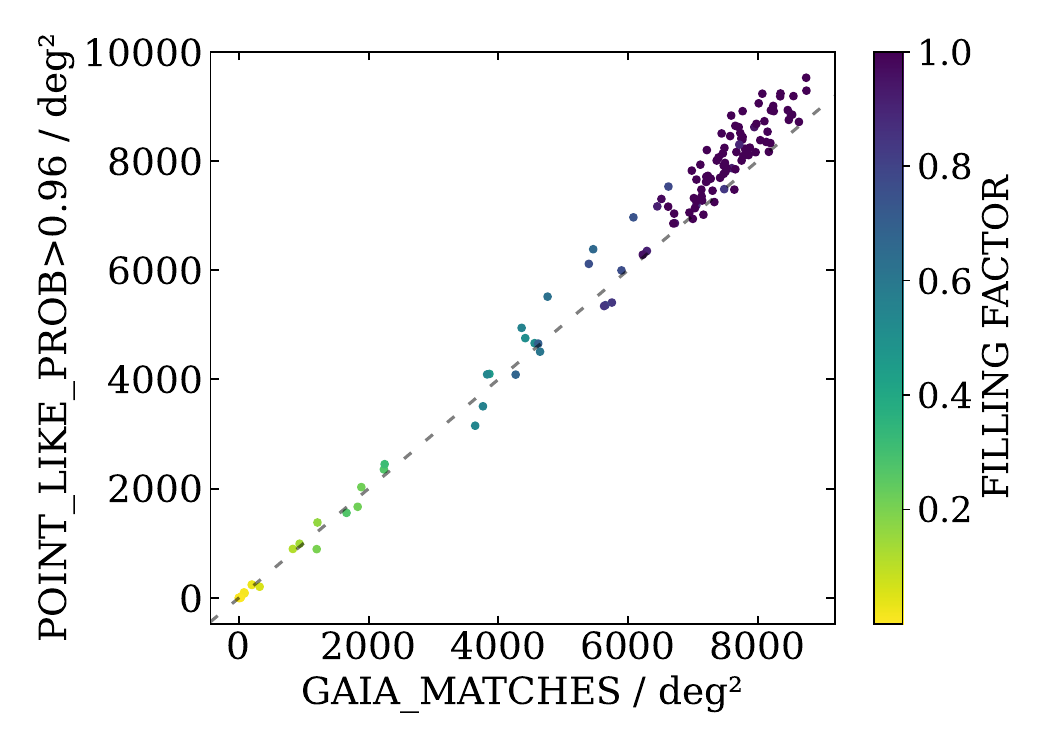}
\caption{The number of objects with a high point-like probability compared to the number of detections with a good match in the {\it Gaia} DR3 catalogue for each tile of EDF-N patch 64.}
\label{fig:pointlike}
\end{figure}
In order to identify possible problems with the point-like probability estimation in a patch, we compare for each tile the number of objects with a high point-like probability ($\verb|POINT_LIKE_PROB| > 0.96$) to the number of objects matching the {\it Gaia} DR3 catalogue. As shown in Fig.\ \ref{fig:pointlike}, these two numbers are expected to be comparable.
\subsection{\label{subsect:psfvalidation}PSF validation}
\begin{figure}[t!]
\centering
\includegraphics[width=0.45\textwidth]{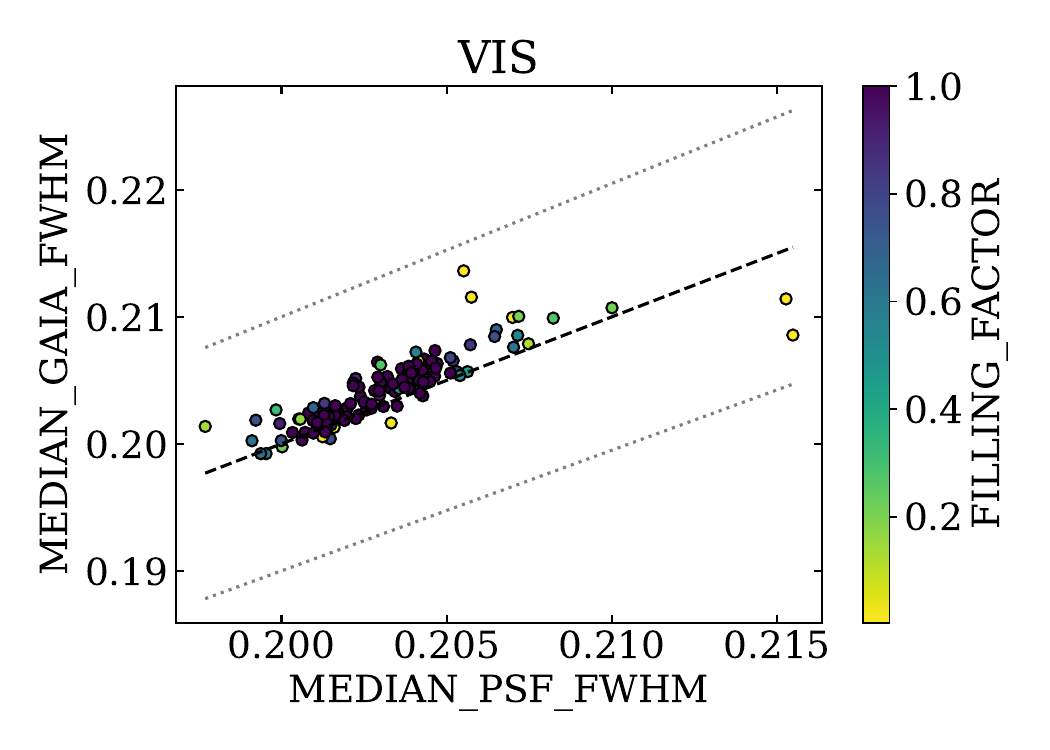}
\caption{Median FWHM of the VIS propagated PSF compared to the median PSF of {\it Gaia} stars measured on the VIS MER mosaics for each tile of EDF-N patch 64. Dashed line represents the one-to-one relation and dotted lines $\pm5\%$ difference.}
\label{fig:VISpsfvalidaion}
\end{figure}
The PSF validation compares the agreement between the median FWHM of the propagated PSFs discussed in Sect.\ \ref{subsect:psf_calculation} and the median FWHM of {\it Gaia} stars as measured on the MER mosaics. On a patch, the overall agreement is at the percent level for VIS (see Fig.\ \ref{fig:VISpsfvalidaion}) and NIR and sub-percent for EXT. Tiles with more than 5\% discrepancy are reported as outliers but not invalidated. In general the PSF validation for DR1 looks as shown in Fig.\ref{fig:VISpsfvalidaion}. However, for some bands and patches the comparison shows an offset in the one-to-one relation indicating an issue with the PSF model used for the photometry.
\subsection{\label{subsect:depth}Image depth}
\begin{figure}[t!]
\centering
\includegraphics[width=0.45\textwidth]{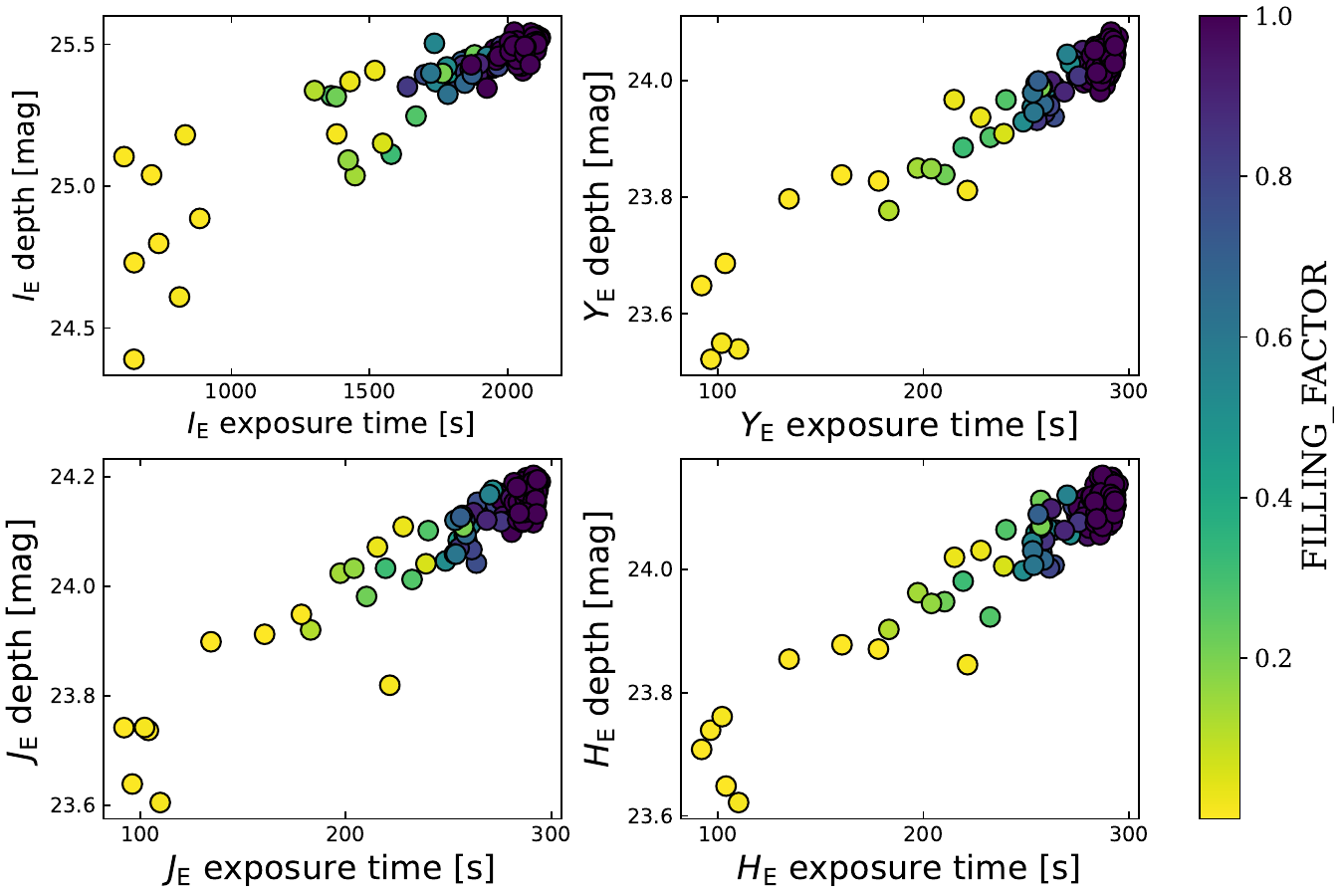}
\caption{The $5\,\sigma$ depth in the \Euclid filters for all tiles in the EDF-N patch 64.}
\label{fig:validation_depth}
\end{figure}
In data validation we determine the depth of all aligned images by measuring the flux in randomly selected apertures of $2^{\prime\prime}$ diameter while skipping over apertures that contain detected objects \citep{Q1-TP004}. We use the median of these flux values to determine the $5\,\sigma$ depth in AB magnitudes. Figure \ref{fig:validation_depth} shows this characteristic depth in \IE, \YE, \JE, and \HE as a function of the mean exposure time computed from the exposure time maps such as shown in Fig.\ \ref{figure:vis_expt_TILE101837492} for all tiles of the EDF-N patch 64.  As expected the tiles at the border of the patch with a low \verb|FILLING_FACTOR| have a lower exposure time and correspondingly lower depth. Given the uncertainties in both quantities the image depth scales as expected.

\subsection{\label{subsect:photometry_validation}Photometry validation}\begin{figure*}[t!]
\sidecaption
\includegraphics[width=.7\textwidth]{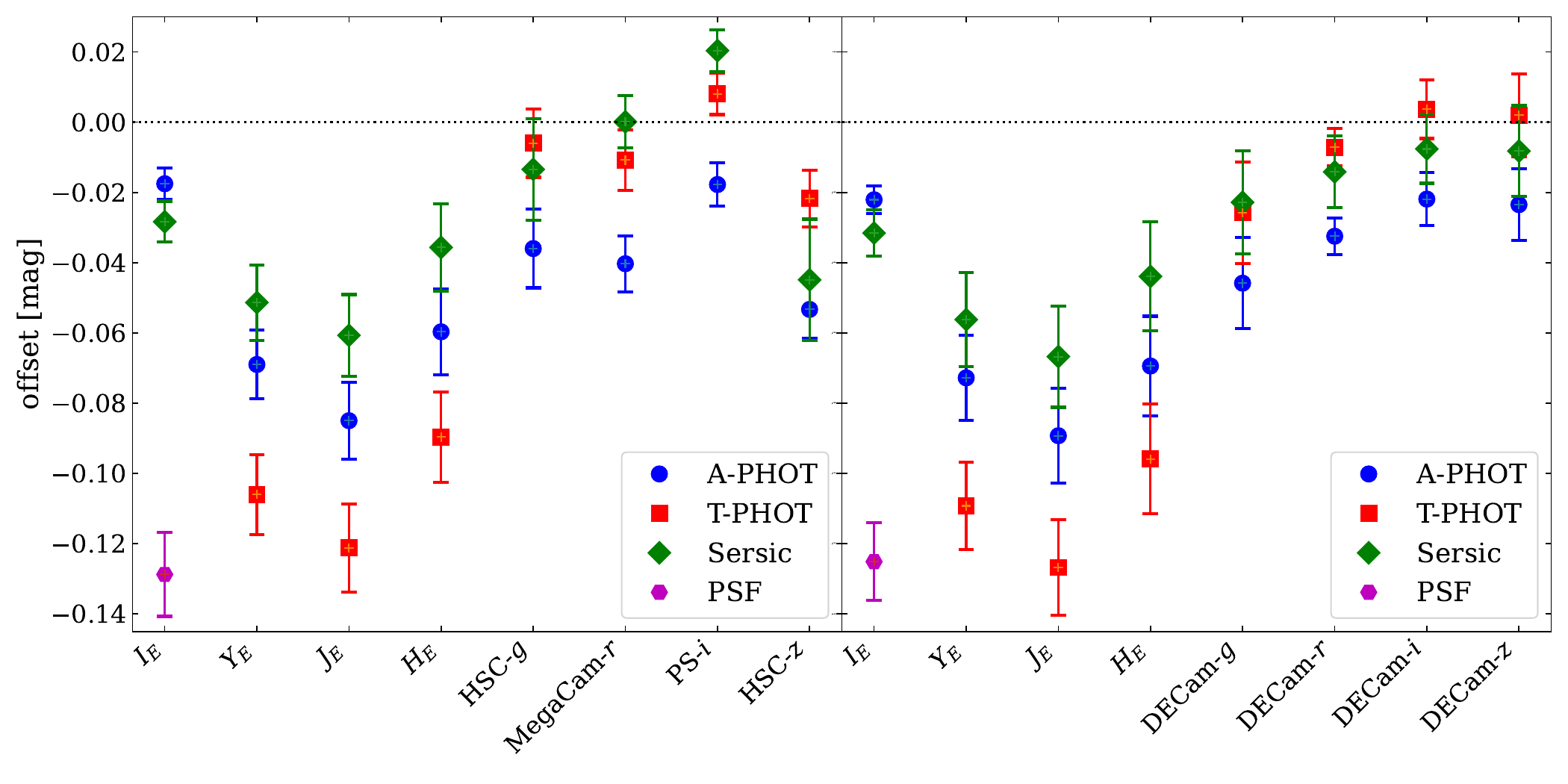}
\caption{Mean magnitude offsets with respect to the expected brightness from \textit{Gaia} for all photometric measurements in DR1 (aperture photometry \texttt{A-PHOT}, template fitting photometry \texttt{T-PHOT}, S\'ersic fitting \texttt{Sersic} and PSF fitting) and bands, as derived from the validation pipeline for the tiles of EDF-N patch 64 and EDF-S patch 126 on the left and right panels, respectively. The error bars represent the standard deviations of the specific samples used to compute the mean magnitude offset. The PSF fitting offset and the difference between A-PHOT and T-PHOT are mostly due to a bug which was found after DR1 processing.}
\label{fig:phot_validaion}
\end{figure*}
To validate our photometry we compare our measured values for point-like sources  with the values expected from \textit{Gaia} DR3 photometry. To do that, we use transformation functions that translate the measured values from \textit{Gaia} $G$, BP, and RP into each of the photometry bands used in DR1  (\IE, \YE, \JE, \HE, and the various EXT bands). For the EXT data we are also using in each band the measured transformation function that was used in the absolute photometric calibration of the EXT data \citep{Q1-TP001, 2020ASPC..527..701G}. For the \Euclid filters we created transformation functions using synthetic stellar spectra from \cite{EP-Serrano} in the suitable range $0< {\rm BP}-{\rm RP}<2.5$. For the typically more than a thousand \textit{Gaia} matches per wide tile we do a careful selection to remove all objects with a potentially compromised photometry and compute the median offset between the measured and expected object brightness for each band photometric method.

Figure \ref{fig:phot_validaion} shows, for all bands, the mean offset and standard deviation for the tiles in EDF-N patch 64 in the left panel and EDF-S patch 126 in the right panel. The band-specific offsets in Fig.\ \ref{fig:phot_validaion} are typical for DR1 data with a nominal EWS coverage. The mean offset in \IE is about $-0.02\pm0.004$, which is an improvement compared to Q1. The offsets in the \Euclid NIR bands are, depending on the photometric method, in the range $-0.036$ to $-0.126$ with a scatter of typically $0.015$. This offset, determined using theoretical transformation functions, is larger than in Q1. The mean offsets in the EXT bands are typically in  the range of a few percent and rather reflect the specific setup and absolute photometric calibration of the various surveys than the capabilities of the measurements in the OU-MER pipeline. The difference between \texttt{A-PHOT} and \texttt{T-PHOT} estimates in the NIR bands are mainly caused by a software bug in the \texttt{T-PHOT} procedure which was found after the DR1 processing and affects mainly bright, point-like sources. The same template fitting algorithm is involved in the PSF photometry computation, explaining the large offsets in PSF fitting ($-0.13\,\mathrm{mag}$). We hence suggest to use those measurements with caution, in particular the PSF photometry.

\subsection{\label{subsect:alt_val}Alternative approaches to the photometry validation}
\subsubsection{Comparison to published extragalactic deep catalogues}
As an alternative to using {\it Gaia} data as ground truth or measuring characteristic quantities, ~\cite{DR1-TP010} compares the photometry
in our deep COSMOS catalogues (see Table \ref{tab:field_overview})
and the COSMOS2025 catalogue \citep{2026A&A...707A.391S} for extended sources.
In general, a good agreement is found in all four \Euclid bands and for three photometric methods \texttt{A-PHOT}, \texttt{T-PHOT} and S\'ersic fitting. All methods slightly overestimate the flux with respect to COSMOS2025, with \texttt{T-PHOT} providing a better agreement in the NIR bands. However, possible systematics induced by the adopted techniques amount to the same order of magnitude as the measured bias.

\subsubsection{Source injection}
As described in Euclid Collaboration: Kong et al. (in prep.), a source injection pipeline, \textit{Cornifer}, was developed to validate the output quantities from the MER PF. In this project, synthetic galaxies and stars with known properties are injected into observed VIS, NIR, and EXT imaging data. The injected sources are designed to mimic the characteristics of the real sources. The modified input images then go through the standard MER PF.
The photometric offset measured through source injection is smaller than that obtained from comparisons with external datasets, such as Gaia. A comparison of the injected and measured brightness shows no major issues in the recovered photometry below the $5\,\mathrm{\%}$ level.

\section{\label{subsubsect:relative_photometry}Estimating the global homogeneity of the MER photometry}
In Sect.\ \ref{subsect:photometry_validation} we used the \textit{Gaia} photometry as an absolute standard to validate the MER photometry in each processed tile. In this section we measure the offset between the \textit{Gaia} and the MER photometry over the entire DR1 area to asses the homogeneity of the MER photometry. Likewise, we interpret the scatter between the \textit{Gaia} and the \Euclid MER flux in the various bands as a measurement of the relative flux errors. Colour variations associated to different stellar populations, as well as differential extinction, are also expected to contribute to the observed scatter, so one should consider these relative flux errors as upper limits.

Starting with the cross-matched \textit{Gaia} sources in the MER DR1 final catalogues we apply a series of selection criteria to remove objects with a bad or compromised \textit{Gaia} or MER photometry:
\begin{itemize}
    \item we select objects that are very likely stars by using $\texttt{CLASSPROB\_DSC\_COMBMOD\_STAR} > 0.99$, which is a \textit{Gaia} quantity measuring the probability of the source being a single star; 
    \item we select single stars using $\texttt{NON\_SINGLE\_STAR} = 0$;
    \item we select objects with sufficient signal-to-noise ratio (S/N) in all \textit{Gaia} bands (${\rm S/N}_{G}>500$, ${\rm S/N}_\mathrm{BP}>25$ and ${\rm S/N}_\mathrm{RP}>35$);
    \item following \cite{Riello2021} and their definitions for the flux excess $C^{\ast}$ and its error $\sigma_{C^{\ast}}$ (equations 6 and 18), we select objects with consistent \textit{Gaia} flux measurements $\abs{C^{\ast}} < \sigma_{C^{\ast}}(G)$;
    \item to avoid saturation in VIS, which in A-PHOT and T-PHOT affects also the photometry in all low-resolution bands, we select VIS-detected sources (using $\texttt{VIS\_DET} = 1$);
    \item to avoid objects affected by incomplete coverage, saturation, ghosts, or other cosmetic defects we de-select objects with holes in the central region ($\ang{;;1.5} \times \ang{;;1.5}$ for VIS, $\ang{;;2.1} \times \ang{;;2.1}$ for the NISP bands, and $\ang{;;5.1} \times \ang{;;5.1}$ for the EXT bands) in the imaging data;
    \item finally, we select only sources below the band-specific saturation limit and with a high ${\rm S/N}$ detection (see Table~\ref{tab:relative_photometry} for the precise magnitude limits applied to each band).
\end{itemize}

\begin{figure}[h!]
\centering
\includegraphics[width=0.9\linewidth]{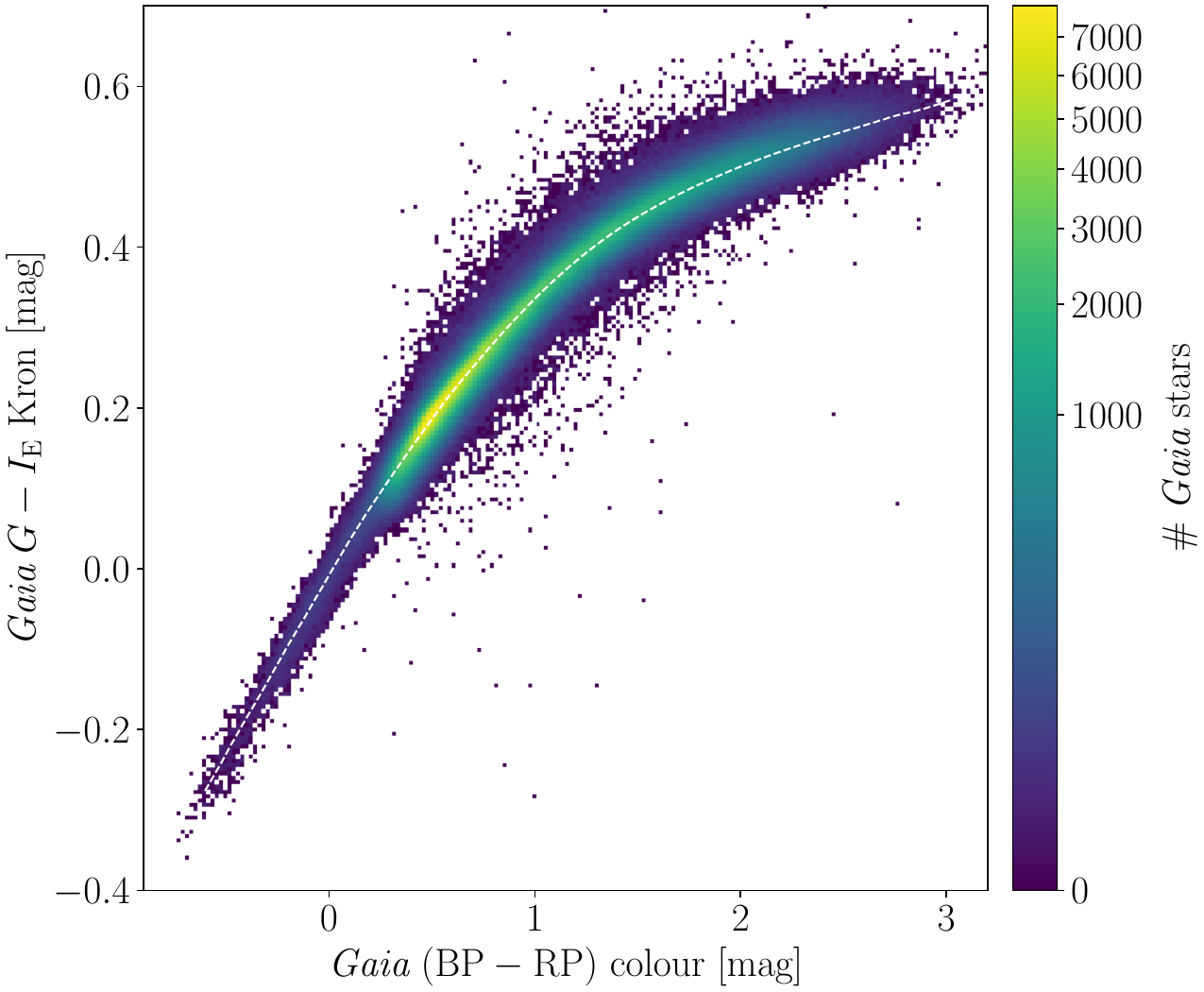}
\caption{The difference between the \textit{Gaia} $G$-band magnitude and the MER VIS Kron magnitude as a function of the \textit{Gaia} $BP - RP$ colour for a sample of 1.9 million \textit{Gaia} stars in the EWS DR1 survey. All magnitudes are in the AB photometric system. The colour bar has been chosen to highlight low-density regions and outliers. The dashed white curve shows the result of applying a smoothed median fit to the data. As reported in Table~\ref{tab:relative_photometry}, the scatter along the curve has NMAD around $15$~mmag in the vertical direction.}
\label{fig:color_color_plot}
\end{figure}

These selections result, depending of the \Euclid band under consideration (see Table~\ref{tab:relative_photometry}), in sample sizes of 0.2 to 1.9 million \textit{Gaia} stars over the entire EWS released in DR1.
\begin{table}[t]
    \centering
    \caption{Relative photometry errors in the MER EWS DR1 catalog.}    
    \begin{tabular}{lccc}
        \hline
        \hline
        Band & Sample size & Magnitude range & NMAD \\
          & $(\times 10^6$ sources) & (mag) & (mmag) \\
        \hline
        \IE &  1.90  & (21.5, 16.2) & 15 \\
        \YE &  1.05  & (22.0, 16.8) & 35 \\
        \JE &  0.95  & (22.0, 16.8) & 40 \\
        \HE &  1.10  & (22.5, 16.7) & 47 \\
        DECam-$g$ &  1.41  & (20.0, 15.7) & 27 \\
        DECam-$r$ &  1.46  & (19.5, 15.8) & 22 \\
        DECam-$i$ &  1.40  & (18.7, 16.0) & 21 \\
        DECam-$z$ &  1.31  & (18.5, 16.0) & 26 \\
        MegaCam-$u$ &  0.07  & (19.5, 15.5) & 97 \\
        MegaCam-$r$ &  0.22  & (19.5, 17.6) & 26 \\
        HSC-$g$ &  0.35  & (20.0, 17.2) & 33 \\
        HSC-$z$ &  0.19  & (20.5, 17.2) & 29 \\
        PS-$i$ &  0.22  & (17.5, 14.0) & 20 \\
        \hline
    \end{tabular}
    \tablefoot{The normalized median absolute deviation (NMAD) values were calculated using the residuals to the fitted median curves to the MER aperture (\textsc{A-PHOT}) photometry measurements and the \textit{Gaia} DR3 magnitudes. See Sect.~\ref{subsubsect:relative_photometry} for more details.}
    \label{tab:relative_photometry}
\end{table}
As illustrated in Fig.\ \ref{fig:color_color_plot} for the \IE band (VIS Kron flux), the selected objects follow a tight relation in the colour-colour diagram $G-\IE$ vs. $\mathrm{BP} - \mathrm{RP}$, with few outliers only. The white dashed curve shows a smoothed median fit to the data.

The residuals from these curves measure the homogeneity and stability of the MER photometry over the EWS data released in DR1. Since the residuals are based on a fit to the median of the measured values, they only give an estimate of the relative error in the MER photometry. Table
~\ref{tab:relative_photometry} lists the residuals NMAD for each \Euclid band in the Euclid DR1 EWS. The values range from 15\,mmag in \IE up to 97\,mmag in the UNIONS CFIS $u$ band.

We should note that these error estimates in Table \ref{tab:relative_photometry} also include the \textit{Gaia} relative photometric error, which is estimated as $\leq$3\,mmag \citep{Riello2021}. Our method also assumes that the {\it Gaia} photometry can be used to estimate the photometry in the \Euclid bands, which might not always be the case for all bands (i.e. MegaCam-$u$). Finally, differential extinction can also contribute to the scatter in the colour-colour plots.
\section{\label{sect:psf_discussion}Discussion of the PSF treatment in DR1}
Usually the PSF model, which is provided over a limited area for a combination of imaging throughput (combination of mirror, instrument, filter,...), is given such that
\[
 S = \sum_{i} {\rm psf}_i = 1.0
\]
with ${\rm psf}_i$ the value of the PSF at pixel $i$. For runtime and memory reasons, the MER Q1 and DR1 pipelines are not able to digest PSFs with stamp sizes larger than \ang{;;5}. If provided with larger input PSFs, we cut out from the centre the maximum feasible size, and for the stamp images it is then $S_{\rm stamp}<1.0$. This means that not all flux in the PSF is collected; for example, $S_{\rm stamp}=0.9$ would indicate that 90\% of the flux is collected inside the extracted PSF stamp region, while 10\% the flux falls outside. In the Q1 processing we had re-normalized the PSF stamps and forced $S=1.0$ for all PSF related processing. In the Q1 photometry validation this procedure resulted in significant offsets of up to $10\%$ in bands with larger PSFs in the Pan-STARRS $i$ band, which is generally the band with the larger PSF sizes. To correct these offsets, it was decided not to re-normalise any PSF in DR1.

As a result of this decision, the resulting flux $F$, computed for an object, is changed. Comparing the flux values using normalized (n) and non-normalized (nn) PSFs the results are to the first order
\begin{itemize}
    \item for both \texttt{A-PHOT} and \texttt{T-PHOT} in the target band $B$
    \begin{equation}F^{\rm nn}_{B} = F^{\rm n}_{B}/(S^{\rm nn}_{B}/S^{\rm nn}_{D})\end{equation}
    \item for S\'ersic in band $B$
    \begin{equation}F^{\rm nn}_{B} = F^{\rm n}_{B}/S^{\rm nn}_{B}\end{equation}
    \item for the VIS PSF fitting
    \begin{equation}F^{\rm nn}_{\IE} = F^{\rm n}_{\IE}/S^{\rm nn}_{\IE}\end{equation}
\end{itemize}
with the detection band $D$ (either \IE or NIR). The Kron photometry in the detection band (column \verb|FLUX_DETECTION_TOTAL|) does not depend on the PSF model and its normalization.

The MER PF gets as input PSF models with variable shapes and normalisation values $S$ across the FOV for each input band. The forward processing of these PSF models in MER as described in Sect.\ \ref{subsect:psf_calculation} and \cite{Q1-TP004} section 5 results in MER PSF models with variable shape and normalization $S^{\rm nn}_{B}(x,y)$ across the tile area. To provide typical values for representative tiles, Table \ref{tab:psf_norm} lists as key statistics for the normalization values, the median, NMAD, and the quantiles corresponding to $2.0\,\sigma$ for all bands of tiles $102020534$ and $102159192$. While all input data delivered by OU-NIR and OU-VIS to the MER PF have the same PSF models \citep[see][]{DR1-TP002, DR1-TP003} in the \Euclid bands, the normalization values for the MER coadds still change from tile to tile since the mapping, orientation, and layering of the \Euclid observations on each tile are different. For the EXT bands the variable observing conditions lead to tile-specific normalization values.

Using the non-normalised PSFs is not the best choice in all
cases. e.g., large extended sources,
but by not normalizing the PSF stamps, we managed to solve photometry issues in DR1, such as the offset for Pan-STARRS $i$ in Q1. Note that, while the Pan-STARRS photometry has globally improved, this dataset suffered from defects in local, isolated areas, which led to its exclusion in an additional, separate processing of the photometric redhifts, as explained in \cite{DR1-TP011}.

\begin{figure}[h!]
\centering
\includegraphics[width=0.99\linewidth]{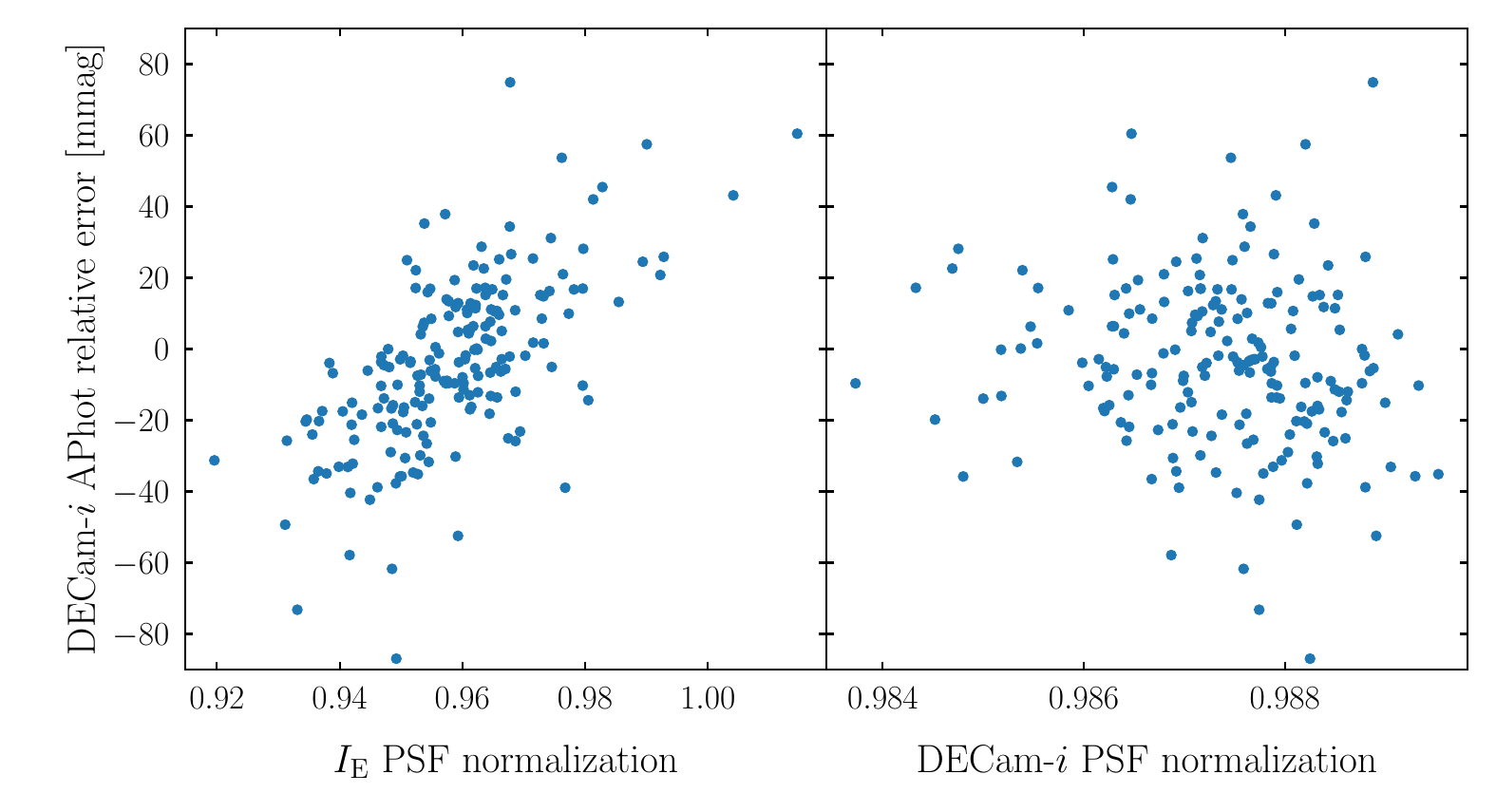}
\caption{The relative photometric offset in DECam-$i$ for the \textit{Gaia} stars in tile $102020534$ as function of the PSF normalization in \IE (left) and DECam-$i$ (right).}
\label{fig:decam_aphot_norm}
\end{figure}

A careful analysis of the \textit{Gaia} photometric offsets from Sect.\ \ref{subsubsect:relative_photometry} does show a strong dependency on the PSF normalization value. This is illustrated in Fig.\ \ref{fig:decam_aphot_norm}, which shows the \texttt{A-PHOT} offset in tile $102020534$ for DECam-$i$ as a function of the \IE and DECam-$i$ PSF normalization in the left and right panel, respectively. The strong dependency on the PSF normalization in \IE is evident. Extending to larger areas of DR1 and other bands confirms that the \IE PSF normalization is currently the limiting factor of the photometric accuracy in DR1, according to the equations above.
\begin{table}[t]
    \centering
    \caption{Median, NMAD, and the quantiles of the PSF normalization values that correspond to $2.0\,\sigma$ for all VIS, NIR, and EXT bands in the tile s $(102)020534$ (EDF-S) and $(102)159192$ (EDF-N).}
    \begin{tabular}{lcccccccc}
        \hline
        \hline
        Band & Median & NMAD &$q2.77,q97.23$\\
        \hline
020534/\IE & 0.957& 1.13e-02 & ($0.936,0.980$) \\
020534/\YE & 0.985& 7.41e-03 & ($0.970,1.000$) \\
020534/\JE & 0.984& 6.76e-03 & ($0.969,0.999$) \\
020534/\HE &  0.984& 5.26e-03 & ($0.971,0.994$) \\
020534/DECam-$g$ & 0.974& 2.89e-03 & ($0.969,0.979$) \\
020534/DECam-$r$ & 0.982& 3.12e-03 & ($0.977,0.988$) \\
020534/DECam-$i$ & 0.987& 1.09e-03 & ($0.985,0.989$) \\
020534/DECam-$z$ & 0.980& 1.89e-03 & ($0.976,0.984$) \\
159192/\IE & 0.961& 1.13e-02 & ($0.938,0.983$) \\
159192/\YE & 0.987& 6.58e-03 & ($0.971,1.001$) \\
159192/\JE &  0.987& 6.31e-03 & ($0.971,1.001$) \\
159192/\HE & 0.986& 4.56e-03 & ($0.973,0.996$) \\
159192/MegaCam-$u$ &  1.000& 2.94e-04 & ($0.999,1.000$) \\
159192/HSC-$g$ &  0.996& 6.91e-04 & ($0.994,0.997$) \\
159192/MegaCam-$r$ & 0.999& 2.99e-04 & ($0.999,1.000$) \\
159192/PS-$i$ & 0.880& 3.72e-03 & ($0.868,0.887$) \\
159192/HSC-$z$ & 0.999& 8.24e-04 & ($0.997,1.000$) \\
%
        \hline
    \end{tabular}
    \label{tab:psf_norm}
\end{table}

Similar to our results, also \cite{DR1-TP002} identify the \IE PSF normalization as the main contributor to the error in their PSF-based photometry. 
We note that for \IE based colours this dependency cancels out for both \texttt{A-PHOT} and \texttt{T-PHOT}.
\section{\label{sect:deepwide}Comparison of objects in EWS and EDF coverage}
In the \Euclid survey design some sky areas are observed multiple times.
Comparing the processing results from different data sets on the same tile allows an assessment of the statistical properties for detection and deblending as well as the accuracy  and general systematics of the object properties measured on the same objects with different input data.
\subsection{\label{subsect:deepwide_detection}Wide versus deep object detection and deblending}
\begin{figure}
    \centering
    \includegraphics[width=0.48\textwidth]{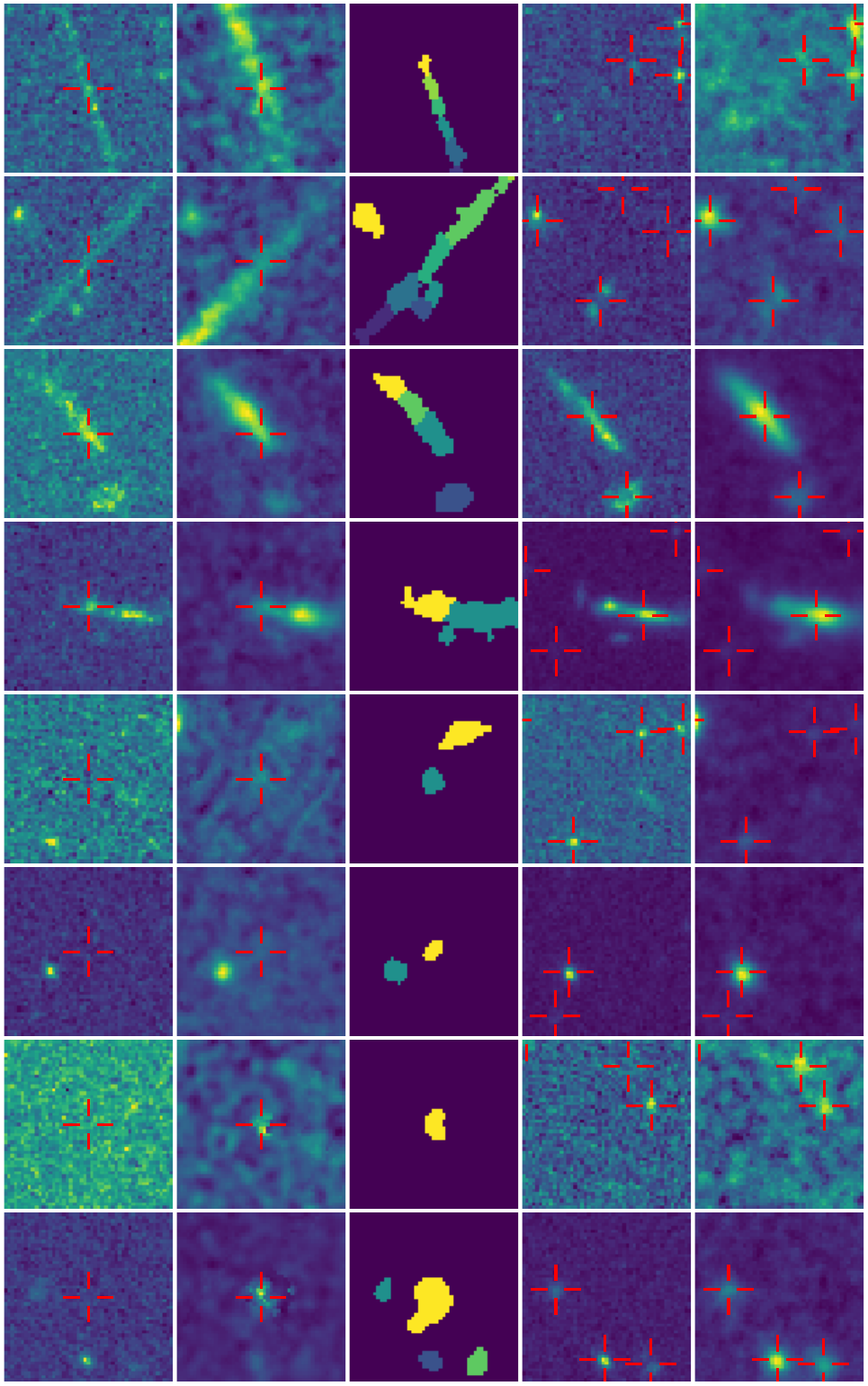}
    \caption{Examples of wide objects without a counterpart in the deep catalogues. From left to right the columns show the $5^{\prime\prime}\times5^{\prime\prime}$ large cutout images in wide VIS detection, wide NIR detection, wide segmentation, deep VIS detection and deep NIR detection. The crosshairs show the target object position on the wide cutouts and all detected objects on the deep cutouts. The first two rows show VIS `objects' from diffraction spikes, rows 3 and 4 mismatches due to different deblending, rows 5 and 6 diffuse NIR detections, and rows 7 and 8 NIR detections created by unmasked bad pixels.}
    \label{figure:wide_vs_deep_objs}
\end{figure}
For this comparison, we selected the EWS processed tile 102020534
(with exposure times of about $2.5,\,0.35,\,0.35,\,0.35\,\mathrm{ks}$ in \IE, \YE, \JE, \HE),
which is located at EDF-S. We built a matching deep catalogue by combining the objects lists of the nine overlapping EDF-S tiles
(with an exposure time of about $27.0,\,2.8,\,2.8,\,2.8\,\mathrm{ks}$ in \IE, \YE, \JE, \HE).
and then selecting the objects according to the core area of the Wide tile. 

Within a maximum distance of $0\overset{\prime\prime}{.}3$
we have $96\,704$ wide objects from the total list of $114\,810$ that find a cross-match in the deep list.
In this tile, $54\%$ of the unmatched wide objects are VIS detected and $46\%$ are NIR detected. Considering that only about $20\%$ of the objects are NIR detected, these sources are clearly overrepresented in the sample of unmatched wide sources.

A close inspection of the imaging data and the various object lists reveal that this seemingly large amount of unmatched wide objects comes from a quite limited number of features which are illustrated in Fig.\ \ref{figure:wide_vs_deep_objs}:
\begin{itemize}
\item diffraction spikes are different in the wide and deep images and the false positive detections along the wide spikes can not be matched in deep images. Note that these objects are marked correspondingly in the \verb|DET_QUALITY_FLAG| (rows 1 and 2);
\item in the wide and deep data the same sky positions are recorded with different S/N, which leads to very different deblending results and many unmatched objects in the wide/deep comparison (rows 3 and 4);
\item some unmatched NIR-detected objects originate from rather diffuse features that are not confirmed in the deep imaging (rows 5 and 6);
\item unrecognized cosmics and bad pixels in the NIR input images create quite a number of false positive NIR detections. Visually they can be recognized by their sharp edges (rows 7 and 8).
\end{itemize}
While the number of isolated false VIS detections is small, the comparably large number of diffuse false positives in NIR indicates too shallow detections limits there. The wide vs. deep object matching also indicates an over-deblending for the wide and an under-deblending for the deep survey data. The optimization of the detection and deblending parameters and especially a differentiation of the parameter for deep and wide is an active research area in preparing for DR2.

\begin{figure}
    \centering
    \includegraphics[width=0.48\textwidth]{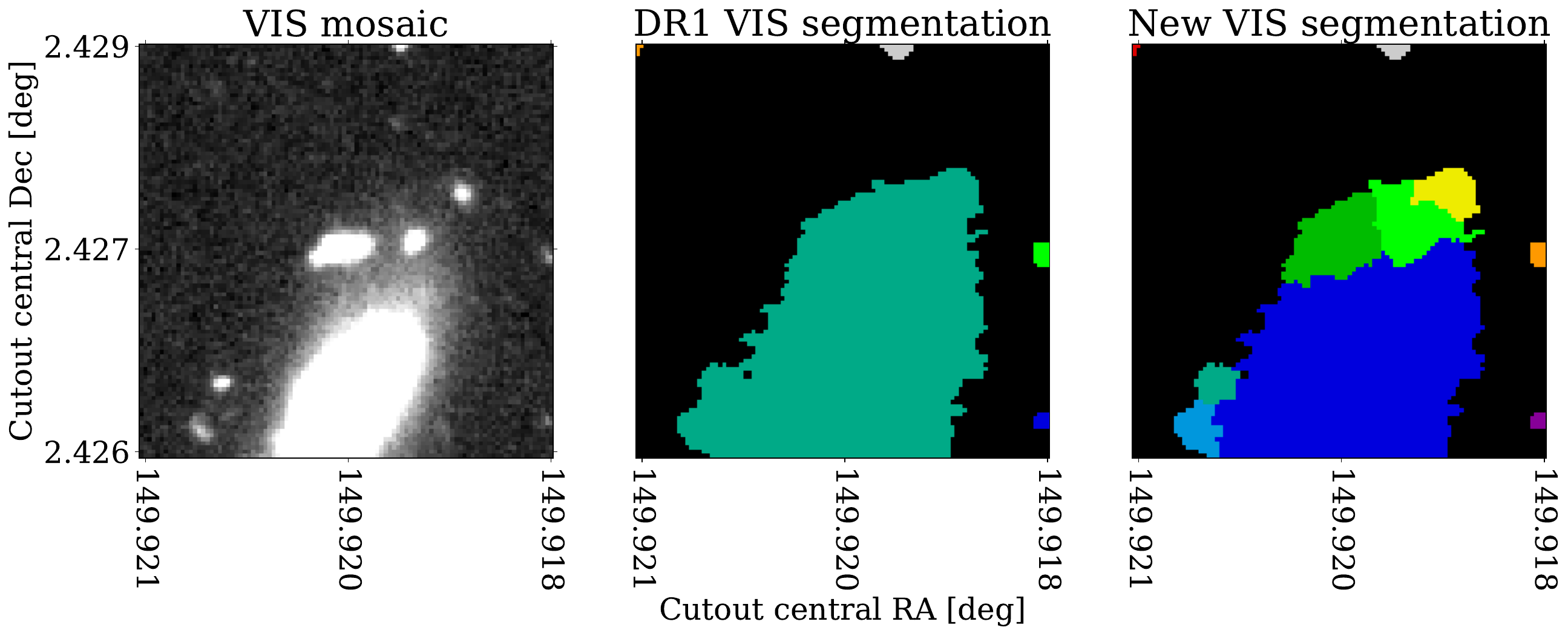}
    \includegraphics[width=0.48\textwidth]{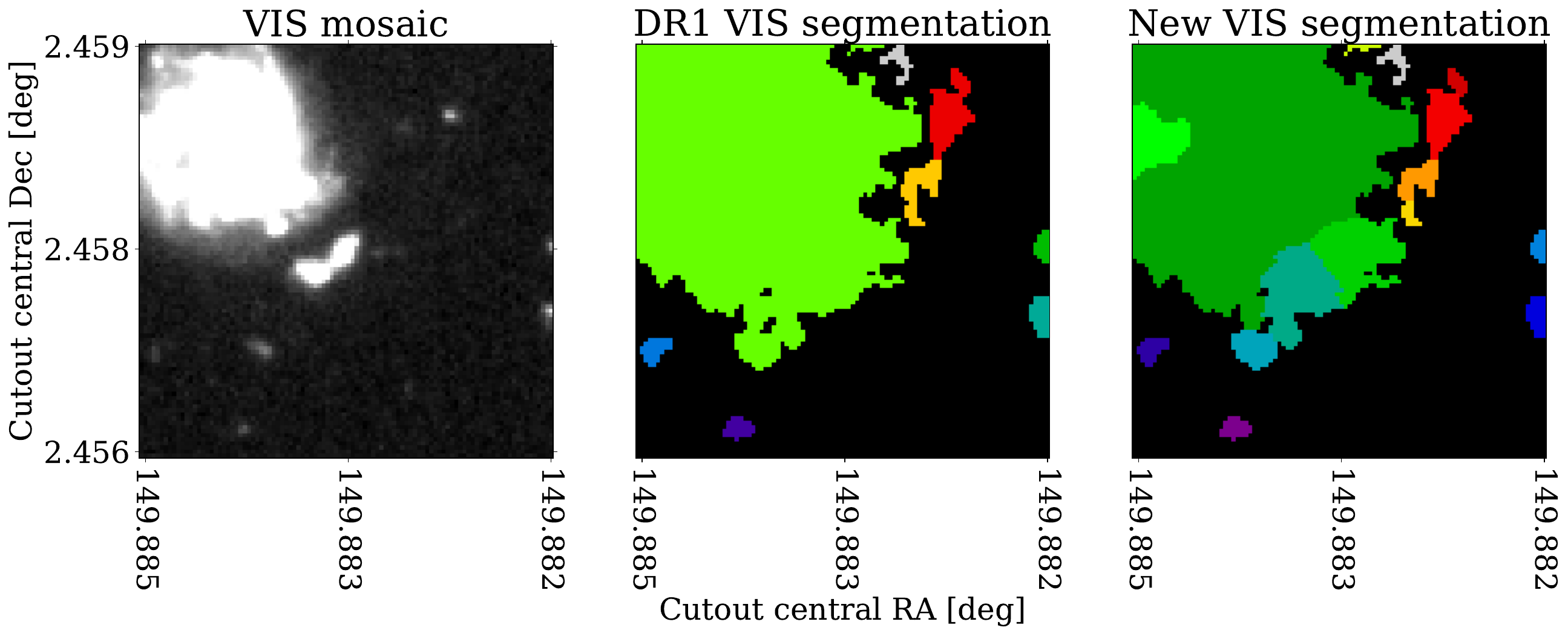}
    \includegraphics[width=0.48\textwidth]{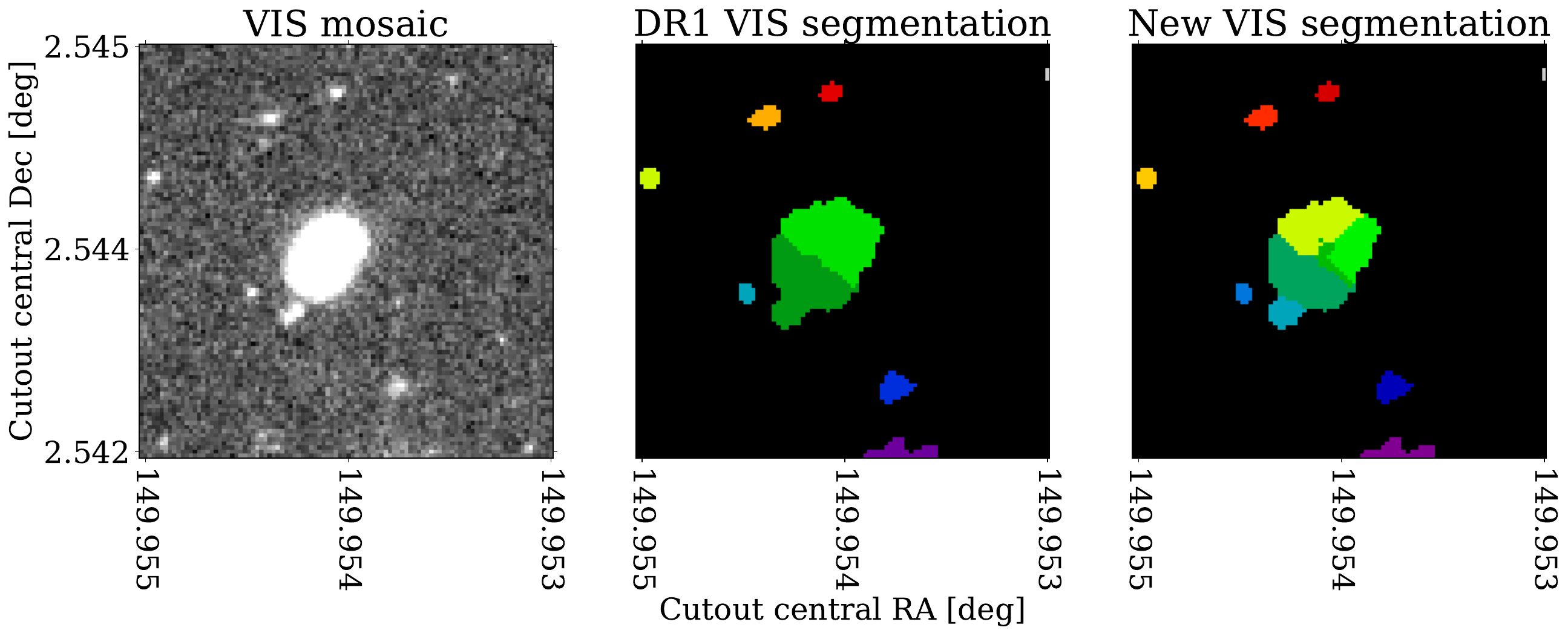}
    \caption{Comparison of the DR1 nominal segmentation of the VIS mosaic and the segmentation obtained with a different configuration of the deblending algorithm. The left columns shows the VIS mosaic produced by the MER pipeline, the central column shows the nominal DR1 segmentation, while the right column shows the segmentation map obtained with a different configuration applied to mitigate the mismatch in MER detections between the wide and deep processing. In the central and right columns, sources are color-coded according to their $\texttt{SEGMENTATION\_MAP\_ID}$. Note that in some cases (see bottom panels), the different deblending parametrisation could lead to over-deblended clusters of pixels.}
    \label{figure:deblending_wide_vs_deep}
\end{figure}

In Fig. \ref{figure:deblending_wide_vs_deep} we report a comparison between the nominal deep DR1 segmentation and the segmentation obtained with a different tuning of the deblending configuration. In particular, we set a smaller value of the deblender scale factor parameter \cite[see][]{Tramacere2016}, enhancing smaller features in the image. It must be noted that, even if a different parametrisation of the deblending algorithm can lead to better segmentation of most of the critical cases, it could also lead to over-deblending in others, see again Fig. \ref{figure:deblending_wide_vs_deep}.

\subsection{\label{subsect:wide_deep_phot}Photometry}
\begin{figure}
    \centering
    \includegraphics[width=0.49\textwidth]{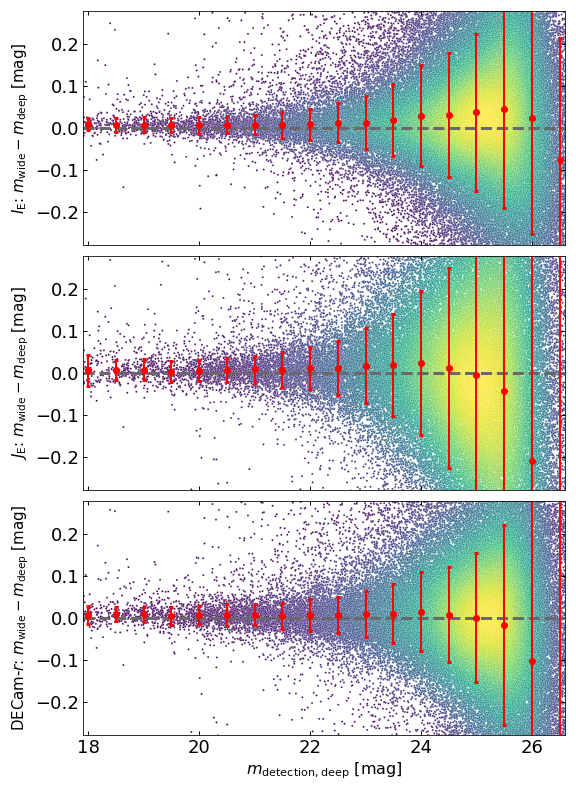}
    \caption{The comparison of the deep and wide \texttt{A-PHOT} photometry of the cross-matched objects of tile 102020534. The panels show the difference between brightness in wide and deep for  \IE, \JE, and DECam-$r$ (from top to down) versus the deep detection flux. The continuous line connects the median values and the NMAD in binned intervals.}
    \label{figure:photometry_wide_vs_deep}
\end{figure}
To investigate the consistency in the MER photometry between EWS and EDF we compare the flux values of the cross-matched objects in the VIS, NIR, and EXT bands. As examples Fig.\ \ref{figure:photometry_wide_vs_deep} shows the difference between the deep and wide \texttt{A-PHOT} AB-magnitudes for \IE (detection), \JE, and DECam-$r$ for the EDF-S for all objects in tile 102020534 (panels upper, middle, and lower, respectively). In all panels the $x$-axis value is the deep detection flux, which places all objects at the same position on the abscissa. The continuous line displays the median and the errors bars the NMAD values in bins of $0.5\,\mathrm{mag}$.

Over a large dynamic range in detection brightness the offset in all three bands is constant and in the range of $0.01\,\mathrm{mag}$ and below. Such small offsets can be explained by according differences in the photometric calibration of the deep and wide images. The trend to negative values in $m_{\rm wide}-m_{\rm deep}$ at faint magnitudes is caused by the Malmquist bias \cite[e.g.][]{2005MNRAS.362..321B}. From the shallower wide data, only the objects that are scattered to brighter flux values due to noise fluctuations enter this sample of objects cross-matched with deep sources. Inconsistencies in either the wide or the deep data, for example due to problems in the background modeling, would create a significant signature in Fig.\ \ref{figure:photometry_wide_vs_deep}.

\begin{figure}[h!]
\centering
\includegraphics[width=0.49\textwidth]{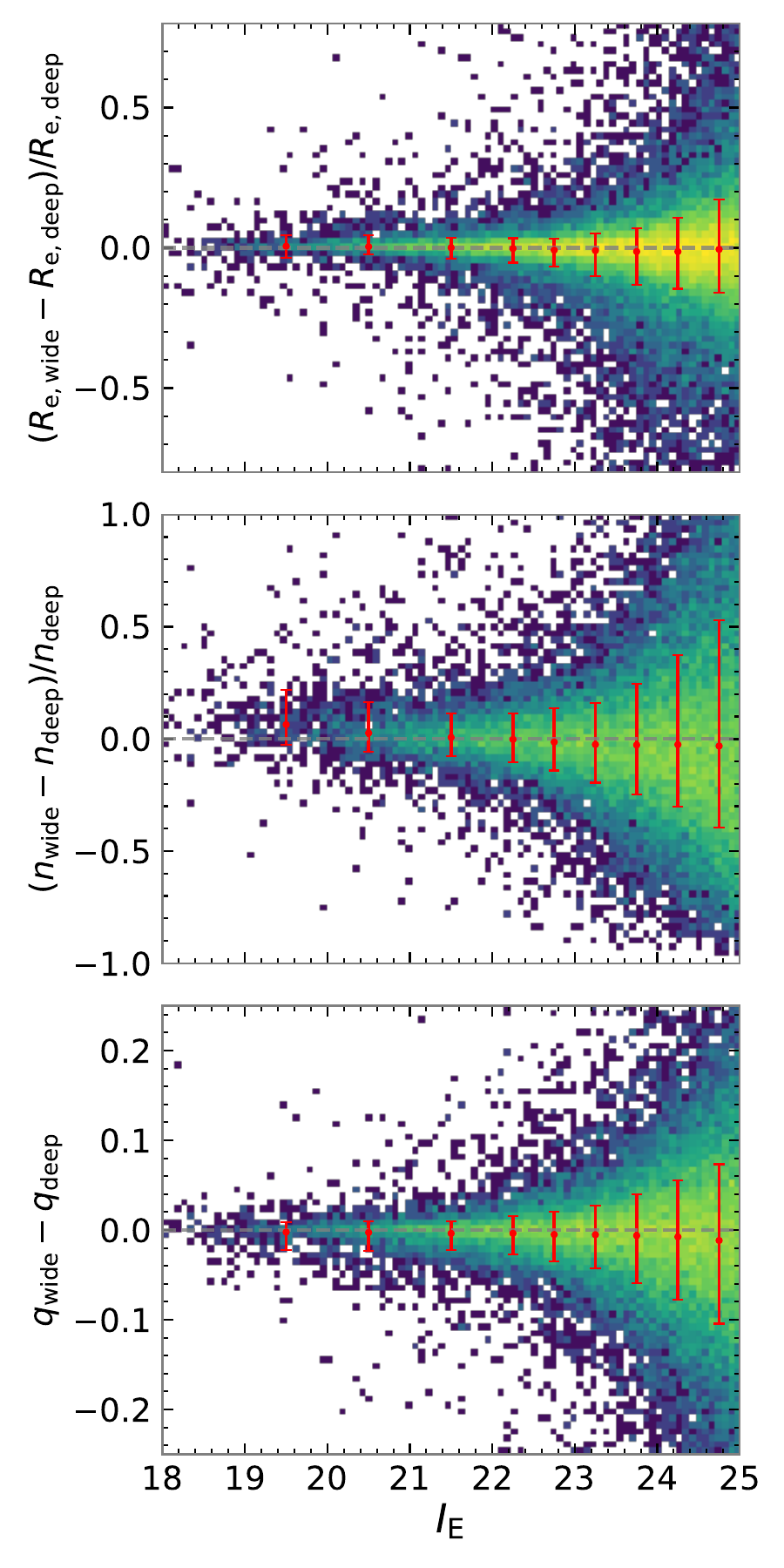}
\caption{Difference in the best-fit S\'ersic parameters in the wide and deep images, as a function of \IE (deep detection) magnitude (S\'ersic $R_{\rm e}$, $n$, and $q$ from upper to lower). There is overall agreement between the measures, confirming that S\'ersic fits provide a robust characterization of morphology, which is not too impacted by image depth.}
\label{fig:morpho_deep_vs_wide}
\end{figure}
\subsection{\label{subsect:object_morph}Object morphology}
With deeper imaging, light profiles of sources extend into the outskirts that were previously below the surface brightness limit, and substructures within galaxies can also be revealed. The characterization of galaxy morphology is therefore dependent on the image depth (as well as on its resolution). For an exhaustive analysis of the impact of image quality on a series of morphological measurements, we refer the reader to \cite{2026OJAp....960562S}. Here we will focus on the impact of image depth on the products of the morphology catalogues described in Sect.\  \ref{subsect:morphology}, by comparing the values measured in the wide and deep images. 

We apply the cleaning steps described in \cite{Q1-SP040} to limit the sample used throughout this section to $56\,194$ valid galaxies, among which $48\,641$ have successful S\'ersic fits in both wide and deep processing (the condition for the S\'ersic index $n < 5.45$ for Q1 is now $n < 7.9$ to account for the new maximum value allowed for $n$). We note that, for objects with $\IE < 23$, which is the recommended limit in the Euclid Morphological Challenge (to remain at a non-significant bias level and $<10\%$ scatter, \citealp[]{Bretonniere-EP26}), only 6.1\% of the galaxies are removed as a result of a poor fit, against 13.4\% at all magnitudes.

\begin{figure*}[h!]
\centering
\includegraphics[width=1.0\linewidth]{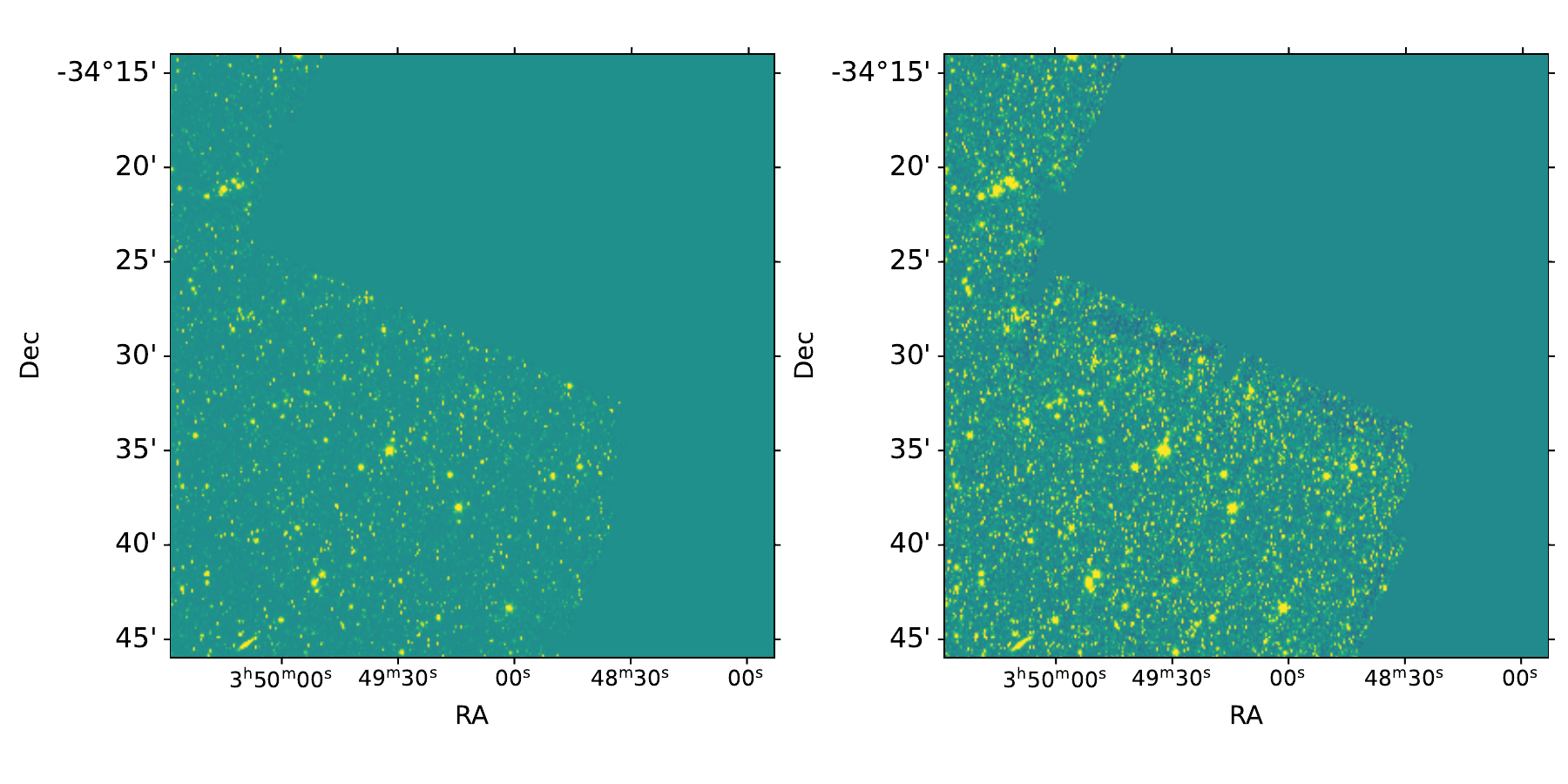}
\caption{The coadded \IE (left) and \JE (right) images of the EWS border tile $102036162$. About $52\%$ of the tile core area is covered.}
\label{fig:border_tile}
\end{figure*}
Figure \ref{fig:morpho_deep_vs_wide} shows the comparison between the effective radius $R_{\rm e}$, the Sérsic index $n$, and the axis ratio $q$
of the best-fit profile derived in the VIS wide and deep images, as a function of the object magnitude (using the \IE deep detection magnitude). S\'ersic parameters are supposed to be robust measurements that are not dependent on image depth and this is confirmed here by the null or negligible median difference in all three parameters. Indeed, the relative difference in $R_{\rm e}$ has an absolute median value below the percent level for \IE$< 23.5$, with 68\% of the objects within a $\pm$10\% difference. For the S\'ersic index, there is a subtle difference for the bright objects in the first two bins, which have a slightly higher $n$ value in the wide than in the deep image, with median relative difference of 6.4\% and 2.6\%. The scatter also increases with magnitude more strongly than for $R_{\rm e}$, For the axis ratio, the absolute median difference remains below 0.012 at all magnitudes (and 0.005 for $\IE < 23.5$). The difference distribution however exhibits a mild but systematic asymmetry, with a consistently larger upper-tail spread than lower-tail spread, as quantified by the distance to the median for the 16th and 84th percentiles of the distribution, with (median$- P_{16}$)/($P_{84} -$median) ranging from 1.09 to 1.81.
We also looked at the differences between the wide and deep parameters as a function of the considered parameter itself (e.g. $n_{\rm wide}-n_{\rm deep}$ versus $n_{\rm deep}$) and found no noticeable trend. The differences measured here are much smaller than the physically meaningful variations usually investigated in studies of the size evolution of galaxies with redshift, or used in the classification of galaxies between late ($n \sim 1$) and early-type ($n \sim 4$) morphologies \citep{Q1-SP040}. This means the S\'ersic parameters from the MER PF provide valuable information for morphological studies at least down to the brightness limit mentioned above.

\section{\label{sect:overview}Overview of the MER DR1 data}
The data selected for the Q1 release were carefully picked from very homogeneous observations of the three EDFs \citep[see][]{Q1-TP001}, and consequently the released MER data are very homogeneous (coverage, depth, etc.) as well. DR1 contains data taken over a much longer time that have experienced a much larger range of observing conditions. While \cite{DR1-TP001} describes the DR1 data selection and the observing conditions, this section discusses the properties and features of the released MER data.
\subsection{\label{subsect:mer_ews}MER EWS data}
%
%
%
\subsubsection{\label{subsubsect:border_tiles}Partially covered tiles}
Since the MER PF runs on tiles which are defined independent from the \Euclid observations (see Sect.\ \ref{subsect:tiling}), there are border tiles that are not fully covered with \Euclid data. Figure \ref{fig:border_tile} displays the \IE and \JE coadded images of the border tile $102036162$ in the left and right panels, respectively. The coverage of this tile is $52\%$, as measured via the fraction of \texttt{HEALPix} indices containing detected objects. $7803$ ($88\%$) of the 8887 tiles released in DR1 have a coverage $>80\%$. Note also that the border area is covered by a decreasing number of image dithers which results in a decreasing S/N.
\subsubsection{\label{subsubsect:inhomo_tiles}Tiles with inhomogeneous \Euclid coverage}
\begin{figure*}[h!]
\centering
\includegraphics[width=0.49\linewidth]{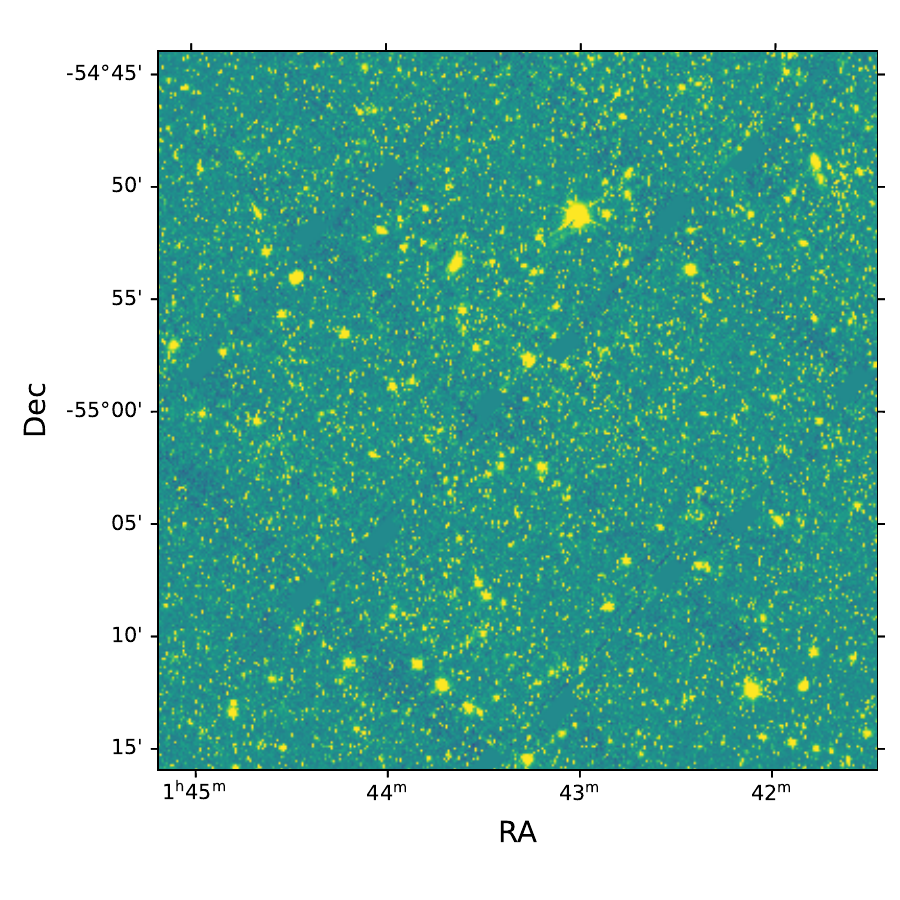}
\includegraphics[width=0.49\linewidth]{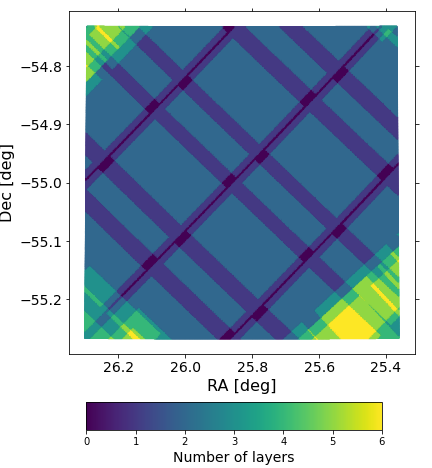}

\caption{{\bf Preliminary }The coadded \JE image of tile $102015130$ (left), which has a shallow coverage, as shown on the layer image (right).}
\label{fig:shallow_tile}
\end{figure*}
The MER PF is run on tiles that are covered with VIS, NIR, and EXT data. Due to adverse conditions such as proton showers coming from the Sun, some \Euclid data is deteriorated and the quality criteria in the VIS and NIR PF's are not met. If this is the case for part of the data, the resulting imaging is shallower than intended. This is shown in Fig. \ref{fig:shallow_tile} for the \JE-band coadd of $102015130$. As can be seen on the layer image in the right panel, most of the image is covered with only two data layers instead of the usual four.

\begin{figure*}[h!]
\centering
\includegraphics[width=1.0\linewidth]{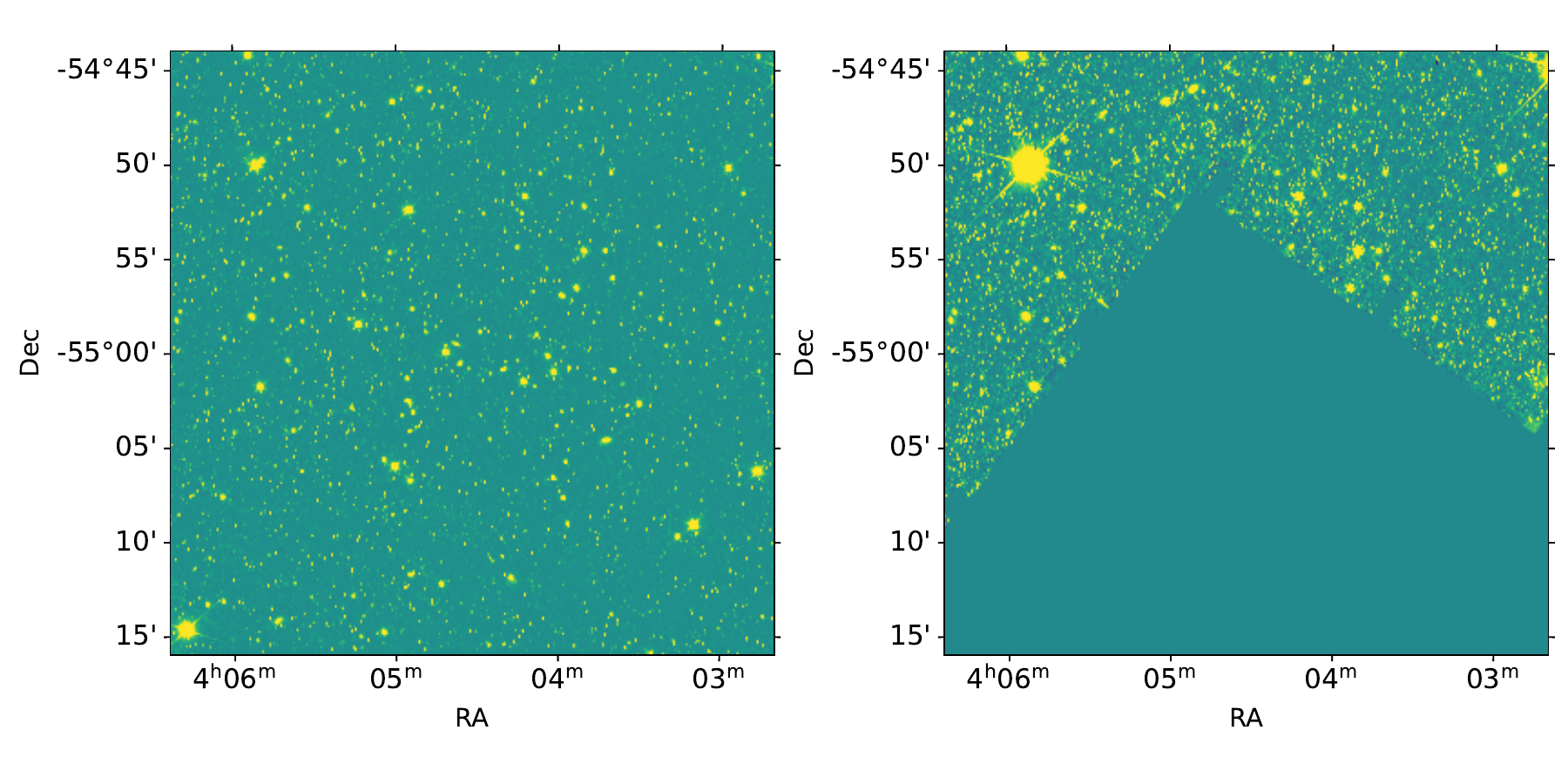}
\caption{The coadded \IE (left) and \JE (right) images of the EWS tile $102015171$. About $52\%$ of the tile core area is covered.}
\label{fig:inhomo_tile}
\end{figure*}

More severe data deterioration can also lead to an incomplete coverage in the \Euclid bands, which means that part of the tile is not covered in some bands. As an example, Fig.\ \ref{fig:inhomo_tile} displays for tile $102015171$ the \IE image on the left panel, which is fully covered, and the \JE image on the right, which is filled to about $50\%$. For this tile the coverage in \YE and \HE is the same in \JE. Defining a tile as incomplete if the coverage in a band is more than $10\%$ smaller than the combined coverage in VIS and NIR, then there are $34$, $405$, $560$, and $510$ incompletely covered tiles in \IE, \YE, \JE, and \HE, respectively.
%
\subsection{\label{sect:mer_edf_plus}MER data for EDS, calibration and extra fields}
\begin{table*}[t]
    \caption{An overview on MER processing results for the all fields of DR1}
    \centering
\begin{threeparttable}
\begin{tabular}{llrrrrrrl}
 \hline
 \hline
Field&tiling&number&exposure&exposure&exposure&exposure&source&EXT data\\
 & mode &of tiles\tnote{a}&time \IE\tnote{b}&time \YE\tnote{b}&time \JE\tnote{b}&time \HE\tnote{b}&density\tnote{c}& \\
\hline 
EWS & wide & 8388 & 2274.0 & 314.1 & 312.7 & 314.1 & 116.2 &PS-$i$ HSC-$g/z$ MegaCam-$u/r$ DECam-$g/r/i/z$\\
EDF-N & deep & 474 & 20827.7 & 2845.3 & 2813.1 & 2841.5 & 218.2 &HSC-$g/r/r2/i/i2/z$\\
EDF-S & deep & 547 & 20265.2 & 2809.2 & 2821.1 & 2821.7 & 216.7 &DECam-$g/r/i/z$\\
EDF-F & deep & 222 & 4018.7 & 559.5 & 558.8 & 557.4 & 129.8 &DECam-$g/r/i/z$ HSC-$g/r2/i/i2/z$\\
CDFS & deep & 15 & 45847.4 & 6215.6 & 6144.2 & 6147.6 & 244.2 &DECam-$g/r/i/z$ HSC-$g/r2/i/i2/z$\\
COSMOS & deep & 46 & 51853.2 & 6312.5 & 6338.5 & 6310.5 & 232.4 &DECam-$g/r/i/z$ HSC-$g/r/r2/i/i2/z$\\
AEGIS & deep & 27 & 28953.0 & 4033.2 & 4034.1 & 4033.9 & 245.6 & \\
SELFCAL\tnote{d} & wide & 161 & 7804.4 & 1043.7 & 1042.7 & 1043.1 & 152.6 &MegaCam-$u/r$ PS-$i$ HSC-$g/z$\\
F-014 & wide & 137 & 4929.0 & 662.7 & 670.1 & 670.5 & 136.5 &MegaCam-$u/r$ PS-$i$ HSC-$g/z$\\
PN & wide & 10 & 2179.7 & 288.7 & 288.6 & 288.7 & 315.0 & \\
 SSO-ECLIPTIC & wide & 82 & 9502.4 & 1250.7 & 1253.6 & 1230.4 & 128.5 & \\
EMDS-HSC-HL-3 & wide & 54 & 9845.0 & 1059.8 & 996.4 & 1056.6 & 158.2 &HSC-$g/r2/i2/z$\\
EMDS-HSC-HL-4 & wide & 50 & 8477.9 & 1404.3 & 1341.6 & 1404.3 & 165.2 & HSC-$g/r/r2/i2/z$\\
ILS-WAVES-S & wide & 160 & 4207.7 & 541.8 & 538.1 & 546.7 & 116.6 & \\
\hline
\end{tabular}
\begin{tablenotes}
\item[a] With a data coverage $>0.3$.
\item[b] Median of the tile exposure times in \rm{[s]}.
\item[c] In \rm{[arcmin$^{-2}$]}.
\item[d] For the MER processing the self-calibration data was sub-divided in 16 chunks covering the same field.
\end{tablenotes}
\end{threeparttable}
    \label{tab:field_overview}
\end{table*}
In addition to EWS data, the DR1 also contains some additional fields with generally much deeper data compared to the EWS. Next to the three EDFs (EDF-South, Fornax and North with 10, 2 and 11 EWS visits) there are the photo-$z$ calibration fields CDFS, COSMOS, and the colour gradient field AEGIS. Then there are the three instrument calibration fields \citep[see][]{DR1-TP026} self-calibration (SELFCAL), contamination (F-014) and wavelength dispersion (PN), and four fields (SSO-ECLIPTIC, EMDS-HSC-HL-3, EMDS-HSC-HL-4 and ILS-WAVES-S) that were observed during the phase diversity calibration (PDC) thermal stabilization campaign \citep[see][]{DR1-TP001}. These additional fields span a wide range in terms of size and observing time and all have been processed with the MER PF as presented in this paper. Table \ref{tab:field_overview} lists the main features of the MER processed data for EWS and the additional fields.

\begin{figure}[h!]
\centering
\includegraphics[width=0.45\textwidth]{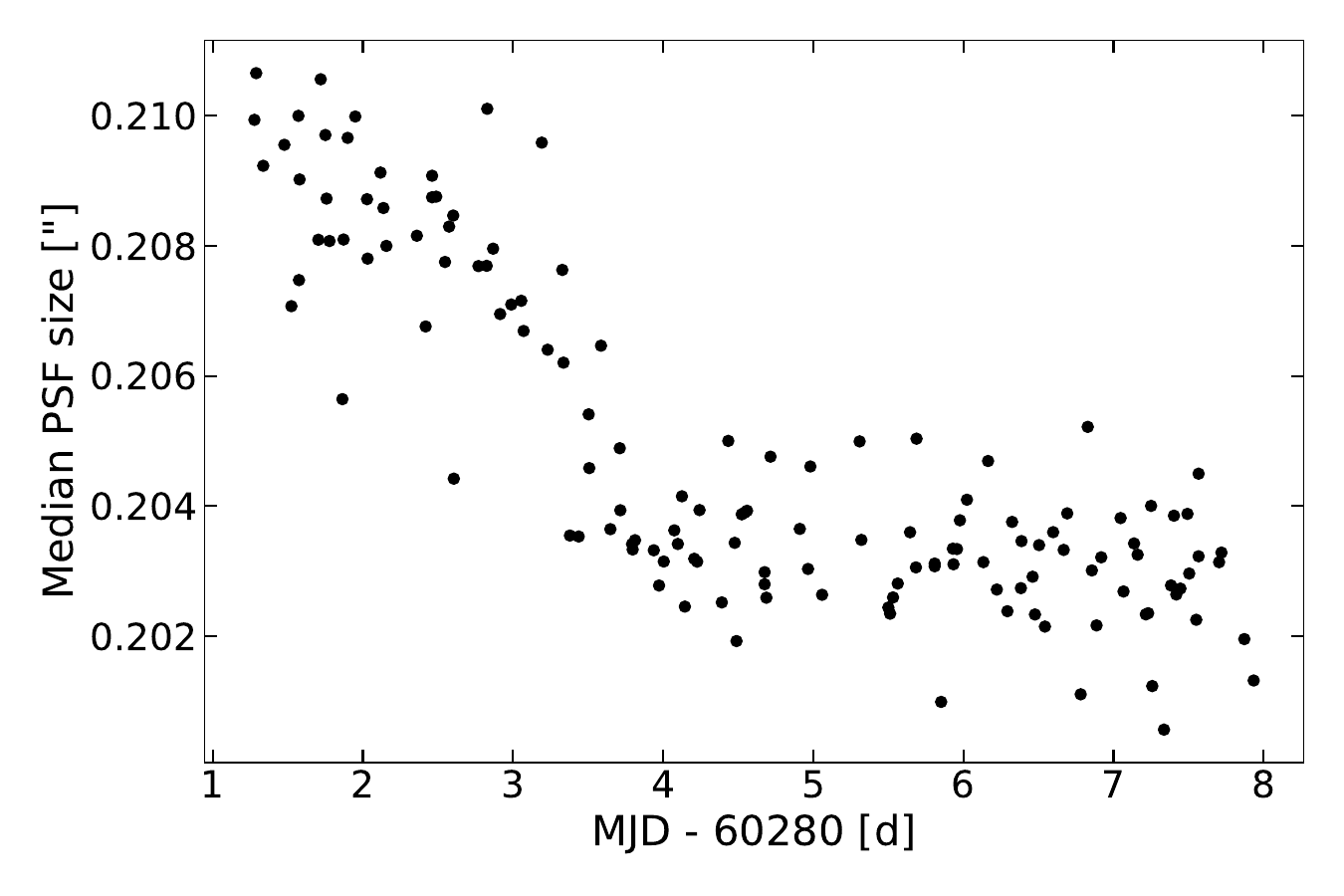}
\caption{The PSF size of the {\it Gaia} sources in the tiles of ILS-WAVES-S as a function of  observing time.}
\label{fig:ils_psf}
\end{figure}

When validating the field ILS-WAVES-S we found that the PSF size as measured on the {\it Gaia} stars in \IE (see Sect.\ \ref{subsect:psfvalidation}) for each tile changed from 
$0\overset{\prime\prime}{.}210$ at the beginning of the observations to about 
$0\overset{\prime\prime}{.}203$ at the middle of the observation period (see Fig.\ \ref{fig:ils_psf}), which is caused by the final relaxation of the VIS instrument after the de-focused PDC campaign \cite[see][]{Neumann26}.

\section{\label{sect:conclusions}Conclusions and outlook}

In this paper we have presented an overview of the OU-MER pipeline that has been used to process the data for the Euclid DR1. The photometric and morphological measurements which represent the output of the OU-MER pipeline are being used by the downstream OUs in the \Euclid SGS (OU-PHZ, OU-SIR, OU-SHE, and OU-LE3) for photometric and spectroscopic redshift estimates, shear measurements, and ultimately the measurements of cosmological quantities. Their results are presented in a series of dedicated papers \citep[see][for a detailed overview]{DR1-TP001}.

DR1 data come with some caveats and known issues that had been identified by members of the SGS and the Euclid Consortium in general. A description of how to work with OU-MER products in the DR1 dataset as well as the photometry and morphology cookbooks can be found in the DR1 Explanatory Supplement\footnote{\url{http://po.pages.euclid-sgs.uk/ExplanatorySupplement/release-DR1-Foundation/}}. We also offer additional metadata such as tile indices, \Euclid exposure times, detection numbers and similar for all MER data and link to python scripts to enhance the object catalogues.

The OU-MER pipeline is evolving with time with the goal to improve its output products. In addition to the dedicated processing for data releases such as Q1 or DR1, these improved versions are being tested in the \Euclid SGS on-the-fly (OTF) processing \citep{DR1-TP033} which is continuously run on new incoming data from the \Euclid satellite. For DR2 we are targeting improvements in coadding of the \Euclid images, the object deblending and the identification of spurious sources.

The pipeline validation described in Sect.\ \ref{sect:validation}, the photometry validation with outside catalogues in \cite{DR1-TP010}, and feedback from the downstream OUs tell us that our data, and in particular our photometry, is of sufficient quality to support DR1 science.
As part of the continuous pipeline development for the OTF processing and with dedicated tests we are aiming to improve the consistency of our photometry to meet the tighter constraints of the next and more extended Data Release 2.
\begin{acknowledgements}
\AckEC \AckDRone See \cite{DR1cite}. We acknowledge the usage of the HyperLeda database (http://leda.univ-lyon1.fr).
\end{acknowledgements}

\bibliography{Euclid, mer_v1}
%
%

%
\begin{appendix}
  \onecolumn 
%
%
\end{appendix}
\end{document}